\documentclass[a4paper,twoside]{article}

\usepackage{times}
\usepackage{a4wide}

\input{fleqn.clo}
\PassOptionsToPackage{fleqn}{amsmath}
\PassOptionsToPackage{fleqn}{amstex}

\usepackage[american]{babel}
\usepackage{xcolor}
\usepackage{graphicx}
\usepackage{sidecap}
\usepackage[format=plain,indention=0.5cm]{caption}

\usepackage{multicol}

\usepackage{ifthen}
\newboolean{final}
\setboolean{final}{true}
\newboolean{preprint}
\setboolean{preprint}{true}

\newboolean{notfinal}
\setboolean{notfinal}{true}
\iffinal
\setboolean{notfinal}{false}
\fi
\newboolean{notpreprint}
\setboolean{notpreprint}{true}
\ifpreprint
\setboolean{notpreprint}{false}
\fi

\newboolean{adpversion}
\setboolean{adpversion}{false}
\newboolean{notadpversion}
\setboolean{notadpversion}{true}

\def\etitle{Canonical Formulation of Spin\\in General Relativity}

\newcommand{\mychapter}[1]{\chapter{#1}}

\newcommand{\mysection}[1]{\section{#1}}

\newcommand{\mysubsection}[1]{\subsection{#1}}
\newcommand{\mysubsubsection}[1]{\subsubsection{#1}}

\def\GR{general relativity}
\def\SR{special relativity}
\def\eqname{Eq.}
\newcommand{\Eq}[1]{Eq.\ (\ref{#1})}

\def\chapname{chapter}
\def\chapsname{chapters}
\def\secname{section}
\def\secsname{sections}
\newcommand{\Chap}[1]{{\chapname} \ref{#1}}
\newcommand{\Sec}[1]{{\secname} \ref{#1}}

\usepackage{xcolor}
\usepackage{amssymb}
\iffinal
\newcommand{\myremark}[1]{}
\fi
\providecommand{\myremark}[1]{{\color{red}$\bullet$}\marginline{\color{red}#1}}

\ifnotfinal

\fi

\usepackage{amsmath}
\usepackage{amsfonts}
\usepackage{amssymb}
\allowdisplaybreaks

\def\nl{\\ & \quad}
\def\nlq{\\ & \qquad \quad \;}
\def\vct#1{\mathbf{#1}}

\def\picSin{((\hat{\mathbf{p}}_1 \times \hat{\mathbf{S}}_1) \cdot \hat{\mathbf{n}}_{1 2})}
\def\piicSin{((\hat{\mathbf{p}}_2 \times \hat{\mathbf{S}}_1) \cdot \hat{\mathbf{n}}_{1 2})}
\def\piicSiin{((\hat{\mathbf{p}}_2 \times \hat{\mathbf{S}}_2) \cdot \hat{\mathbf{n}}_{1 2})}
\def\picSiin{((\hat{\mathbf{p}}_1 \times \hat{\mathbf{S}}_2) \cdot \hat{\mathbf{n}}_{1 2})}
\def\picSipii{((\hat{\mathbf{p}}_1 \times \hat{\mathbf{S}}_1) \cdot \hat{\mathbf{p}}_2)}
\def\SiSii{(\hat{\mathbf{S}}_1 \cdot \hat{\mathbf{S}}_2)}
\def\SiSi{\hat{\mathbf{S}}_1^2}
\def\pipii{(\hat{\mathbf{p}}_1 \cdot \hat{\mathbf{p}}_2)}
\def\Sipi{(\hat{\mathbf{S}}_1 \cdot \hat{\mathbf{p}}_1)}
\def\Siipii{(\hat{\mathbf{S}}_2 \cdot \hat{\mathbf{p}}_2)}
\def\Sipii{(\hat{\mathbf{S}}_1 \cdot \hat{\mathbf{p}}_2)}
\def\Siipi{(\hat{\mathbf{S}}_2 \cdot \hat{\mathbf{p}}_1)}
\def\Sin{(\hat{\mathbf{S}}_1 \cdot \hat{\mathbf{n}}_{1 2})}
\def\Siin{(\hat{\mathbf{S}}_2 \cdot \hat{\mathbf{n}}_{1 2})}
\def\pin{(\hat{\mathbf{p}}_1 \cdot \hat{\mathbf{n}}_{1 2})}
\def\piin{(\hat{\mathbf{p}}_2 \cdot \hat{\mathbf{n}}_{1 2})}
\def\pipi{\hat{\mathbf{p}}_1^2}
\def\piipii{\hat{\mathbf{p}}_2^2}

\def\pa{\partial}
\def\dd{\mathrm{d}}
\def\dD{\mathrm{D}}

\def\Riem{R}

\DeclareMathOperator{\Order}{\mathcal{O}}

\setboolean{adpversion}{true}
\setboolean{notadpversion}{false}

\renewcommand{\mychapter}[1]{\section{#1}}

\renewcommand{\mysection}[1]{\subsection{#1}}

\renewcommand{\mysubsection}[1]{\subsubsection{#1}}
\renewcommand{\mysubsubsection}[1]{\paragraph{#1}}

\def\chapname{section}
\def\chapsname{sections}

\numberwithin{equation}{section}

\xdefinecolor{mylinkcolor}{rgb}{0,0,0.5}
\ifpreprint
\usepackage[
	bookmarksnumbered, bookmarksopen, bookmarksopenlevel=2,
	colorlinks=true, breaklinks=false, % breaklinks must be removed for arxiv
	filecolor=mylinkcolor, citecolor=mylinkcolor, linkcolor=mylinkcolor,
	urlcolor=mylinkcolor, menucolor=mylinkcolor,
]{hyperref}
\fi
\ifnotpreprint
\usepackage[breaklinks=true,bookmarksnumbered,bookmarksopen,bookmarksopenlevel=2]{hyperref}
\fi

\usepackage[square,comma,numbers,sort&compress,merge]{natbib}

\begin{document}

\pagestyle{headings}

\title{\etitle\thanks{Dissertation,
	Friedrich-Schiller-Universit\"at, Jena, 2010 (submitted in June).
	Cite as \href{http://dx.doi.org/10.1002/andp.201000178}{Ann.\ Phys.\ (Berlin) \textbf{523}, 296--353 (2011)}.}}

\author{Jan Steinhoff\thanks{Email:~\textsf{jan.steinhoff@uni-jena.de}}
% 		Phone: +49\,3641\,947\,107,
% 		Fax: +49\,3641\,947\,102}
	\bigskip \\
	Theoretisch--Physikalisches Institut, \\
	Friedrich--Schiller--Universit\"at, \\
	Max--Wien--Platz 1, 07743 Jena, Germany, EU}

% must be removed for arxiv
% \hypersetup{pdftitle={Dissertation: \etitlefl},pdfauthor={Jan Steinhoff}}

\pdfbookmark[2]{Title, Abstract, Contents}{title}
\maketitle

\thispagestyle{empty}

\begin{abstract}
% ~
% 
% \begin{center}
% \includegraphics{abstract_fig_adp}\\
% spinning object with canonical variables
% \end{center}
% 
% ~
% 
% \noindent
The present thesis aims at an extension of the canonical formalism of Arnowitt,
Deser, and Misner from self-gravitating point-masses to objects with spin.
This would allow interesting applications, e.g., within the post-Newtonian (PN) approximation.
The extension succeeded via an action approach to linear order in the single
spins of the objects without restriction to any further approximation.
An order-by-order construction within the PN approximation is possible and
performed to the formal 3.5PN order as a verification.
In principle both approaches are applicable to higher orders in spin.
The PN next-to-leading order spin(1)-spin(1) level was tackled, modeling the
spin-induced quadrupole deformation by a single parameter. %for each object.
All spin-dependent Hamiltonians for rapidly rotating
bodies up to and including 3PN are calculated.
\end{abstract}

\newpage

\tableofcontents
% \newpage

\ifnotadpversion
\automark[chapter]{chapter}
\fi

\mychapter{Introduction}
Though {\GR} has seen and passed many experimental tests, one of its
most fascinating predictions, namely gravitational waves, has not been
observed \emph{directly}. However, observations of certain binary
pulsar signals are in good agreement with the energy loss predicted by {\GR}
due to gravitational waves, see, e.g., \cite{Kramer:Stairs:Manchester:McLaughlin:Lyne:others:2006}.
This indirect observation of gravitational waves originates from Hulse and Taylor
(first found for the binary pulsar PSR B1913+16) and was awarded the Nobel Prize in 1993.
% In the future various predictions
% of {\GR} are expected to be confirmed with unprecedented accuracy
% from observations of the double pulsar system PSR J0737-3039A and B,
% which was only discovered in 2003.
Nowadays there is less doubt that gravitational waves exist, and one aims
at a direct observation with assiduous efforts, both by experiments on Earth,
e.g., LIGO, VIRGO, GEO 600, and by the future space mission LISA \cite{Sathyaprakash:Schutz:2009}.
The direct measurement of gravitational waves is not only interesting,
but would furthermore open up an entirely new spectrum for astronomical observations. Such gravitational wave astronomy
is expected to have great impact on astrophysics and fundamental physics
\cite{Sathyaprakash:Schutz:2009}, possibly starting a new era in these fields.

% \ifnotadpversion
% \thispagestyle{scrplain}
% \fi

Beside the experimental challenge of measuring extraordinarily small
relative chan\-ges in length ($\lesssim 10^{-21}$ detectable by now)
% (by now a precision of more than $10^{21}$ was reached),
there are important problems to be solved on the theoretical side in order
to successfully establish the new field of gravitational wave astronomy.
The theoretical challenge lies within the area of data analysis, for both the noise
dominated \cite{Jaranowski:Krolak:2005} and signal dominated \cite{Vallisneri:2009}
cases. An accurate understanding and knowledge of the expected gravitational wave
signals is a key ingredient to allow faithful astronomical or astrophysical statements from the
data analysis process. An appealing source for gravitational waves is the
inspiral and merger of two compact objects, like black holes and neutron stars.
The advantage of this kind of source is its quite periodic behavior, which can be
studied over long periods of time. However, minute changes in frequency
and amplitude of the gravitational waves need to be predicted in an accurate way. While fully numerical methods
are ideal to study the very late inspiral (or plunge) and merger phases of compact objects,
the post-Newtonian approximation to {\GR} provides a good analytic handle on the
inspiral phase and can give accurate predictions over many orbits.
The post-Newtonian approximation was pushed to high orders for nonspinning objects,
see, e.g., \cite{Blanchet:2006}, and it is desirable to catch up
to these orders for the spinning case.

A successful and efficient way to calculate the \emph{conservative} part of
the dynamics of two compact objects within the post-Newtonian approximation
is based on the canonical formalism of Arnowitt, Deser, and Misner (ADM).
However, this formalism has been coupled so far to nonspinning point-like objects only.
The main goal of the present thesis is to extend this coupling to spinning objects.
Not only this is useful for subsequent applications, but an interesting problem as such (though rather mathematical). To linear order in spin the problem is solved
using an action approach, similar to a treatment of spin-$\frac{1}{2}$ Dirac fields
coupled to gravity given by Kibble \cite{Kibble:1963}. Further, an order-by-order construction of the
canonical formalism with spin is given as a check. This construction is based
on consistency conditions on the formalism. In particular it is
sufficient to rely on a certain form of total linear
and angular momentum expressed in terms of canonical variables in order to
reproduce the result of the action approach to next-to-next-to-leading
order in the post-Newtonian approximation. The assumed form of total linear and angular momentum, i.e.,
the generators of translations and rotations,
guarantees that a great part of the global Poincar\'e algebra is fulfilled.
The connection to the action approach is given by Noether's theorem
on conserved quantities.

Higher orders in spin correspond to quadrupole and even higher multipole corrections.
Both the action approach and an order-by-order construction are in principle
applicable to canonical formulations at higher orders in spin.
However, only the next-to-leading order spin(1)-spin(1) level will be tackled here.
This requires a modeling of the spin-induced quadrupole deformation, described by
a single parameter for each object. This parameter is not only distinct
for black holes and neutron stars, for the latter kind of object it also depends on the
assumed equation of state or on other details of a particular theoretical neutron star model.
If gravitational wave astronomy becomes available with a high enough
precision in the future, one may hope to \emph{measure} this
(and maybe other) neutron star parameter.

The results obtained here within the post-Newtonian approximation cover the
next-to-leading order spin(1)-spin(2) and spin(1)-spin(1) conservative Hamiltonians.
The conservative next-to-leading order spin-orbit Hamiltonian was reproduced.
For maximally rotating bodies all Hamiltonians up to and including the third post-Newtonian order are now known.
A maximal rotating body is defined to have a dimensionless spin (i.e., rescaled by the mass of the object
and identical to the dimensionless Kerr parameter for black holes) of value one,
corresponding to an extremal Kerr black hole.
% NS spin: 1.5 milliseconds = 4200 rad/s, density 4.5x10^17 kg/m^3, mass 1.5 sun = 3.0x10^30 kg, moment of inertia 1.5×10^45 g cm^2 = 1.5×10^38 kg m^2
% => a = 0.3
% Notice that millisecond pulsars (or neutron stars) could have dimensionless spins
% of the order $\frac{1}{10}$, and for black holes even higher spins are possible.
Notice that millisecond pulsars (or neutron stars) and black holes can easily have
dimensionless spins bigger than $\frac{1}{10}$ (a rough approximation
for the sun yields $\frac{1}{5}$ \cite{Shapiro:Teukolsky:1983}).
Thus spins close to maximal ones are expected to be astrophysically relevant.
In this case the next-to-leading order spin Hamiltonians obtained here
are needed for an accurate description of the dynamics during the inspiral phase.
It was found recently in \cite{Reisswig:Husa:Rezzolla:Dorband:Pollney:Seiler:2009} that
spin effects as such and in particular the orientations of the spins have a big
impact on the event rates expected in detectors, especially when spins are close to maximum.

If the fourth post-Newtonian order Hamiltonian for nonspinning objects
could be obtained in the future, the spin Hamiltonians calculated here
would be applicable to an even larger class of binaries (with smaller spins).
Notice that the effective one-body approach for nonspinning objects, see, e.g.,
\cite{Damour:Nagar:2009:2,*Buonanno:etal:2009,*Damour:Nagar:2009:3},
is able to
cover such higher post-Newtonian orders by calibration to numerical relativity
and further provides predictions for the full waveforms, including merger and ringdown phases.
An extension of the effective one-body approach to spinning objects is possible \cite{Damour:2001}.
Subsequent implementation of higher order spin Hamiltonians seems to be interesting,
and was already performed for the next-to-leading order spin-orbit Hamiltonian
\cite{Damour:Jaranowski:Schafer:2008:3,*Pan:etal:2009,*Barausse:Buonanno:2009}.

Now the organization of the present thesis is given, with references to
relevant published work of the author for certain {\chapsname}
(for a short review see also \cite{Steinhoff:Hergt:Schafer:2010:1}).
In \Chap{prelim} spinning objects in special and general relativity are reviewed.
Further, an overview of canonical formulations of {\GR} is given, with
emphasis on the ADM formalism and coupling to nonspinning objects.
In \Chap{action} the action approach to the canonical formulation of self-gravitating
spinning objects to linear order in spin is performed \cite{Steinhoff:Schafer:2009:2}.
An order-by-order construction based on consistency considerations is
performed to next-to-next-to-leading order
in \Chap{symmetry} \cite{Steinhoff:Schafer:Hergt:2008,Steinhoff:Wang:2009}
as a check.
In \Chap{higher} first general quadrupole corrections to the equations of motion
and the stress-energy tensor are given \cite{Steinhoff:Puetzfeld:2009} and then
used to extend the canonical formalism to spin-induced quadrupole deformation
at next-to-leading order \cite{Steinhoff:Hergt:Schafer:2008:1,Steinhoff:Schafer:2009:1,Hergt:Steinhoff:Schafer:2010:1}.
As an application of the formalism, conservative Hamiltonians at next-to-leading order are derived in \Chap{result}.
These are the spin-orbit \cite{Steinhoff:Schafer:Hergt:2008} (derived earlier by Damour, Jaranowski, and Sch\"afer),
spin(1)-spin(2) \cite{Steinhoff:Hergt:Schafer:2008:2}, and spin(1)-spin(1) Hamiltonians,
the latter was first derived for black holes \cite{Steinhoff:Hergt:Schafer:2008:1,Steinhoff:Schafer:2009:1}
and later for compact objects in general (including neutron stars) \cite{Hergt:Steinhoff:Schafer:2010:1}.
Finally, conclusions and outlook are given in \Chap{conc}.

\def\notationintro{Lower case Latin indices from the beginning of the alphabet ($a$, $b$, \dots)
label the individual spinning objects and then consequently take on values from one to the number of objects.
Three different frames are utilized in this thesis, denoted by different indices.
Greek indices ($\alpha$, $\mu$, \dots) refer to the coordinate frame,
upper case Latin indices from the middle of the alphabet ($I$, $J$, \dots) belong to a
local Lorentz frame, and upper case Latin indices from the beginning of the alphabet ($A$, $B$, \dots)
denote the so called body-fixed Lorentz frame.
Lower case Latin indices from the middle of the alphabet ($i$, $j$, \dots) are used for the spatial part
of the mentioned frames and are running through $i = 1, 2, 3$.
In order to distinguish the three frames when splitting them into
spatial and time part, we write $a = (0), (i)$ for Lorentz indices
(or $a = (0), (1), (2), (3)$ in more detail),
$A = [0], [i]$ for the body-fixed frame, and $\mu = 0, i$ for the coordinate frame.
Indices appearing twice in a product are implicitly summed over its index range, except
for label indices of the objects.
Round and square brackets are also used for index symmetrization
and antisymmetrization, respectively, e.g., $A^{(\mu\nu)} \equiv \frac{1}{2} (A^{\mu\nu} + A^{\nu\mu})$.
Partial derivatives are denoted by $\partial_{\mu}$ or by a comma as an index $~_{,\mu}$.
Similarly, the 4-dimensional covariant derivative is written as $~_{||\mu}$ and
the induced 3-dimensional one as $~_{;i}$.
A 3-dimensional vector is also written in boldface, e.g., $\vct{x}$.
The signature of spacetime is taken to be $+2$.
Units are such that the speed of light $c$ and the gravitational
constant $G$ are equal to one.}
{\notationintro} Other symbols are defined in this thesis on their first occurrence.
For convenience also a summary of defined symbols %(and associated conventions)
is given in the \hyperref[symbols]{appendix}.

\ifnotadpversion
\automark[section]{chapter}
\fi

% \newpage
\mychapter{Preliminaries\label{prelim}}
This {\chapname} gives a short review of the achievements regarding spin in the theory of relativity
as well as the canonical formulation of {\GR}.
Emphasis is put on the problems to be solved if one aims at a
canonical formulation of self-gravitating spinning objects in the pole-dipole approximation.

\mysection{Spin in Special Relativity\label{spinSR}}
Spin already has very interesting properties in {\SR}. Its canonical structure
is obtained here as a consequence of the Poincar\'e algebra by introducing
the spin as a specific part of the total angular momentum.

\mysubsection{Center, Spin, and Mass Dipole}
The 4-dimensional total linear momentum $P^{\mu}$ and total angular momentum $J^{\mu\nu} = - J^{\nu\mu}$
of a physical system are conserved quantities due to Poincar\'e invariance.
The 4-dimen\-sional total spin tensor $S^{\mu\nu}$ can then be defined by
\begin{equation}
J^{\mu\nu} = Z^{\mu} P^{\nu} - P^{\mu} Z^{\nu} + S^{\mu\nu} \label{Sdef} \,.
\end{equation}
That is, spin is the difference of total angular momentum and its orbital part.
However, a different choice for the yet arbitrary center
$Z^{\mu}$ of the system will result in a different spin $S^{\mu\nu}$ (with $J^{\mu\nu}$ being unchanged).
This just expresses the dependence of angular momenta
on the choice of a reference point.
Separating time and space components
\begin{equation}
J^{ij} = Z^i P^j - P^i Z^j + S^{ij} \,, \qquad
J^{i0} = Z^i E - P^i t + S^{i0} \label{centerMotion} \,,
\end{equation}
one infers that the spin transforms as
\begin{equation}
S^{ij} \rightarrow S^{ij} + \delta Z^i P^j - P^i \delta Z^j \,, \qquad %\label{Sijtrans}
S^{i0} \rightarrow S^{i0} + \delta Z^i E \,, \label{Si0trans}
\end{equation}
under a change of the center $Z^i \rightarrow Z^i - \delta Z^i$.
$E \equiv P^0$ is the total energy and $t \equiv Z^0$ the time coordinate.
Notice that $J^{i0}$ is the total mass dipole of the system at $t=0$ relative to the coordinate origin,
so (\ref{centerMotion}) tells us that $S^{i0}$ is the mass dipole relative to the center $Z^i$.
This explains the transformation property (\ref{Si0trans}).
One may also describe the 3-dimensional spin $S^{ij}$ as the
flow dipole and $S^{\mu\nu}$ as the 4-dimensional dipole
moment of the system relative to the center $Z^i$.

By its definition (\ref{Sdef}), $S^{\mu\nu}$ transforms as a tensor under Lorentz boosts,
with interesting consequences. In classical mechanics the center of mass, i.e.,
the center for which the mass dipole vanishes, is independent of the reference frame.
In {\SR} such a center can in general not be found. Under a Lorentz boost all components
of $S^{\mu\nu}$ transform, so if the mass dipole $S^{i0}$ vanishes in one reference
frame, it will only be zero in all others if the system has no spin, $S^{\mu\nu}=0$.
A nice graphic interpretation is given by figure \ref{centerfig}.
Notice that a spinning system in {\SR} has a minimal extension of the order $S / M$
orthogonal to the axis of rotation \cite{Moller:1949,Fleming:1965}.
Here $S$ is the spin length, $2 S^2 = S^{\mu\nu} S_{\mu\nu}$,
and $M$ is the rest mass of the system, $M^2 = - P^{\mu} P_{\mu}$.
In {\GR}, $S / M$ is the radius coordinate of the ring singularity of Kerr spacetime \cite{Kerr:1963}.

\ifadpversion
\begin{SCfigure}[1.7]
\includegraphics{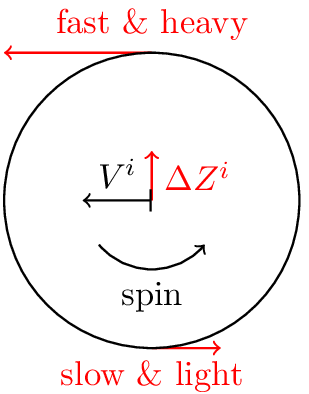}
\hspace{0.5cm}
\fi
\ifnotadpversion
\begin{SCfigure}[2.9]
\begin{tikzpicture}[semithick]
\draw[->,red] (0,1.5) node[above]{fast \& heavy} -- (-1.5,1.5);
\draw[->,red] (0,-1.5) node[below]{slow \& light} -- (0.7,-1.5);
\draw[->,red] (0,0) -- (0,0.25) node[right]{$\Delta Z^i$} -- (0,0.5);
\draw (0,0) circle (1.5);
\draw[|->] (0,0) -- (-0.35,0) node[above,black]{$V^i$} -- (-0.7,0);
\draw[->] (-140:0.7) arc (-140:-90:0.7) node[below]{spin} arc (-90:-40:0.7);
\end{tikzpicture}
\fi
\caption{If a spinning spherical symmetric object moves with a velocity $V^i$ to the left,
its upper hemisphere moves faster with respect to the reference system
than its lower hemisphere. Thus the upper hemisphere has a higher relativistic
mass than the lower one --- the object acquires a mass dipole $E \Delta Z^i$. \cite{Fleming:1965}\label{centerfig}}
\end{SCfigure}

However, by virtue of (\ref{Si0trans}) one can always choose the center $Z^i$ such
that the mass dipole $S^{i0}$ vanishes in one specific reference frame characterized by
a timelike vector $f_{\mu}$. That is, the center is then the center of mass as
observed in this frame. It holds
\begin{equation}\label{genSSC}
S^{\mu\nu} f_{\nu} = 0 \,,
\end{equation}
which is the so called spin supplementary condition. This condition fixes the center
and ensures that the spin tensor $S^{\mu\nu}$ has three independent components only.
Basically three important such conditions can be found in the literature \cite{Pryce:1948,Fleming:1965},
\begin{align}
f_{\mu} &= P_{\mu} \,, 				&\text{or}& & S^{\mu\nu} P_{\nu} &= 0 \,, \label{covSSC} \\
f_{\mu} &= - \delta_{\mu}^0 \,, 		&\text{or}& & \tilde{S}^{\mu0} &= 0 \,, \label{comSSC} \\
f_{\mu} &= P_{\mu} - M \delta_{\mu}^0 \,, 	&\text{or}& & \hat{S}^{\mu\nu} P_{\nu} - M \hat{S}^{\mu0} &= 0 \,. \label{NWSSC}
\end{align}
In the following, we will indicate
center and spin belonging to the second condition \cite{Pryce:1948,Moller:1949} by a tilde, $\tilde{Z}^i$ and $\tilde{S}^{\mu\nu}$,
a hat relates to the third condition \cite{Pryce:1948,Newton:Wigner:1949}, $\hat{Z}^i$ and $\hat{S}^{\mu\nu}$,
while center and spin of the first condition \cite{Fokker:1929,*Synge:1935}
are just denoted by $Z^i$ and $S^{\mu\nu}$.
We call $Z^i$ the center of inertia, $\tilde{Z}^i$ the center of mass, and
$\hat{Z}^i$ the center of spin \cite{Fleming:1965}.
Notice that the first condition is manifestly covariant, and is called
covariant spin supplementary condition here. A different covariant condition is discussed in \Sec{spinGR}.
The third condition is called canonical spin supplementary condition, which will be explained in the following.

\mysubsection{Poincar\'e Algebra}
The Poincar\'e group is one of the most important groups in physics.
Its generators $P^{\mu}$ and $J^{\mu\nu}$ obey
the Poisson bracket realization of the well-known Poincar\'e algebra
\begin{align}
	\{ P^{\mu} , P^{\nu} \} &= 0 \,, \qquad
	\{ P^{\mu} , J^{\rho\sigma} \} = - \eta^{\mu\rho} P^{\sigma} + \eta^{\mu\sigma} P^{\rho} \,, \\
	\{ J^{\mu\nu} , J^{\rho\sigma} \} &= - \eta^{\nu\rho} J^{\mu\sigma} + \eta^{\mu\rho} J^{\nu\sigma}
		+ \eta^{\sigma\mu} J^{\rho\nu} - \eta^{\sigma\nu} J^{\rho\mu} \,,
\end{align}
where $\eta^{\mu\nu}$ is the Minkowski metric.
Splitting space and time one gets, see, e.g, \cite{Weinberg:1995},
\begin{gather}
\left\{P_{i},P_{j}\right\}=0 , \quad
\left\{P_{i},E\right\}=0 , \quad
\left\{J_{i},E\right\}=0 , \quad
\{G_{i},P_{j}\}=E\delta_{ij} , \quad
\{G_{i},E\}=P_{i} \,, \label{poinc1} \\
\left\{J_{i},P_{j}\right\}=\epsilon_{ijk}P_{k} , \quad
\left\{J_{i},J_{j}\right\}=\epsilon_{ijk}J_{k} , \quad
\{J_{i},G_{j}\}=\epsilon_{ijk}G_{k} , \quad
\{G_{i},G_{j}\}=-\epsilon_{ijk}J_{k} , \label{poinc2}
\end{gather}
with the total angular momentum vector $J_i = \tfrac{1}{2} \epsilon_{ijk} J^{jk}$
and the 3-dimensional Levi-Civita symbol $\epsilon_{ijk}$.
The boost vector $J^{i0}$ has an explicit
dependence on time $t$, which was split off as
\begin{equation}\label{boost}
	J^{i0} = G^i - P^i t \,.
\end{equation}
This defines the vector $G^i$, which is related to the
spin supplementary condition $\tilde{S}^{\mu0} = 0$ with center of mass $\tilde{Z}^i$
by $G^i = \tilde{Z}^i E$, cf.\ (\ref{centerMotion}).

Notice that in {\GR} total linear and angular momentum can be defined
for asymptotically flat spacetimes as global quantities by certain surface integrals.
% over the whole manifold.
In this case all considerations of this and the following {\secname} remain valid in full {\GR},
see \Sec{GPoincare}.

\mysubsection{Canonical Structure}
Using $\tilde{Z}^i = G^i / E$, $\tilde{S}_{ij} = J_{ij} - \tilde{Z}^i P_j + P_i \tilde{Z}^j$
and the Poincar\'e algebra (\ref{poinc1}, \ref{poinc2}),
the Poisson brackets between $P_i$, $\tilde{Z}^i$ , and $\tilde{S}_{ij}$
follow as
\begin{gather}
\{ \tilde{Z}^i , P_j \} = \delta_{ij} \,, \qquad
\{ \tilde{Z}^i, \tilde{Z}^j \} = - \frac{\tilde{S}_{ij}}{E^2} \,, \qquad
\{ \tilde{S}_{ij} , \tilde{Z}^k \} = \frac{P_i \tilde{S}_{kj}}{E^2} + \frac{P_j \tilde{S}_{ik}}{E^2} \,, \label{PBcom1} \\
\{ \tilde{S}_{ij} , \tilde{S}_{kl} \} =
	\mathcal{P}_{ki} \tilde{S}_{jl} - \mathcal{P}_{kj} \tilde{S}_{il}
	- \mathcal{P}_{li} \tilde{S}_{jk} + \mathcal{P}_{lj} \tilde{S}_{ik} \,, \label{PBcom2}
\end{gather}
all other zero, where
\begin{equation}\label{Pcal}
\mathcal{P}_{ij} = \delta_{ij} - \frac{P_i P_j}{E^2} \,, \qquad
\mathcal{P}_{ij}^{-1} = \delta_{ij} + \frac{P_i P_j}{M^2} \,,
\end{equation}
and $\delta_{ij}$ is the Kronecker symbol.

Now we proceed to the canonical spin supplementary condition (\ref{NWSSC}), which
can be written as $(E+M) \hat{S}^{i0} = \hat{S}^{ij} P_j$. From (\ref{Si0trans}) and $\tilde{S}^{i0} = 0$
we get
\begin{equation}
	\hat{Z}^i - \tilde{Z}^i = \delta Z^i = - \frac{\hat{S}^{i0}}{E} = \frac{P_k \hat{S}_{ki}}{E (E + M)} \,.
\end{equation}
Having $\delta Z^i$, \Eq{Si0trans} relates $\hat{S}^{ij}$ and $\tilde{S}^{ij}$ by
\begin{equation}
\tilde{S}_{ij} = \hat{S}_{ij} + \frac{P_j P_k \hat{S}_{ki}}{E (E+M)}
       - \frac{P_i P_k \hat{S}_{kj}}{E (E+M)} \,.
\end{equation}
Contraction with $P_i$ leads to $E P_i \tilde{S}_{ij} = M P_i \hat{S}_{ij}$. Finally, in terms of $\tilde{S}_{ij}$ one has
\begin{equation}
\hat{Z}^i = \tilde{Z}^i + \frac{P_k \tilde{S}_{ki}}{M (E + M)} \,, \qquad
\hat{S}_{ij} = \tilde{S}_{ij} + \frac{P_i P_k \tilde{S}_{kj}}{M (E+M)}
       - \frac{P_j P_k \tilde{S}_{ki}}{M (E+M)} \,.
\end{equation}
The Poisson brackets (\ref{PBcom1}, \ref{PBcom2}) transform into\footnote{
Notice that Poisson brackets with $M$ were calculated according
to its definition $M^2 = E^2 - P_i P_i$.}
\begin{equation}
\{ \hat{Z}^i , P_j \} = \delta_{ij} \,, \qquad
\{\hat{S}_{ij}, \hat{S}_{kl}\} = \delta_{ik} \hat{S}_{jl} - \delta_{jk} \hat{S}_{il}
	- \delta_{il} \hat{S}_{jk} + \delta_{jl} \hat{S}_{ik} \,, \label{PBNW}
\end{equation}
all other zero. Thus $\hat{Z}^i$, $P_j$, and $\hat{S}_{ij}$ are \emph{canonical variables}.
This realization is due to Pryce \cite{Pryce:1935,Pryce:1948}.
Newton and Wigner further showed that $\hat{Z}^i$ is the only center with this property \cite{Newton:Wigner:1949}.

Similarly, we can proceed to the covariant spin supplementary condition (\ref{covSSC}) by
\begin{equation}\label{covToNW}
Z^i = \hat{Z}^i + \frac{P_k \hat{S}_{ki}}{M (E + M)} \,, \qquad
S_{ij} = \hat{S}_{ij} + \frac{P_i P_k \hat{S}_{kj}}{M (E + M)} - \frac{P_j P_k \hat{S}_{ki}}{M (E + M)} \,,
\end{equation}
and find the Poisson brackets,
\begin{gather}
\{ Z^i, Z^j \} = \mathcal{P}_{ik} \mathcal{P}_{jl} \frac{S_{kl}}{M^2} \,, \qquad
\{ S_{ij} , Z^k \} = \frac{\mathcal{P}_{im} \mathcal{P}_{jn}}{M^2} ( P_m S_{nk} + P_n S_{km} ) \,, \\
\{ Z^i , P_j \} = \delta_{ij} \,, \qquad
\{ S_{ij} , S_{kl} \} =
	\mathcal{P}_{ki}^{-1} S_{jl} - \mathcal{P}_{kj}^{-1} S_{il}
	- \mathcal{P}_{li}^{-1} S_{jk} + \mathcal{P}_{lj}^{-1} S_{ik} \,,
\end{gather}
all other zero.

To conclude, there are several possibilities for spin supplementary conditions
and centers, however, only (\ref{NWSSC}) leads to canonical variables, (\ref{PBNW}).
This is an important fact for a canonical formulation of
spin in {\GR}.

\mysection{Spin in General Relativity\label{spinGR}}
It is well-known that spin in {\GR} leads to certain gravitomagnetic effects,
see, e.g., \cite{Schafer:2004}. In this {\secname}
the pole-dipole approximation for compact objects is introduced,
providing an analytic description of spin in {\GR}.

\mysubsection{Gravitational Skeleton}
In electrostatics, the multipole approximation of a charge density $\rho$,
\begin{equation}\label{rhoexp}
\rho(\vct{x}) = \left( q - q^i \partial_i + \frac{1}{2!} q^{ij} \partial_i \partial_j - \dots \right) \delta(\vct{x}) \,,
\end{equation}
can be obtained from a Taylor series of its Fourier transform in the form
\begin{equation}
\rho(\vct{k}) = \left( q + i q^i k_i + \frac{1}{2!} i^2 q^{ij} k_i k_j + \dots \right) (2\pi)^{-3/2} \,,
\end{equation}
by the well-known transition formulas for the Dirac delta distribution $\delta(\vct{x}) \leftrightarrow (2\pi)^{-3/2}$
and partial coordinate derivative $\partial_i \leftrightarrow - i k_i$.
Here $\vct{x} = (x^i)$ are the spatial coordinates and $\vct{k} = (k_i)$
the corresponding ones in Fourier space.
The quantities $q$, $q^i$, and $q^{ij}$ are the electric monopole,
dipole, and quadrupole. The potential $\phi$ follows as
\begin{equation}
\phi = - 4 \pi \Delta^{-1} \rho
	= \left( q - q^i \partial_i + \frac{1}{2!} q^{ij} \partial_i \partial_j - \dots \right) \frac{1}{|\vct{x}|} \,,
\end{equation}
where $\Delta = \partial_i \partial_i$ is the Laplacian and $\Delta^{-1}$ its inverse operator
(with the usual boundary conditions).
In most textbooks, the multipole approximation is derived directly for the potential
or the field. Notice that the multipole approximation breaks down at high values of
$\vct{k}$, i.e., in the ultraviolet, or at small values of $\vct{x}$ in the potential.
This is the reason for the divergent self-energy of the approximated charge density (\ref{rhoexp}).

Now the multipole approximation is applied to the stress-energy tensor $T^{\mu\nu}$.
As it is desirable to have a manifestly covariant approximation scheme, we write
\begin{equation}\label{Texp}
\sqrt{-g} T^{\mu\nu} = \int \dd \tau \bigg[
	t^{\mu\nu} \delta_{(4)}
	- ( t^{\mu\nu\alpha} \delta_{(4)} )_{||\alpha}
	+ \frac{1}{2!} ( t^{\mu\nu\alpha\beta} \delta_{(4)} )_{||(\alpha\beta)}
	- \dots
\bigg] \,.
\end{equation}
Here $\tau$ is the proper time of a representative worldline $z^{\rho}(\tau)$,
$g$ the determinant of the 4-dimensional metric $g_{\mu\nu}$,
$\delta_{(4)} = \delta(x^{\rho} - z^{\rho}(\tau))$, and $t^{\mu\nu\dots}$ are 4-dimensional
covariant multipole moments.
If one performs the $\tau$ integration in (\ref{Texp}) by eliminating the time part of $\delta_{(4)}$
and writes the covariant derivatives as partial derivatives and Christoffel symbols, then
(\ref{Texp}) indeed takes on the form of (\ref{rhoexp}).
Equation (\ref{Texp}) in substance is Mathisson's \emph{gravitational skeleton} \cite{Mathisson:1937,*Mathisson:2010}, but
in the form given by W.\,M.\ Tulczyjew \cite{Tulczyjew:1959}. Interestingly enough Mathisson
unknowably used a test-function formulation of the delta distribution, years before this
formulation was used by Laurent Schwartz for his mathematically rigorous \emph{Th\'eorie des Distributions}
\cite{Schwartz:1950}.

The divergent self-interactions already present in electrostatics become more severe
if the field equations are nonlinear.
% Then no straightforward split of the field
% into the field generated by some object and an external field by means of the
% superposition principle is possible.
If the distributional stress-energy tensor (\ref{Texp}) is used as a source for
a nonlinear field equation, products of distributions will appear, which lack a
mathematical definition. However, this problem can be overcome,
as in quantum field theory, by a regularization and renormalization program. In
particular, dimensional regularization \cite{tHooft:Veltman:1972,*Bollini:Giambiagi:1972:1}
is most useful for theories involving gauge freedoms, like {\GR}.
Dimensional regularization has been employed successfully in post-Newtonian calculations
\cite{Damour:Jaranowski:Schafer:2001,Damour:Jaranowski:Schafer:2008:2,Blanchet:Damour:EspositoFarese:2004}
to a high order of nonlinearity. However, many treatments of multipole approximations
in {\GR} avoid these problems by considering (\ref{Texp}) for test bodies only,
which by definition are neglected as a source of the gravitational field.

The relation between source multipoles related to $T^{\mu\nu}$ used here and
field multipoles \cite{Thorne:1980} was considered in \cite{Ohashi:2003}.
Only for linear theories like electrostatics this relation is straightforward.

\mysubsection{Pole-Dipole Approximation}
The stress-energy tensor (\ref{Texp}) must fulfill
\begin{equation}\label{Tvar}
{T^{\mu\nu}}_{||\nu} = 0 \,.
\end{equation}
This corresponds to Mathisson's \emph{variational equations of mechanics} \cite{Mathisson:1937,*Mathisson:2010}
and imposes certain conditions on the multipole moments. In the pole-dipole
approximation only monopole $t^{\mu\nu}$ and dipole $t^{\mu\nu\alpha}$
are kept in (\ref{Texp}). Evaluating (\ref{Tvar}) one sees that
$t^{\mu\nu}$ and $t^{\mu\nu\alpha}$ can be expressed in terms of a vector $p^{\mu}$
and an antisymmetric tensor $S^{\mu\nu}$, which have to fulfill
the dynamic equations
\begin{equation}
\frac{\dD S^{\mu\nu}}{\dd \tau} = 2 p^{[\mu} u^{\nu]} \,, \qquad
\frac{\dD p_{\mu} }{\dd \tau} =
	- \frac{1}{2} \Riem_{\mu\rho\beta\alpha}^{\text{(4)}} u^{\rho} S^{\beta\alpha} \,, \label{eom}
\end{equation}
with $u^{\mu} = \frac{\dd z^{\mu}}{\dd \tau}$, $\dD$ the 4-dimensional covariant differential, and
$\Riem_{\mu\rho\beta\alpha}^{\text{(4)}}$ the 4-dim\-en\-sio\-nal Riemann tensor defined by
\begin{equation}\label{Rdef}
a_{\mu||\alpha\beta} - a_{\mu||\beta\alpha} = \Riem^{(4)}_{\nu\mu\alpha\beta} a^{\nu} \,,
\end{equation}
for an arbitrary $a_{\mu}$.
The stress-energy tensor can be written as
\begin{equation}\label{PDset}
\sqrt{-g} T^{\mu\nu} = \int \dd \tau \bigg[
	u^{(\mu} p^{\nu)} \delta_{(4)}
	- \left( S^{\alpha(\mu} u^{\nu)}  \delta_{(4)} \right)_{||\alpha}
\bigg] \,.
\end{equation}
$p^{\mu}$ and $S^{\mu\nu}$
are the linear momentum and spin of the object and now play the role
of monopole and dipole moment.
Their equations of motion were already derived by Mathisson \cite{Mathisson:1937,*Mathisson:2010}
within his manifestly covariant formalism, albeit restricted to a specific
spin supplementary condition. In the general form (\ref{eom}) they were first
given by Papapetrou \cite{Papapetrou:1951}, however, his method was not
manifestly covariant. W.\,M.\ Tulczyjew gave a derivation of (\ref{eom}) as well as
of the stress-energy tensor (\ref{PDset}) in a manifestly covariant way \cite{Tulczyjew:1959},
using essentially Mathisson's method.
Further important rederivations have been performed in \cite{Tulczyjew:Tulczyjew:1962,*Taub:1964,Dixon:1964}.
Higher multipole corrections will be discussed in \Sec{QuadSec}.

Obviously a spinning object in {\GR} does not follow a geodesic.
For test bodies this effect can be studied numerically, see, e.g.,
\cite{Semerak:1999,Kyrian:Semerak:2007}. % and references therein.
Further, without giving a relation between $p^{\mu}$ and $u^{\mu}$,
the system of equations (\ref{eom}) is not closed.

\mysubsection{Spin Supplementary Condition\label{sscsec}}
A spin supplementary condition (\ref{genSSC}) must be preserved in time.
Using (\ref{eom}) this leads to a relation between $p^{\mu}$ and $u^{\mu}$ \cite{Barausse:Racine:Buonanno:2009},
\begin{equation}\label{PUrel}
p^{\mu} = \frac{1}{- f_{\alpha} u^{\alpha}} \left( - f_{\nu} p^{\nu} u^{\mu} + S^{\mu\nu} \frac{\dD f_{\nu}}{\dd \tau} \right) \,,
\end{equation}
and thus, for a suitable $f_{\nu}$, closes the system of equations (\ref{eom}).
A good spin supplementary condition is the covariant one,
\begin{equation}\label{pSSC}
	S^{\mu\nu} p_{\nu} = 0 \,,
\end{equation}
or $f_{\mu} = p_{\mu}$, which has been suggested in the context of {\GR} in
\cite{Tulczyjew:1959}. Indeed, this condition guarantees existence and uniqueness
of a corresponding worldline $z^{\rho}(\tau)$ \cite{Beiglbock:1967,*Schattner:1979:1,*Schattner:1979:2}.
The mass quantity $m$, $p_{\mu} p^{\mu} = - m^2$, and the spin length $S$,
$2 S^2 = S_{\mu\nu} S^{\mu\nu}$, are conserved for this condition.
A covariant condition has the advantage that the relation
between $p^{\mu}$ and $u^{\mu}$ (\ref{PUrel}) is manifestly covariant.
However, also the noncovariant condition $\tilde{S}^{\mu0} = 0$ was applied
in {\GR} \cite{Corinaldesi:Papapetrou:1951}.

A different covariant condition is given by
\begin{equation}\label{uSSC}
	S^{\mu\nu} u_{\nu} = 0 \,,
\end{equation}
or $f_{\mu} = u_{\mu}$, which was used in both special \cite{Frenkel:1926,*Lanczos:1929}
and general relativity \cite{Mathisson:1937,*Mathisson:2010,Pirani:1956,*Pirani:2009}. While there are no serious
objections to use this condition, as it closes the system of equations (\ref{eom}),
it has some features which are usually not wanted.
The condition (\ref{uSSC}) does not uniquely specify a worldline. Instead, the worldline
depends on the choice of initial conditions and in general performs a kind of classical \emph{Zitterbewegung}
around the worldline defined by (\ref{pSSC}), see \cite{Tulczyjew:1959,Kyrian:Semerak:2007}.
As quadrupole corrections are needed to describe a black hole at the
quadratic level in spin \cite{Thorne:1980}, we will only consider the
pole-dipole approximation at linear order in spin here. Then the conditions
(\ref{pSSC}) and (\ref{uSSC}) are fully equivalent and it holds $p_{\mu} = m u_{\mu}$.

As a generalization of the canonical spin supplementary condition (\ref{NWSSC})
to {\GR} one could take
\begin{equation}\label{canSSC}
\hat{S}^{\mu\nu} p_{\nu} + m \hat{S}^{\mu\nu} n_{\nu} = 0 \,,
\end{equation}
with some timelike unit vector $n_{\nu}$. However, it needs to be proven
if or under which conditions (\ref{canSSC}) leads to canonical variables.

Finally, contraction of the first relation in (\ref{eom}) with $u_{\nu}$ leads to the well-known formula
\begin{equation}\label{PUnossc}
p^{\mu} = - u_{\nu} p^{\nu} u^{\mu} - \frac{\dD ( S^{\mu\nu} )}{\dd \tau} u_{\nu} \,.
\end{equation}
This relation, however, does not close the system of equations (\ref{eom}), it
just is a component of (\ref{eom}).

\mysection{Canonical Formulation of General Relativity\label{cangrav}}
In this {\secname} the canonical formalism of ADM
\cite{Arnowitt:Deser:Misner:1959,*Arnowitt:Deser:Misner:1960:1,Arnowitt:Deser:Misner:1962,*Arnowitt:Deser:Misner:2008}
is introduced. Possible couplings to matter are reviewed, and point-masses
are treated in detail. Finally alternatives to the ADM approach are discussed.

\mysubsection{The ADM Formalism}
The Einstein-Hilbert action of {\GR} $W_G$ is given by a spacetime integral over
the Lagrangian density $\mathcal{L}_G$ as
\begin{equation}\label{EHaction}
W_G[g_{\mu\nu}] = \int \dd^4 x \, \mathcal{L}_G \,, \qquad
\mathcal{L}_G = \frac{1}{16\pi} \sqrt{-g} \Riem^{(4)} \,,
\end{equation}
where $\Riem^{(4)}$ is the 4-dimensional Ricci scalar. Alternatively the action
can be varied with respect to the tetrad field $e_{I\mu}$ instead of $g_{\mu\nu}$, see \Sec{mincoup}.
In order to find a canonical form of this action it is convenient to
perform a splitting of spacetime into a stack of 3-dimensional
hypersurfaces with constant time coordinate $t$. In these coordinates
the unit normal vector $n^{\mu}$, $n_{\mu} n^{\mu} = -1$, of the hypersurfaces
has the components
\begin{equation}\label{ndef}
	n_{\mu} = (-N, 0, 0, 0) \,, \quad \text{or} \quad n^{\mu} = \frac{1}{N} (1, - N^i) \,,
\end{equation}
where $N$ is the lapse function and $N^i$ the shift vector. With the help of the projector\footnote{Notice that $0 = n_{\mu} \gamma^{\mu\nu} = -N \gamma^{0\nu}$
and thus $\gamma^{0\nu} = 0$ for our choice of the time coordinate.}
\begin{equation}\label{nprojector}
\gamma^{\mu\nu} = g^{\mu\nu} + n^{\mu} n^{\nu}
	= \left( \begin{array}{cc} 0 & 0 \\ 0 & \gamma^{i j} \end{array} \right) \,,
\end{equation}
this splitting can be constructed in a geometrical way, see, e.g.,
\cite{Gourgoulhon:2007}. The 3-dimensional hypersurfaces have an induced
metric $g_{ij} = \gamma_{ij}$, with $\gamma_{ik} \gamma^{kj} = \delta_{ij}$,
a Riemann tensor $\Riem_{ijkl}$, a Ricci tensor $\Riem_{ij}$, and a Ricci scalar $\Riem$.
These quantities are intrinsic geometric objects of the hypersurfaces,
whereas the extrinsic curvature
\begin{equation}\label{Kdef}
K_{ij} \equiv - n_{(i||j)} %= - N \Gamma^0_{ij}
	= \frac{1}{2 N} \left( - \gamma_{ij,0} + 2 {N^k}_{;(i} \gamma_{j)k} \right) \,,
\end{equation}
depends on their embedding in spacetime.

Applying this splitting of spacetime to the Lagrangian density $\mathcal{L}_G$ leads to
\begin{equation}\label{EH31}
\mathcal{L}_G = \frac{1}{16\pi} N \sqrt{\gamma} \left[ R + K_{ij} K^{ij}
	- ( \gamma_{ij} K^{ij} )^2 \right] + (\text{td}) \,,
\end{equation}
where $(\text{td})$ denotes a total divergence, which is neglected for now.
Instead of varying with respect to the ten independent components of $g_{\mu\nu}$,
we now use $\gamma_{ij}$, $N$, and $N^i$. Notice that no time derivatives of $N$ and $N^i$
appear. In order to obtain a canonical formulation we have to introduce the field momentum
\begin{equation}\label{pfield}
\pi^{ij} = 16\pi \frac{\partial \mathcal{L}_G}{\partial \gamma_{ij,0}}
	= \sqrt{\gamma} (\gamma^{ij}\gamma^{kl} - \gamma^{ik}\gamma^{jl})K_{kl} \,,
\end{equation}
where (\ref{Kdef}) was used. This can be inverted as
\begin{equation}
K_{ij} = \frac{1}{2 \sqrt{\gamma}} ( \gamma_{ij}\gamma_{kl} - 2 \gamma_{ik}\gamma_{jl} ) \pi^{kl} \,.
\end{equation}
The Legendre transformed Lagrangian density then reads
\begin{gather}
\mathcal{L}_G = \frac{1}{16\pi} \pi^{ij} \gamma_{ij,0}
	- N \mathcal{H}^{\text{field}}
	+ N^i \mathcal{H}^{\text{field}}_i
	+ (\text{td}) \,, \label{WADM} \\
\mathcal{H}^{\text{field}} = - \frac{1}{16\pi\sqrt{\gamma}} \left[ \gamma \Riem
	- \gamma_{ij} \gamma_{k l} \pi^{ik} \pi^{jl}
	+ \frac{1}{2} \left( \gamma_{ij} \pi^{ij} \right)^2 \right] \,, \qquad
\mathcal{H}^{\text{field}}_i = \frac{1}{8\pi} \gamma_{ij} \pi^{jk}_{~~ ; k} \,, \label{HHifield}
\end{gather}
and the action is additionally varied with respect to $\pi^{ij}$ now. Notice that $N$ and $N^i$
play the role of Lagrange multipliers after Legendre transformation, the corresponding
constraints are the vanishing of $\mathcal{H}^{\text{field}}$ and $\mathcal{H}^{\text{field}}_i$.

A subsequent gauge fixing is subtle as it requires a fine-tuning of the action, see, e.g., \cite{Hanson:Regge:Teitelboim:1976}.
As shown in \cite{Arnowitt:Deser:Misner:1962,*Arnowitt:Deser:Misner:2008,DeWitt:1967,Regge:Teitelboim:1974} by different methods, see also \cite{Schwinger:1963:1},
one must replace the total divergence in (\ref{WADM}) by $- \frac{1}{16\pi} \mathcal{E}_{i,i}$
for asymptotically flat spacetimes, where $\mathcal{E}_i = \gamma_{ij,j} - \gamma_{jj,i}$.
This is related to the total energy $E$ of asymptotically flat spacetimes by
\begin{equation}\label{Esurf}
E = \frac{1}{16\pi} \oint \dd^2 s_i \mathcal{E}_i \,,
\end{equation}
where $\oint \dd^2 s_i$ denotes an integral over the asymptotic boundary
of a spatial hypersurfaces at fixed time.
% This ADM energy does not vanish and may thus be neglected in the action if one
% considers \emph{local} variations only (which is sufficient to get the \emph{local} field equations),
% but not if the action principle should also be valid for the global solution \cite{Regge:Teitelboim:1974}.
This ADM energy will turn out to be the generator of time evolution after gauge fixing.
For further discussion of boundary terms in the action of {\GR},
also for the case of not asymptotically flat spacetimes, see, e.g., \cite{York:1986}.
However, for asymptotically flat spacetimes the gravitational Hamiltonian may be written as
\begin{equation}\label{Dham}
H_G = \int \dd^3 x \, (N \mathcal{H}^{\text{field}}
	- N^i \mathcal{H}^{\text{field}}_i) + E[\gamma_{ij}] \,.
\end{equation}
Indeed, the action has the canonical structure momentum $\pi^{ij}$ times velocity $\gamma_{ij,0}$
minus Hamiltonian $H_G$. Variation thus results in Hamilton's equations
\begin{equation}
\frac{\partial \pi^{ij}}{\partial t} =
	- 16\pi \frac{\delta H_G}{\delta \gamma_{ij}} \equiv \{ \pi^{ij}, H_G \} \,, \qquad
\frac{\partial \gamma_{ij}}{\partial t} =
	16\pi \frac{\delta H_G}{\delta \pi^{ij}} \equiv \{ \gamma_{ij}, H_G \} \,,
\end{equation}
where $\delta$ denotes the variational derivative here and
the equal-time Poisson brackets are given by
\begin{equation}
\{ \gamma_{ij}({\bf x}), \pi^{kl}({\bf x}') \}
	= 16\pi \delta_{k(i} \delta_{j)l} \delta({\bf x} - {\bf x}') \,.
\end{equation}
Before gauge fixing, the surface term $E$ has no impact on these field
equations, which could be obtained from local variations\footnote{However, one
should not constrain to local variations for asymptotically flat spacetimes \cite{Regge:Teitelboim:1974}.}.
As further explained in \Sec{Dirac}, the gauge fixing is accompanied with solving the constraints $\mathcal{H}^{\text{field}} = 0$
and $\mathcal{H}^{\text{field}}_i = 0$, so $H_G$ then turns into the ADM energy $E$.
To make this more concrete, we choose the ADM transverse-traceless gauge conditions
\begin{equation}\label{ADMTTcond}
\partial_j ( \gamma_{ij} - \tfrac{1}{3} \gamma_{kk} \delta_{ij} ) = 0 \,, \qquad
\pi^{ii} = 0 \,,
\end{equation}
in which the transverse-traceless decomposition of $\gamma_{ij}$ and $\pi^{ij}$ may be
written as
\begin{align}
	\gamma_{ij} &= \left( 1 + \frac{\phi}{8} \right)^4 \delta_{ij} + h^{\text{TT}}_{ij} \,, \label{gdecomp} \\
	\pi^{ij} &= \tilde{\pi}^{ij} + \pi^{ij\text{TT}} \,, \label{pidecomp}
\end{align}
where $h^{\text{TT}}_{ij}$ and $\pi^{ij\text{TT}}$ are transverse-traceless,
e.g, $h^{\text{TT}}_{ii} = h^{\text{TT}}_{ij,j}=0$, and the longitudinal $\tilde{\pi}^{ij}$ is
related to a vector potential $\tilde{\pi}^i = \Delta^{-1} \pi^{ij}{\!}_{,j}$ by
\begin{equation}
\tilde{\pi}^{ij} = \tilde{\pi}^i{}_{, j} + \tilde{\pi}^j{}_{, i}
	- \frac{1}{2} \delta_{ij} \tilde{\pi}^k{}_{, k}
	- \frac{1}{2} \Delta^{-1} \tilde{\pi}^k{}_{, ijk} \,.
\end{equation}
The advantage of this gauge is that in (\ref{gdecomp}) there is a trace term but no
longitudinal part related to a vector potential, while in (\ref{pidecomp}) it is
the other way around. Because of the orthogonality of the individual parts of the transverse-traceless decomposition,
the kinetic term $\pi^{ij} \gamma_{ij,0}$ in the action turns into $\pi^{ij\text{TT}} h^{\text{TT}}_{ij,0}$.
Then only the transverse-traceless parts remain dynamical variables.
Now the four field constraints can be solved for the four nondynamical variables $\phi$ and $\tilde{\pi}^i$
in terms of $h^{\text{TT}}_{ij}$ and $\pi^{ij\text{TT}}$.
An analytic solution for $\phi$ and $\tilde{\pi}^i$, however,
can in general only be given in some approximation scheme.
Notice that ADM introduced two slightly different gauges \cite{Arnowitt:Deser:Misner:1962,*Arnowitt:Deser:Misner:2008},
the one used here was actually seldom used by ADM themselfes.
However, the gauge used here is better for applications,
as the form of the trace term in (\ref{gdecomp}) is adapted
to the Schwarzschild metric in isotropic coordinates
(with obvious advantages for perturbative expansions).
The action turns into
\begin{equation}
W_G[h^{\text{TT}}_{ij}, \pi^{ij\text{TT}}] =  \frac{1}{16\pi} \int \dd^4 x \, \pi^{ij\text{TT}} h^{\text{TT}}_{ij,0}
	- \int \dd t \, H_{\text{ADM}} \,,
\end{equation}
where the ADM Hamiltonian $H_{\text{ADM}}$ is just the ADM energy $E$ expressed
in terms of the gauge-reduced canonical variables $h^{\text{TT}}_{ij}$ and $\pi^{ij\text{TT}}$,
\begin{equation}
H_{\text{ADM}} = E[h^{\text{TT}}_{ij}, \pi^{ij\text{TT}}]
	= - \frac{1}{16\pi} \int \dd^3 x \, \Delta \phi[ h^{\text{TT}}_{ij}, \pi^{ij\text{TT}} ] \,.
\end{equation}
Notice that the surface integral (\ref{Esurf}) was written as a volume integral now
and the asymptotic behavior of $\phi$ was used.
The action must be varied only with respect to the independent components of $h^{\text{TT}}_{ij}$
and $\pi^{ij\text{TT}}$, which is ensured with the help of the transverse-traceless projector
\begin{equation}\label{TTproj}
\begin{split}
\delta^{\text{TT}kl}_{ij} &=
\tfrac{1}{2} [
	(\delta_{ik}-\Delta^{-1}\pa_{i}\pa_{k})(\delta_{jl}-\Delta^{-1}\pa_{j}\pa_{l})
	+(\delta_{il}-\Delta^{-1}\pa_{i}\pa_{l})(\delta_{jk}-\Delta^{-1}\pa_{j}\pa_{k}) \nl
	-(\delta_{kl}-\Delta^{-1}\pa_{k}\pa_{l})(\delta_{ij}-\Delta^{-1}\pa_{i}\pa_{j})
] \,.
\end{split}
\end{equation}
The Poisson brackets after gauge fixing correspondingly read
\begin{equation}
\{ h^{\text{TT}}_{ij}({\bf x}), \pi^{kl\text{TT}}({\bf x}') \}
	= 16\pi \delta^{\text{TT}kl}_{ij} \delta({\bf x} - {\bf x}') \,.
\end{equation}

\mysubsection{Matter Couplings\label{matcoup}}
Point-masses are the simplest kind of matter that can be coupled to {\GR}.
Its contribution to the action is just
\begin{equation}
W_M[g_{\mu\nu}, z^{\mu}] = \int \dd \tau \, L_M \,, \qquad
L_M = - m \sqrt{- g_{\mu\nu}(z^{\rho}) u^{\mu} u^{\nu}} \,.
\end{equation}
This action is invariant under a change of the parameter $\tau$, which simplifies
the variation as no constraint of the form $u_{\mu} u^{\mu} = -1$ is needed.
$m$ is assumed to be a constant.
Variation of the action leads to the equations of motion
\begin{equation}
\frac{\dD}{\dd \tau} \left[ \frac{u^{\mu}}{\sqrt{- u_{\rho} u^{\rho}}} \right] = 0 \,.
\end{equation}
These equations only have a unique solution if a gauge for $\tau$ is chosen.
The Einstein field equations now have a source $T^{\mu\nu}$,
\begin{equation}
\Riem^{\mu\nu}_{(4)} - \frac{1}{2} g^{\mu\nu} \Riem_{(4)} = 8\pi T^{\mu\nu} \,, \qquad
\text{with} \; \sqrt{-g} T^{\mu\nu} \equiv 2 \frac{\delta W_M}{\delta g_{\mu\nu}} \,,
\end{equation}
and $\Riem^{\mu\nu}_{(4)}$ the 4-dimensional Ricci tensor. The singular stress-energy tensor density reads explicitly
\begin{equation}
\sqrt{-g} T^{\mu\nu} = \int \dd \tau
	\frac{m u^{(\mu} u^{\nu)}}{\sqrt{- u_{\rho} u^{\rho}}} \delta_{(4)} \,.
\end{equation}
The 4-dimensional momentum is introduced as
\begin{equation}
p_{\mu} = \frac{\partial L_M}{\partial u^{\mu}} = m \frac{u_{\mu}}{\sqrt{- u_{\rho} u^{\rho}}} \,.
\end{equation}
It obviously holds\footnote{Due to Euler's theorem, this actually holds for any
Lagrangian which is a homogeneous function of degree one in
the velocity $u^{\mu}$. This in turn is required by reparametrization invariance.}
\begin{equation}
L_M = \frac{\partial L_M}{\partial u^{\mu}} u^{\mu} = p_{\mu} u^{\mu} \,.
\end{equation}
Thus a Legendre transformation leads to a vanishing canonical (i.e., defined as usual) Hamiltonian.
Its place is taken by the mass-shell constraint
\begin{equation}\label{mshell}
p_{\mu} p^{\mu} + m^2 = 0 \,,
\end{equation}
which has to be added to the action via a Lagrange multiplier $\lambda(\tau)$,
as further explained in the next {\secname}.
This constraint is a consequence of the inability to express $u_{\mu}$ uniquely in terms of $p_{\mu}$,
which in turn is due to invariance under reparametrization, or gauging, of $\tau$.
Indeed, it is a common feature of reparametrization invariant actions
that the canonical Hamiltonian vanishes and the time evolution
is instead generated by certain constraints. As seen in the last {\secname}
this also holds for {\GR}, whose action is invariant under
reparametrizations of spacetime, or general coordinate transformations.

Up to now the matter action was transformed into\footnote{Fields within the matter action are always taken
at the position $z^{\mu}$ from now on.}
\begin{equation}\label{ppaction}
W_M[g_{\mu\nu}, z^{\mu}, p_{\mu}, \lambda] = \int \dd \tau \, (p_{\mu} u^{\mu} - H_{M\tau}) \,, \qquad
H_{M\tau} = \lambda (g^{\mu\nu} p_{\mu} p_{\nu} + m^2) \,.
\end{equation}
Notice that the Hamiltonian $H_{M\tau}$ generates an evolution with respect
to the arbitrary parameter $\tau$.
Further the variation $\delta p_{\mu}$ leads to $u_{\mu} = 2 \lambda p_{\mu}$
and from the mass-shell constraint (\ref{mshell}) one thus has
$\lambda = \frac{1}{2m} \sqrt{- u_{\mu} u^{\mu}}$. It may be
checked that the equations of motion for $p_{\mu}$ and the
stress-energy tensor are equivalent to the ones above, which justifies the
Legendre transformation in the presence of constraints.
More on constrained Hamiltonian dynamics is discussed in the next {\secname}.
By solving the constraint and applying the gauge choice $\tau = z^0 \equiv t$, or $u^0 = 1$, the action is expressed
in terms of the independent variables $p_i$ and $z^i$. It holds
\begin{equation}
p_0 = ( {\gamma_0}^{\mu} - n_0 n^{\mu}) p_{\mu}
	= g_{0i} \gamma^{ij} p_j + N np \,,
\end{equation}
where $np \equiv n^{\mu} p_{\mu}$. From the constraint we get
\begin{equation}
( \gamma^{\mu\nu} - n^{\mu} n^{\nu} ) p_{\mu} p_{\nu} + m^2 = 0
\qquad \Rightarrow \qquad np = - \sqrt{m^2 + \gamma^{ij} p_i p_j} \,.
\end{equation}
Further we have
\begin{equation}
0 = n_i = n^{\mu} g_{\mu i} = \frac{1}{N} ( g_{0i} - N^j g_{ji} )
\qquad \Rightarrow \qquad g_{0i} = \gamma_{ij} N^j \,.
\end{equation}
Putting all together we arrive at
\begin{equation}\label{Lpm31}
W_M[\gamma_{ij}, N, N^i, z^i, p_i] = \int \dd t \, (p_i \dot{z}^i - H_M) \,, \qquad
H_M = - p_0 = - N np - N^i p_i \,,
\end{equation}
where a dot $\dot{~}$ denotes the total time derivative $\frac{\dd}{\dd t}$.
The original Hamiltonian $H_{M\tau}$ vanishes by virtue of the constraint.
Variation of the matter variables $z^i$ and $p_i$ results in Hamilton's
equations with $H_M$ as the matter part of the Hamiltonian. Thus $z^i$ and $p_i$ have the
Poisson brackets
% \begin{equation}
$\{ z^i, p_j \} = \delta_{ij}$,
% \end{equation}
all other zero. As in the last {\secname}, the variables $\gamma_{ij}$, $\pi^{ij}$, $N$, and $N^i$
are now used for the gravitational field.

The gauge fixing procedure is analogous to the last {\secname}, there are just certain
matter corrections to the field constraints following from the $N$- and $N^i$-variations,
\begin{equation}
\mathcal{H} \equiv \mathcal{H}^{\text{field}} + \mathcal{H}^{\text{matter}} = 0 \,, \qquad
\mathcal{H}_i \equiv \mathcal{H}^{\text{field}}_i + \mathcal{H}^{\text{matter}}_i = 0 \label{Consts} \,,
\end{equation}
where
\begin{equation}
\mathcal{H}^{\text{matter}} = - np \, \delta = \sqrt{m^2 + \gamma^{ij} p_i p_j} \, \delta \,, \qquad
\mathcal{H}^{\text{matter}}_i = p_i \delta \,,
\end{equation}
with $\delta = \delta(\vct{x} - \vct{z})$.
The first relation in (\ref{Consts}) is called the Hamilton constraint, while
the second one is the momentum constraint.
The ADM Hamiltonian $H_{\text{ADM}}$ still results from the ADM energy by solving the field constraints
using the gauge conditions (\ref{gdecomp}, \ref{pidecomp}), but now also depends on the matter variables
$z^i$ and $p_i$, which have entered via source terms of the constraints.
All field \emph{and matter} interaction terms in the action,
\begin{equation}\label{PMaction}
W = \frac{1}{16\pi} \int \dd^4 x \, \pi^{ij\text{TT}} h^{\text{TT}}_{ij,0}
	+ \int \dd t \bigg[ p_i \dot{z}^i - H_{\text{ADM}} \bigg] \,,
\end{equation}
are contained in the ADM Hamiltonian, or the ADM energy. This is a unique
feature of {\GR}, and still holds for couplings to other matter and even
other \emph{fields} \cite{DeWitt:1967}.

Finally we review the most important couplings of matter and fields to gravity that have
received a canonical formulation, see also \cite{Isenberg:Nester:1980}.
Besides for point-masses \cite{Arnowitt:Deser:Misner:1960:3,*Kimura:1961,Arnowitt:Deser:Misner:1962,*Arnowitt:Deser:Misner:2008},
such canonical formulations were found for fluids
\cite{Kunzle:Nester:1984,*Holm:1985,*Bao:Marsden:Walton:1985},
massive scalar fields \cite{Schwinger:1963:1,Schwinger:1963:2}, spin-$\frac{1}{2}$ Dirac fields
\cite{Dirac:1962,Kibble:1963,Deser:Isham:1976,Geheniau:Henneaux:1977,Nelson:Teitelboim:1978},
and gauge spin-1 fields, including Maxwell
\cite{Arnowitt:Deser:Misner:1960:5,Schwinger:1963:2} and Yang-Mills \cite{Teitelboim:1980}.
Problematic from a canonical point of view are derivative-coupled theories
\cite{Isenberg:Nester:1980}, like Dirac fields and also pole-dipole objects.
It is thus fortunate that the sought-for canonical formulation of
pole-dipole objects will be seen to resemble to Dirac fields coupled to gravity,
for which a canonical formulation was found.
Though the classical spin of pole-dipole objects is not restricted in its size,
we consider pole-dipole objects only at linear order in spin here. This means the spin is treated
as an infinitesimal quantity and thus formally takes on the smallest (nonzero) classical value,
which seems to give rise to similarities to the minimal (nonzero) quantized spin $\frac{1}{2}$ of Dirac fields.
Thus the achievements on canonical formulations of Dirac fields coupled to gravity
served as a very useful guide here, in particular the paper of Kibble \cite{Kibble:1963}.
However, an additional problem to be solved for spinning objects  in {\GR}
concerns the canonical spin supplementary condition.
So far the canonical formulation of spinning objects was found for
test-bodies in an external gravitational field \cite{Kunzle:1972}, see also the very
recent work in \cite{Barausse:Racine:Buonanno:2009}. %, but not coupled to a dynamical gravitational field.

\mysubsection{Other Formalisms and Constrained Hamiltonian Dynamics\label{Dirac}}
Before gauge fixing, {\GR} possesses a canonical formulation in the presence of
the constraints $\mathcal{H} = 0$ and $\mathcal{H}_i = 0$. There exists a
general framework to handle such a constrained Hamiltonian dynamics,
which was developed most notably by Dirac as a general route to canonical quantization
\cite{Dirac:1950,*Dirac:1951,*Dirac:1958:1,*Dirac:1964},
see also \cite{Hanson:Regge:1974,Hanson:Regge:Teitelboim:1976,Henneaux:Teitelboim:1992}.
Further important work was done by Bergmann and his collaborators, but focused
on {\GR} and its canonical quantization
\cite{Bergmann:1949,*Bergmann:Brunings:1949,*Bergmann:Penfield:Schiller:Zatzkis:1950,*Anderson:Bergmann:1951,*Bergmann:Goldberg:1955,*Anderson:1958}.
Though Dirac also considered the canonical formulation of {\GR}
\cite{Dirac:1958:2,*Dirac:1959},
his approach is formulated in a very general way.
Early work on this subject can even be traced back to Rosenfeld \cite{Rosenfeld:1930};
for a historical review, see, e.g, \cite{Salisbury:2007,Pons:2005}.
A particular important achievement of ADM for canonical general relativity
was the identification of the ADM Energy as the Hamiltonian after gauge fixing
\cite{Arnowitt:Deser:Misner:1960:4,Arnowitt:Deser:Misner:1959,*Arnowitt:Deser:Misner:1960:1}.
Yet another canonical treatment of {\GR} was given by Schwinger \cite{Schwinger:1963:1}.
%based on his action principle \cite{Schwinger:1951,Schwinger:1953:1}.
This formulation is similar to the ADM one, essentially only different variables were used
and many more such reformulations are possible. A further very appealing formulation
was given by Ashtekar \cite{Ashtekar:1986,*Ashtekar:1987},
in whose variables the gravitational
constraints considerably simplify, and which forms the basis of loop quantum gravity, see, e.g., \cite{Rovelli:2008}.

We will now summarize some of the results of Rosenfeld, Dirac, and Bergmann
on constrained Hamiltonian dynamics. In the last {\secname} the mass-shell
constraint (\ref{mshell}) manifests the inability to uniquely express the
velocity $u^{\mu}$ in terms of the corresponding momentum $p_{\mu}$.
The standard route to a Hamiltonian seems to be impassable in such a situation as the Legendre
transformation can not be applied in its usual way. The solution, however, is simple.
The Legendre transformation may formally be performed as usual if one adds the emerging
constraints via Lagrange multipliers to the action. The additional degrees
of freedom introduced by these multipliers correctly parametrize the ambiguity
present in the relation between velocities and momenta. Further, it can be shown
that the dynamics of the transformed action is equivalent to the dynamics of the
original action. The constraints arising at this stage are entitled as \emph{primary}
and the Hamiltonian is called the \emph{total Hamiltonian} (or Dirac Hamiltonian),
as it includes the primary constraints via Legendre multipliers.

The next step in the analysis of constrained Hamiltonian dynamics consists of
evaluating the consistency requirement that all primary constraints must
be preserved under the time evolution given by the total Hamiltonian.
Of course, some of the resulting consistency conditions can be identically fulfilled
or lead to contradictions (then the dynamics must be considered as inconsistent).
Moreover some conditions are restrictions for the Lagrange multipliers appearing
in the total Hamiltonian. Due to linearity of the total Hamiltonian in the Lagrange multipliers,
these restrictions are actually \emph{linear} equations.
Further, one might also obtain new (independent) constraints from the consistency conditions.
Such new constraints are called \emph{secondary} constraints. For these new constraints the same
consistency requirement applies, and one is eventually lead to further conditions on the Lagrange multipliers and/or to
further secondary constraints %\footnote{Sometimes these constraints are also called tertiary constraint.}
and so on. Finally, one ends up with a complete set of constraints and
linear equations for the Lagrange multipliers.
% primary, secondary, tertiary, quaternary, quinary, senary, septenary, octonary, nonary, denary

The linear equations for the Lagrange multipliers can be used to eliminate certain
linear combinations of these multipliers from the equations of motion.
The usual situation known from courses on classical mechanics is that all
multipliers are uniquely fixed.
However, in the general case some combinations of Lagrange multipliers could remain unfixed and
thus remain as arbitrary degrees of freedom in the equations of motion. The
interpretation is that these degrees of freedom are \emph{physically irrelevant} %indistinguishable
and correspond to a \emph{gauge freedom} of the theory. That is, the corresponding
independent Lagrange multipliers can be chosen at will, interpreted as choosing a gauge.
Hamiltonian formulations of gauge theories will inevitably involve constraints.

The Lagrange multipliers enter the total Hamiltonian together with the primary
constraints. Instead of characterizing the gauge freedom of a theory by
undetermined combinations of Lagrange multipliers, one can give a description in
terms of corresponding primary constraints. For this purpose it is useful to
introduce the notion of \emph{first class} and \emph{second class} constraints.
First class constraints are defined to have vanishing Poisson brackets with
all other constraints. A constraint that is not first class is called
second class. In addition to being first or second
class, the constraints can still be primary or secondary,
and one thus has four categories of constraints now.
An important fact is that the number of independent primary first class constraints is equal to
the number of unfixed Lagrange multipliers in the equations of motion
and thus to the number of gauge degrees of freedom.

Not only the primary first class constraints but also all secondary first class constraints
are related to gauge symmetries \cite{Castellani:1982,Gomis:Henneaux:Pons:1990}
(at least under certain reasonable conditions), see also \cite{Pons:2005}.
To be more precise, all first class constraints, primary as well as secondary,
appear in the generators of gauge symmetries on phase space.
The algebra of first class constraints is therefore related to the algebra
of gauge symmetry generators of the theory. For {\GR}, the algebra
of first class constraints reads \cite{Schwinger:1963:1,DeWitt:1967}
(at least for the vacuum case and for coupling to point-masses)
\begin{align}
	\{ \mathcal{H}(\vct{x}) , \mathcal{H}(\vct{x}') \} =&
		- \left[ \mathcal{H}_i(\vct{x}) \gamma^{i j}(\vct{x})
		+ \mathcal{H}_i(\vct{x}') \gamma^{i j}(\vct{x}') \right] \partial_j \delta(\mathbf{x} - \mathbf{x}') \,, \label{constalg1} \\
	\{ \mathcal{H}_i(\vct{x}) , \mathcal{H}(\vct{x}') \} =&
		- \mathcal{H}(\vct{x}) \, \partial_i \delta(\mathbf{x} - \mathbf{x}') \,, \label{constalg2} \\
	\{ \mathcal{H}_i(\vct{x}) , \mathcal{H}_j(\vct{x}') \} =&
		- \mathcal{H}_j(\vct{x}) \, \partial_i \delta(\mathbf{x} - \mathbf{x}')
		- \mathcal{H}_i(\vct{x}') \, \partial_j \delta(\mathbf{x} - \mathbf{x}') \,. \label{constalg3}
\end{align}
If one goes to the constraint surface by $\mathcal{H} = 0 = \mathcal{H}_i$, then
the right-hand sides vanish. Thus $\mathcal{H}$ and $\mathcal{H}_i$ are indeed
first class. In order to relate this algebra to 4-dimensional diffeomorphism
invariance one should include lapse and shift as well as corresponding momenta
into phase space \cite{Castellani:1982}. It should be noted that though
the total Hamiltonian of {\GR} (\ref{Dham}) is composed of the first class
constraints and looks quite similar to the generator of gauge transformations,
the time evolution given by this Hamiltonian is not just a gauge effect, see, e.g,
\cite{Pons:Salisbury:Sundermeyer:2010}.

One can elaborate more on the distinction between first and second class
constraints. Obviously one can recombine the whole set of constraints into
some equivalent set. We consider the case that such a recombination brings
as many constraints as possible from the second class into the first class.
One can then show by a reductio ad absurdum that the matrix
% \begin{equation}
$c_{ab} = \{ \psi_a, \psi_b \}$,
% \end{equation}
where $\psi_a$ are the constraints that remain second class after
recombination\footnote{The indices $a$ and $b$ label constraints in this {\secname}.},
is invertible, $\det ( c_{ab} ) \neq 0$. The Dirac bracket $\{ A, B \}^{*}$ between two
phase space functions $A$ and $B$ is then defined by
\begin{equation}\label{DB}
\{ A, B \}^{*} = \{ A, B \} - \{ A, \psi_a \} (c^{-1})_{ab} \{ \psi_b, B \} \,.
\end{equation}
This bracket satisfies the laws known from the Poisson bracket. Further, it leads
to the correct equations of motion together with the total Hamiltonian.
The Dirac bracket can thus be used as a substitute for the Poisson bracket.
However, whereas one may use the constraints only after all Poisson brackets were
calculated\footnote{For a more detailed exposition it is useful to introduce the concepts of weak and strong equality.},
the second class constraints $\psi_a = 0$ can be used
\emph{before} an application of the Dirac bracket without changing the result
(e.g., one has $\{ A, \psi_a \}^{*} = 0$ for all $A$ and $\psi_a$).
If one restricts to use the Dirac bracket instead of the Poisson
bracket, one can use the second class constraints $\psi_a = 0$ to solve for
certain phase space variables and eliminate them from all
quantities. Then one has performed an actual reduction of the degrees of freedom. Within this formalism,
gauge conditions are constraints added by hand that bring all (or just some) first class
constraints into the second class. The reduction of degrees of freedom via
gauge fixing then follows with the help of the Dirac bracket in a straightforward way.

\mychapter{Action Approach\label{action}}
In this {\chapname} an extension of the ADM formalism for point-masses
to the pole-dipole approximation is obtained linear in spin.
The derivation is based on a corresponding extension of the point-mass action.

\mysection{Action of the Spherical Top}
It is remarkable that equations of motion (\ref{eom}) and stress-energy tensor (\ref{PDset})
in the pole-dipole approximation are independent of the specific object, i.e., are the same
for black holes and neutron stars. It is thus expected that \emph{any} specific action for a
spinning object coupled to {\GR} will contain (\ref{eom}) and (\ref{PDset})
to some approximation (e.g., linear in spin). The pole-dipole action found here will be based on the
simplest spinning object imaginable --- the spherical top.

\mysubsection{Newtonian Case\label{nTop}}
The spherical top is well-known in classical mechanics. However, we review it
here in a way that allows an easy transition to the special relativistic treatment in \cite{Goenner:Westpfahl:1967,Hanson:Regge:1974}.
We consider in this {\secname} a top with its center of mass resting at the coordinate origin. The center of mass motion
can be added easily. The top can be described as a rigid body consisting of many point-masses
labeled by an index $a$, with positions $z_a^i$ and masses $m_a$.
In terms of body-fixed (constant) coordinates $z_a^{[i]}$ it holds
% \begin{equation}
$z_a^i(t) = \Lambda^{[j]i}(t) z_a^{[j]}$,
% \end{equation}
with a time-dependent rotation matrix $\Lambda^{[j]i}$, $\Lambda^{[k]i} \Lambda^{[k]j} = \delta_{ij}$.
Here and in the following we will indicate 3-dimensional indices in the body-fixed coordinate system
by square brackets. The rotation matrix can be expressed in terms of three independent
angle variables, $\Lambda^{[i]j} = \Lambda^{[i]j}(\varphi_1, \varphi_2, \varphi_3)$, e.g., the Euler angles.
The antisymmetric\footnote{The antisymmetry immediately follows from
the time derivative of $\Lambda^{[k]i} \Lambda^{[k]j} = \delta_{ij}$.}
angular velocity tensor is given by
% \begin{equation}
$\Omega^{ij} = \Lambda^{[k]i} \dot{\Lambda}^{[k]j}$. 
% \end{equation}
A spherical top is completely characterized by \emph{one} moment of inertia $I$,
it holds $2 \sum_a m_a z_a^{[i]} z_a^{[j]} = I \delta_{ij}$.
The Lagrangian of the free spherical top then reads
\begin{equation}
L(\Lambda^{[i]j}, \Omega^{ij}) = \frac{1}{2} \sum_a m_a \dot{z}_a^i \dot{z}_a^i = \frac{1}{4} I \Omega^{ij} \Omega^{ij} = \frac{1}{2} I \Omega^i \Omega^i \,,
\end{equation}
where $\Omega^i = \frac{1}{2} \epsilon_{ijk} \Omega^{jk}$ is the usual
angular velocity vector.
The spin is the generalized momentum of the angular velocities of the form
\begin{equation}
S_{ij} = 2 \dfrac{\partial L}{\partial \Omega^{ij}} = I \Omega_{ij} \,.
\end{equation}
Legendre transformation leads to
\begin{equation}\label{cltop}
L = \frac{1}{2} S_{ij} \Omega^{ij} - H \,,
\end{equation}
with the Hamiltonian $H(\Lambda^{[i]j}, S_{ij}) = \frac{1}{4 I} S_{ij} S_{ij}$. This specific Hamiltonian is
actually independent of $\Lambda^{[i]j}$.

In order to derive the general Euler-Lagrange equations of the Lagrangian (\ref{cltop}) we are not
varying the independent angle variables, but instead use
$\delta \theta^{ij} = \Lambda^{[k]i} \delta \Lambda^{[k]j}$ as independent variations.
Notice that $\delta \theta^{ij}$ is antisymmetric, and thus indeed corresponds
to three independent variations of the angle variables. The result is
\begin{equation}\label{topHam}
\Omega^{ij} = \Lambda^{[k]i} \dot{\Lambda}^{[k]j} = 2 \frac{\partial H}{\partial S_{ij}} \,, \qquad
\dot{S}_{ij} = 2 S_{k[i} \Omega_{j]k}
	- \Lambda^{[k]i} \frac{\partial H}{\partial \Lambda^{[k]j}} + \Lambda^{[k]j} \frac{\partial H}{\partial \Lambda^{[k]i}} \,.
\end{equation}
These are Hamilton's equations for $\Lambda^{[k]j}$ and $S_{ij}$. The Poisson brackets fulfill
\begin{equation}
\dot{A} = \{ A, H \} + \frac{\partial A}{\partial t} \,,
\end{equation}
for a general quantity $A$. Comparing with (\ref{topHam}) we can read off
\begin{gather}
\{ \Lambda^{[i]j}, \Lambda^{[k]l} \} = 0 \,, \qquad
\{ \Lambda^{[i]j}, S_{kl} \} = \Lambda^{[i]k} \delta_{lj} - \Lambda^{[i]l} \delta_{kj} \,, \label{nrPB1} \\
\{ S_{ij}, S_{kl} \} = \delta_{ik} S_{jl} - \delta_{jk} S_{il}
	- \delta_{il} S_{jk} + \delta_{jl} S_{ik} \,. \label{nrPB2}
\end{gather}
Alternatively one could use canonical variables based on the angle variables,
\begin{equation}
\{ \varphi_i, p_j^{\varphi} \} = \delta_{ij}, \qquad \text{with} \; p_i^{\varphi} = \frac{\partial L(\varphi_j, \dot{\varphi}_k)}{\partial \dot{\varphi}_i} \,,
\end{equation}
as in most textbooks.
Further, if the Hamiltonian is independent of $\Lambda^{[i]j}$, which will always be
the case in the following, the spin length is a constant and it is possible to describe
each spin by only two independent canonical variables instead of six contained in
$S_{ij}$ and $\Lambda^{[i]j}$, see, e.g., \cite{Bel:Martin:1980,*Wu:Xie:2010,Damour:Jaranowski:Schafer:2008:1}.
However, we prefer the variables $S_{ij}$ and $\Lambda^{[i]j}$ here.

\mysubsection{Special Relativistic Case\label{nrTop}}
In the relativistic case there are no rigid bodies. However, one can define a top
in a purely mathematical way \cite{Hughes:1961,*Itzykson:Voros:1972,Goenner:Westpfahl:1967,Hanson:Regge:1974}
as a worldline with a Lorentz matrix $\Lambda_{A\mu}$,
% \begin{equation}
$\eta^{AB} \Lambda_{A\mu} \Lambda_{B\nu} = \eta_{\mu\nu}$,
% \end{equation}
such that $\Lambda_{A\mu}$ is a pure rotation,
\begin{equation}
\Lambda_{A\mu} = \left( \begin{array}{cc} -1 & 0 \\ 0 & \Lambda_{[i]j} \end{array} \right) \,,
\end{equation}
in some frame defined by $f_{\mu}$. (Upper case Latin indices from the beginning
of the alphabet refer to the body-fixed frame and have the values $A = [0], [i]$.)
This can be formulated as
\begin{equation}\label{Lssc}
\Lambda_{[0]\mu} = \frac{f_{\mu}}{\sqrt{-f_{\nu} f^{\nu}}} \,, \quad
\text{or} \quad \eta^{[0] A} = - \frac{\Lambda^{A \mu} f_{\mu}}{\sqrt{-f_{\nu} f^{\nu}}} \,.
\end{equation}
It holds
\begin{equation}
\Omega^{\mu\nu} = {\Lambda_{A}}^{\mu} \frac{\dd \Lambda^{A\nu}}{\dd \tau} \,, \qquad
S_{\mu\nu} = 2 \frac{\partial L(u^{\mu}, \Omega^{\mu\nu})}{\partial \Omega^{\mu\nu}} \,, \qquad
p_{\mu} = \frac{\partial L(u^{\mu}, \Omega^{\mu\nu})}{\partial u^{\mu}} \,,
\end{equation}
see, e.g, \cite{Hanson:Regge:1974} or the next {\secname}.
The spin supplementary condition belonging to (\ref{Lssc}) reads
\begin{equation}\label{covSSC2}
S_{\mu\nu} f^{\mu} = 0 \,.
\end{equation}
It will be seen in \Sec{ASSC} in which sense this belongs to (\ref{Lssc}).
Notice that only three relations of (\ref{Lssc}) are independent, e.g., one could
equivalently require $\Lambda^{[i] \mu} f_{\mu} = 0$ only.
The same holds for (\ref{covSSC2}).

% In \cite{Hanson:Regge:1974} the Lagrangian was required to directly produce
% (\ref{covSSC2}) as a constraint on the momenta $S_{\mu\nu}$. This is not easy to ensure as the momenta
% are partial derivatives of the Lagrangian, and one thus gets partial differential equations for the Lagrangian.
% As we are only interested in the linear order in spin,
% this would not pose a severe problem here.
% However, the constraint (\ref{Lssc}) can not be produced in such a way.
% We will therefore not require that the action automatically leads to a spin
% supplementary condition here.
There are many ways to implement the conditions (\ref{Lssc}) and (\ref{covSSC2})
in an action approach, see, e.g., \cite{Frenkel:1926,Goenner:Westpfahl:1967,Hanson:Regge:1974}.
We require here that (\ref{Lssc}) and (\ref{covSSC2}) are preserved under the time evolution given by the
action and try to directly construct such an action.
An alternative, rather indirect, approach would be to add the supplementary conditions
to some action with the help of Lagrange multipliers. As well known from
classical mechanics, this modifies the dynamics by constraint forces,
which ensure that the supplementary conditions are preserved in time.
However, one should carefully check the consistency, in
particular one should be able to find a solution for the Lagrange multipliers.
Also no further (secondary) constraints should appear, which would be physically unacceptable
(we want to have exactly three independent rotational degrees of freedom).
Finally, the Lagrange multipliers can be eliminated from
the action, leading to a dynamics which preserves the constraints
and thus to the action we try to find directly here.

\mysubsection{Minimal Coupling to Gravity\label{mincoup}}
The next logical step is a minimal coupling of the special relativistic spherical top
defined in the last {\secname} to gravity. Such a coupling was already treated
in \cite{Westpfahl:1969:2} based on the developments in \cite{Goenner:Westpfahl:1967}.
In \cite{Bailey:Israel:1975} even nonminimal couplings leading to higher multipole corrections
were considered\footnote{This obviously goes beyond a spherical top, however, the formalism stays the same.}.
Notice that \cite{Bailey:Israel:1980} is not a further development of \cite{Bailey:Israel:1975},
but is a completely different action approach. More recently
yet another approach was given in \cite{Porto:2006} with focus
on an application to the post-Newtonian approximation.

The matter variables $\Lambda_{A\mu}$ have the problem that they fulfill
\begin{equation}\label{BIconstraint}
\Lambda_{A\mu} {\Lambda^A}_{\nu} = g_{\mu\nu} \,.
\end{equation}
That is, $\Lambda_{A\mu}$ is not independent under variation of the metric.
For the Dirac field, one has a similar problem
with the gamma matrices $\gamma_{\mu}$, as it holds
% \begin{equation}
$\gamma_{\mu} \gamma_{\nu} + \gamma_{\nu} \gamma_{\mu} = 2 g_{\mu\nu}$.
% \end{equation}
This problem can be overcome by writing ${\Lambda^A}_{\mu} = \Lambda^{AI} e_{I\mu}$
and treating $\Lambda^{AI}$ and the tetrad field $e_{I\mu}$ as independent variables.
From
\begin{equation}\label{Lcond}
\Lambda_{AI} {\Lambda^A}_{J} = \eta_{IJ} \,,
\quad \text{or} \quad
\gamma_I \gamma_J + \gamma_J \gamma_I = 2 \eta_{IJ} \,,
\end{equation}
it is obviously now consistent that $\Lambda_{AI}$ and $\gamma_I$ are constant
under variations of the tetrad field $e_{I\mu}$. We have three bases involved,
a body-fixed basis, a local Lorentz basis (denoted by upper case Latin indices
from the middle of the alphabet), and a coordinate basis. The field equations
are obtained by an unconstrained variation of $e_{I\mu}$. The metric
$g_{\mu\nu} = e_{I\mu} {e^I}_{\nu}$ as well as the connection are not varied independently.
For the variation of $\Lambda_{AI}$ one has to take into account the condition (\ref{Lcond}).

In \cite{Westpfahl:1969:2,Bailey:Israel:1975,Porto:2006} matter and field degrees
of freedom are not clearly separated in the action.
% the notation $e_{A\mu} = \Lambda_{A\mu}$ is used, which suggests that $e_{A\mu}$ is the
% gravitational tetrad field.
For example, in \cite{Bailey:Israel:1975} the equations of motion for the matter variables
were obtained by adding (\ref{BIconstraint})
as a constraint to the action with the help of Lagrange multipliers,
whereas the field equations were obtained from an unconstrained variation of $\Lambda_{A\mu}$.
However, we need to separate matter and field degrees of freedom here,
which is essential for the canonical reduction in the next {\secsname}.

The covariant angular velocity in the local Lorentz basis can be defined as
\begin{equation}\label{Odef}
\Omega^{IJ} = {\Lambda_{A}}^{I} \frac{\dD \Lambda^{AJ}}{\dd \tau}
	= {\Lambda_{A}}^{I} \left[ \frac{\dd \Lambda^{AJ}}{\dd \tau}
		- {\Lambda^A}_{K} {{\omega_{\mu}}^{KJ}}(z^{\rho}) u^{\mu} \right] \,.
\end{equation}
Here ${{\omega_{\mu}}^{IJ}}$ are the Ricci rotation coefficients,
$e_{I\alpha} e_{J\beta} {\omega_{\mu}}^{IJ} = - \Gamma_{\beta\alpha\mu}^{(4)} + {e^K}_{\alpha , \mu} e_{K \beta}$,
and $\Gamma^{(4)}_{\alpha\mu\nu} = \frac{1}{2} (g_{\alpha\mu,\nu} + g_{\alpha\nu,\mu} - g_{\mu\nu,\alpha})$ is the 4-dimensional Christoffel symbol of first kind.
Notice that the covariant derivative does not act on indices referring
to the body-fixed frame. The matter action shall be of the general form
\begin{equation}\label{PDaction}
	W_M[e_{I\mu}, z^{\mu}, \Lambda^{AI}] = \int \dd \tau \,
		L_M( u^{\mu}, \Omega^{\mu\nu},% g^{\mu\nu}(z^{\rho}),
			g_{\mu\nu}(z^{\rho})) \,.
\end{equation}
The Lagrangian $L_M$ is restricted to
depend on the velocities $u^{\mu}$ and $\Omega^{\mu\nu} = {e_I}^{\mu} {e_J}^{\nu} \Omega^{IJ}$ only, and
not on the ``coordinates'' $z^{\mu}$ and $\Lambda^{AI}$ directly.
This ensures the covariance of the action.
The action shall be invariant under reparametrizations,
so $u^{\mu}$ is not constrained.
If we let the Lagrangian depend on the curvature tensor, we would
include quadrupole corrections, see \Sec{QuadAction}.
An important relation is given by (2c) in \cite{Goenner:Westpfahl:1967} or (9) in \cite{Bailey:Israel:1975},
which reads here
\begin{equation}\label{RPI2}
0 = \frac{\partial L_M}{\partial u^{\alpha}} u^{\beta}
	+ 2 \frac{\partial L_M}{\partial \Omega^{\alpha\nu}} \Omega^{\beta\nu}
%	+ 2 \frac{\partial L_M}{\partial g^{\alpha\nu}} g^{\beta\nu}
	- 2 \frac{\partial L_M}{\partial g_{\beta\nu}} g_{\alpha\nu} \,,
\end{equation}
and is a consequence of $L_M$ being a scalar\footnote{Loosely speaking, one
can read (\ref{RPI2}) as ``the number of upper indices minus the number of lower indices in $L_M$ is zero.''
This is derived in \cite{Bailey:Israel:1975} from an infinitesimal coordinate transformation.}.
Similar to \Sec{nTop}, the Euler-Lagrange equations are obtained with the help of the
antisymmetric variations $\delta \theta^{IJ} = {\Lambda_A}^I \delta \Lambda^{AJ}$, e.g.,
\begin{equation}\label{varOmega}
\delta \Omega^{IJ} = \frac{\dD \delta \theta^{IJ}}{\dd \tau}
	+ 2 {\Omega_K}^{[I} \delta \theta^{J]K}
	- \Riem^{(4)}_{\mu\nu}{}^{IJ} u^{\nu} \delta z^{\mu}
	- \frac{\dD}{\dd \tau} \left( {\omega_{\mu}}^{IJ} \delta z^{\mu} \right)
	- u^{\mu} \delta \omega_{\mu}\!^{IJ} \,. \\
\end{equation}
The $\delta \theta^{IJ}$-variation then leads to
\begin{equation}
\frac{\dD}{\dd \tau} \left[ \frac{\partial L_M}{\partial \Omega^{\mu\nu}} \right]
	= \frac{\partial L_M}{\partial \Omega^{\mu\rho}} {\Omega^{\rho}}_{\nu}
		- \frac{\partial L_M}{\partial \Omega^{\nu\rho}} {\Omega^{\rho}}_{\mu} \,. \label{SEOM}
\end{equation}
The $\delta z^{\mu}$-variation is subtle as it is not manifestly
covariant, see, e.g., the second last term in (\ref{varOmega}). This is
due to the fact that $\Lambda^{AI}$ is held constant for the variation
of the worldline $\delta z^{\mu}$, which is not a covariant process
(e.g., in contrast to a parallel transport of $\Lambda^{AI}$ to the new worldline).
However, using the equations of motion for $\Lambda^{AI}$, (\ref{SEOM}), and the covariance of $L_M$, (\ref{RPI2}),
the result of the $\delta z^{\mu}$-variation reads
\begin{equation}\label{zEOM}
\frac{\dD}{\dd \tau} \left[ \frac{\partial L_M}{\partial u^{\mu}} \right] =
	- \Riem_{\mu\nu}^{(4)}{}^{\alpha\beta} u^{\nu} \frac{\partial L_M}{\partial \Omega^{\alpha\beta}} \,,
\end{equation}
and is manifestly covariant now.
Further, by virtue of (\ref{RPI2}) we can write (\ref{SEOM}) as
\begin{equation}
2 \frac{\dD}{\dd \tau} \left[ \frac{\partial L_M}{\partial \Omega^{\mu\nu}} \right]
	= \frac{\partial L_M}{\partial u^{\mu}} u_{\nu}
		- \frac{\partial L_M}{\partial u^{\nu}} u_{\mu} \,. \label{SEOM2}
\end{equation}
At last, the field equations follow from the $\delta e_{I\mu}$-variation as
\begin{equation}
\Riem^{\mu\nu}_{(4)} - \frac{1}{2} g^{\mu\nu} \Riem_{(4)} = 8\pi T^{\mu\nu} \,, \qquad
\text{with} \; \sqrt{-g} T^{\mu\nu} \equiv {e_I}^{\mu} \frac{\delta W_M}{\delta e_{I \nu}} \,,
\end{equation}
where the left-hand side results from the Einstein-Hilbert part (\ref{EHaction})
and the stress-energy tensor density $\sqrt{-g} T^{\mu\nu}$ reads explicitly
\begin{equation}\label{Aset}
\sqrt{-g} T^{\mu\nu} = \int \dd \tau \left[
	u^{(\mu} g^{\nu)\alpha}
		\frac{\partial L_M}{\partial u^{\alpha}} \delta_{(4)}
	- \left( 2 \frac{\partial L_M}{\partial \Omega^{\alpha\beta}} g^{\alpha\rho} g^{\beta(\mu}
		u^{\nu)} \delta_{(4)} \right)_{||\rho}
\right] \,.
\end{equation}
Here the important relation (\ref{RPI2}) was used again and the antisymmetric part
\begin{equation}
\sqrt{-g} T^{[\mu\nu]} = \int \dd \tau \left[
	- \frac{\dD}{\dd \tau} \left( \frac{\partial L_M}{\partial \Omega^{\alpha\beta}} \right)
	+ 2 \frac{\partial L_M}{\partial \Omega^{\alpha\rho}} {\Omega^{\rho}}_{\beta}
	\right] g^{\alpha[\mu} g^{\nu]\beta} \delta_{(4)} = 0 \,,
\end{equation}
vanishes, see (\ref{SEOM}). Indeed, (\ref{SEOM}) is equivalent to $T^{[\mu\nu]} = 0$.

Comparing (\ref{eom}) and (\ref{PDset}) with
(\ref{zEOM}), (\ref{SEOM2}), and (\ref{Aset}) we get
\begin{equation}
S_{\mu\nu} = 2 \frac{\partial L_M}{\partial \Omega^{\mu\nu}} \,, \qquad
p_{\mu} = \frac{\partial L_M}{\partial u^{\mu}} \,, \label{pdefA}
\end{equation}
as in the special relativistic case.
It should be noted that the given derivation basically follows along the lines
of Bailey and Israel \cite{Bailey:Israel:1975}, but the used variables are
similar to Porto \cite{Porto:2006}, which resembles to \cite{Hanson:Regge:1974}.
However, the variables that are varied here differ from
both \cite{Bailey:Israel:1975} and \cite{Porto:2006}.
The relation between $p_{\mu}$ and $u_{\mu}$ is fixed by (\ref{pdefA}), which means that the action
already implements a specific spin supplementary condition, cf.\ \Eq{PUrel}.
If this would not be the case, then (\ref{pdefA}) should be of the form (\ref{PUnossc}),
which is impossible due to the assumed absence of accelerations in the action.
The approach in \cite{Goenner:Westpfahl:1967,Westpfahl:1969:2} includes
accelerations of the worldline coordinate.
Further, a noncovariant supplementary condition (e.g., the canonical one)
will result in a not manifestly covariant relation between $p_{\mu}$ and $u_{\mu}$, (\ref{pdefA}),
and thus one needs a not manifestly covariant action.
An approach via Lagrange multipliers as discussed in \Sec{nrTop}
seems to be better when using such conditions from the start.
Here we will start with the covariant supplementary conditions and
go over to the canonical ones later by a change of variables.

Now we have to find a suitable reparametrization-invariant Lagrangian.
An intuitive guess is (see \Sec{QuadAction} for more elaborated considerations)
\begin{equation}\label{Lspin}
L_M = \frac{1}{\sqrt{- u_{\rho} u^{\rho}}} \left[ m_0 u_{\mu} u^{\mu}
	+ \frac{I}{4} \Omega_{\mu\nu} \Omega^{\mu\nu} \right] \,,
\end{equation}
where $m_0$ and $I$ shall be constants. Then it holds
\begin{equation}
S_{\mu\nu} = \frac{I \Omega_{\mu\nu}}{\sqrt{- u_{\rho} u^{\rho}}} \,, \qquad
p_{\mu} = \left( m_0 + \frac{1}{4 I} S_{\mu\nu} S^{\mu\nu} \right) \frac{u_{\mu}}{\sqrt{- u_{\rho} u^{\rho}}} \,, \label{SPtop}
\end{equation}
and the dynamical mass $m = \sqrt{- p_{\mu} p^{\mu}}$ is given by
\begin{equation}\label{mtom0}
m = m_0 + \frac{1}{4 I} S_{\alpha\beta} S^{\alpha\beta} \,,
\end{equation}
or $m = m_0$ to linear order in spin. Then (\ref{SPtop}) agrees with (\ref{PUrel}) for $f_{\mu} = p_{\mu}$ at linear order in spin,
which implies that the corresponding spin supplementary condition (\ref{covSSC2}) is preserved in time.
(\ref{Lssc}) only needs to be preserved to zeroth order in spin, which is also the case
(see also \Sec{ASSC}). Due to reparametrization invariance, $L_M$ must be a homogeneous function
of degree one in the velocities and Euler's theorem leads to
\begin{equation}\label{Lmatlin}
L_M = \frac{\partial L_M}{\partial u^{\mu}} u^{\mu} + \frac{\partial L_M}{\partial \Omega^{\mu\nu}} \Omega^{\mu\nu}
= p_{\mu} u^{\mu} + \frac{1}{2} S_{\mu\nu} \Omega^{\mu\nu} \,.
\end{equation}
A Legendre transformation in $u^{\mu}$ and $\Omega^{\mu\nu}$ thus leads to a vanishing
result. Further, the mass-shell constraint (\ref{mshell}) follows
from (\ref{SPtop}), but no constraint on $S_{\mu\nu}$ arises from
(\ref{SPtop}) as opposed to \cite{Hanson:Regge:1974}.
(Indeed, in \cite{Hanson:Regge:1974} the action was constructed such
that the constraint (\ref{covSSC2}) arises directly from the action in this way.)
Similar to \Sec{matcoup} we finally have
\begin{equation}\label{pdaction}
W_M[e_{I\mu}, z^{\mu}, p_{\mu}, S_{\mu\nu}, \Lambda^{AI}, \lambda] = \int \dd \tau \, \left[ p_{\mu} u^{\mu}
	+ \frac{1}{2} S_{\mu\nu} \Omega^{\mu\nu} - H_{M\tau} \right] \,,
\end{equation}
with the function\footnote{Notice that $H_{M\tau}$ is not a Hamiltonian
as $\frac{1}{2} S_{\mu\nu} \Omega^{\mu\nu}$ in (\ref{pdaction}) also contains interaction terms.}
$H_{M\tau}$ containing the mass-shell constraint only,
% \begin{equation}\label{HM}
$H_{M\tau} \! = \! \lambda (g^{\mu\nu} p_{\mu} p_{\nu} + m^2)$.
% \end{equation}
This is the extension of (\ref{ppaction}) to the pole-dipole approximation
at linear order in spin.
% Remember that the variation of $\Lambda^{Aa}$ should be written as the
% independent antisymmetric variations $\delta \theta^{ab} = {\Lambda_A}^a \delta \Lambda^{Ab}$.
We could also add the supplementary conditions,
\begin{equation}\label{supplcondlin}
S_{i\nu} p^{\nu} = 0 \,, \qquad \Lambda^{[i] J} p_J = 0 \,,
\end{equation}
to the action with the help of Lagrange multipliers. However, this
will not change the dynamics as these (independent) conditions are already preserved
in time and their Lagrange multipliers therefore vanish.
Reference \cite{Steinhoff:Schafer:2009:2} immediately started with the
action in the form of \Eq{pdaction} without the detailed
derivation given in this {\secname}.

\mysection{Reduction of the Matter Variables}
Next a fully reduced canonical formalism is derived. For this
the action is put on the constraint surface. That is, all supplementary
conditions, constraints, and gauge conditions are solved in terms of certain
truly independent variables that parametrize the constraint surface.
The equations of motion for this reduced number of variables could
then be obtained by varying the action with respect to these variables.
However, we will transform the action to a new set of reduced variables
such that the equations of motion can easily be seen to resemble to
Hamilton's equations. This allows for an easy identification of
the Hamiltonian and corresponding Poisson brackets. Thus a
fully reduced canonical formalism for spinning objects coupled to
general relativity is found \cite{Steinhoff:Schafer:2009:2}.
Remember that the necessity for a variable transformation to obtain standard canonical Poisson brackets
is already present in the flat space case, see (\ref{covToNW}).
A treatment using Dirac brackets (\ref{DB}) seems to be more complicated,
as one has to consider the brackets for each pair of variables then,
whereas here we only have to handle the action (a single scalar).

The derivation sketched above is very similar to the treatment
of Dirac fields coupled to gravity by Kibble \cite{Kibble:1963}.
In this {\secname} we concentrate on the matter part of the action only.

\mysubsection{Reduced Matter Action}
Similar to \Sec{matcoup} we solve the matter constraints (now including the supplementary
conditions (\ref{supplcondlin})) as
\begin{gather}
	np \equiv n^{\mu} p_{\mu} = - \sqrt{m^2 + \gamma^{ij} p_{i} p_{j}} \,, \label{npDef} \\
	nS_{i} \equiv  n^{\mu} S_{\mu i}
		= \frac{p_{k} \gamma^{kj} S_{ji}}{np} = \gamma_{ij} nS^{j} \,, \label{nSDef} \qquad
	\Lambda^{[j](0)} = \Lambda^{[j](i)} \frac{p_{(i)}}{p^{(0)}} \,, \qquad
	\Lambda^{[0]I} = - \frac{p^I}{m} \,,
\end{gather}
in terms of the independent variables $p_i$, $S_{ij}$, and $\Lambda^{[i](k)}$.
On the constraint surface it holds $H_{M\tau} = 0$.

For simplicity, we will immediately constrain ourselves to the
Schwinger time gauge \cite{Schwinger:1963:1},
\begin{equation}\label{timegauge}
e^{(0) \mu} = - n^{\mu} \,,
\end{equation}
see also \cite{Dirac:1962,Kibble:1963,Nelson:Teitelboim:1978}, as lapse and shift then turn into
Lagrange multipliers in the matter action \cite{Steinhoff:Schafer:2009:2},
like in the ADM formalism for nonspinning objects.
This gauge condition effectively reduces the tetrad $e^{I \mu}$ to a
triad $e^{(i) j}$, it holds
\begin{align}
{e^{(0)}}_i &= 0 = {e_{(i)}}^0 \,, &
{e^{(0)}}_0 &= N = 1 / {e_{(0)}}^0 \,, &
{e^{(i)}}_0 &= N^j {e^{(i)}}_j \,, \\
N^i &=  - N {e_{(0)}}^i \,, &
\gamma_{ij} &= {e^{(m)}}_i e_{(m)j} \,, &
\gamma^{ij} &= {e_{(m)}}^i e^{(m)j} \,.
\end{align}
A further convenient gauge choice is $\tau=z^0=t$ for the yet arbitrary
parameter $\tau$. In terms of the independent variables the matter
Lagrangian (\ref{Lmatlin}) reads explicitly
\begin{equation}\label{Lmat31}
\begin{split}
L_M &= \bigg[ p_{i} + K_{ij} nS^j + A^{kl} e_{(j)k} {e^{(j)}}_{l,i}
		- \bigg( \frac{1}{2} S_{kj}
			+ \frac{p_{(k} nS_{j)}}{np}
		\bigg) \Gamma^{kj}{}_i \bigg] \dot{z}^{i} \nl
+ \frac{nS^i}{2 np} \dot{p}_i
+ \bigg[ S_{(i)(j)} + \frac{nS_{(i)} p_{(j)} -  nS_{(j)} p_{(i)}}{np} \bigg]
		\frac{{\Lambda_{[k]}}^{(i)} \dot{\Lambda}^{[k](j)}}{2} \nl
+ A^{ij} e_{(k)i} e^{(k)}{}_{j,0}
- \int \dd^3 x \, ( N \mathcal{H}^{\text{matter}} - N^i \mathcal{H}^{\text{matter}}_i ) \,,
\end{split}
\end{equation}
with the 3-dimensional Christoffel symbols $\Gamma_{kji}$, the abbreviation $A^{ij}$ defined by
\begin{equation}\label{defA}
\gamma_{ik} \gamma_{jl} A^{kl} = \frac{1}{2} S_{ij} + \frac{nS_i p_j}{2 np} \,,
\end{equation}
and the matter parts of the gravitational constraints given by
\begin{align}
\mathcal{H}^{\text{matter}} &= - np \, \delta
	- K^{ij} \frac{p_i nS_j}{np} \delta - ( nS^k \delta )_{;k} \,, \label{Hcov} \\
\mathcal{H}^{\text{matter}}_i &= (p_i + K_{ij} nS^j ) \delta
	+ \bigg( \frac{1}{2} \gamma^{jk} S_{ik} \delta
		+ \gamma^{jk} \frac{p_{(i} nS_{k)}}{np} \delta \bigg)_{;j} \,. \label{Hicov}
\end{align}
These coincide with the densitized projections
\begin{equation}
\mathcal{H}^{\text{matter}} = \sqrt{\gamma}T_{\mu\nu} n^{\mu}n^{\nu} \,, \qquad
\mathcal{H}^{\text{matter}}_i = - \sqrt{\gamma}T_{i \nu} n^{\nu} \,, \label{Himatter}
\end{equation}
of the stress-energy tensor (\ref{PDset}) at linear order in spin.
For consistency this must of course be the case, as the gravitational constraints
can also be obtained by such projections of the Einstein equations directly, instead
of by varying the action with respect to $N$ and $N^i$. However, in spite of the
simplifying premature (but only partial) gauge fixing (\ref{timegauge}) of the tetrad,
the matter Lagrangian (\ref{Lmat31}) is still complicated compared to the nonspinning case (\ref{Lpm31}).
In particular, the canonical structure is not immediately visible in the used variables.

\mysubsection{Canonical Matter Variables\label{vartrans}}
One already knows from special relativity that the variables in the covariant
spin supplementary condition have quite complicated Poisson brackets.
Thus the complicated structure of the matter action in these variables found
in the last {\secname} is not surprising. We will now try to simplify the structure
of the matter Lagrangian by introducing new variables,
which will turn out to possess standard canonical Poisson brackets.
These new variables are indicated by a hat.
An intuitive guess from the special relativistic case (\ref{covToNW}) is
\begin{equation}
z^i = \hat{z}^i - \frac{nS^i}{m - np} \,, \qquad
	nS_i = - \frac{p_k \gamma^{kj} \hat{S}_{ji}}{m} \,, \label{zredef} \qquad
S_{ij} = \hat{S}_{ij} - \frac{p_i nS_{j}}{m - np} + \frac{p_j nS_i}{m - np} \,,
\end{equation}
belonging to the condition (\ref{canSSC}), as well as
\begin{equation} \label{Lcan}
\Lambda^{[i](j)} = \hat{\Lambda}^{[i](k)} \bigg( \delta_{kj} + \frac{p_{(k)}p^{(j)}}{m (m - np)} \bigg) \,,
\end{equation}
see (3.60c) in \cite{Hanson:Regge:1974}.
These redefinitions replace $A^{kl}$ in (\ref{Lmat31}) by the quantity $\hat{A}^{ij}$ given by
\begin{equation}\label{defAhat}
\gamma_{ik} \gamma_{jl} \hat{A}^{kl} = \frac{1}{2} \hat{S}_{ij} + \frac{m p_{(i} nS_{j)}}{np (m-np)} \,.
\end{equation}
Then the first line of (\ref{Lmat31}) suggests to introduce a new linear momentum for the matter as
\begin{equation}\label{predef}
\hat{p}_i = p_i + K_{ij} nS^j + \hat{A}^{kl} e_{(j)k} {e^{(j)}}_{l,i}
	- \bigg( \frac{1}{2} S_{kj}
		+ \frac{p_{(k} nS_{j)}}{np}
	\bigg) \Gamma^{kj}{}_i \,,
\end{equation}
which reduces to $\hat{p}_i = p_i$ in the special relativistic case.
The matter Lagrangian now turns into (still approximating linear in spin)
\begin{equation}\label{Ltest}
L_{M} = \hat{p}_i \dot{\hat{z}}^i + \frac{1}{2} \hat{S}_{(i)(j)} \hat{\Omega}^{(i)(j)} - H_M \,,
\end{equation}
where $\hat{\Omega}^{(i)(j)} = \hat{\Lambda}_{[k]}{}^{(i)} \dot{\hat{\Lambda}}^{[k](j)}$ and
\begin{equation}\label{Htest}
H_{M} = - \hat{A}^{ij} e_{(k)i} {e^{(k)}}_{j,0}
	+ \int \dd^3 x \, ( N \mathcal{H}^{\text{matter}} - N^i \mathcal{H}^{\text{matter}}_i ) \,.
\end{equation}
Notice that $\hat{\Lambda}^{[i](k)}$ is a 3-dimensional rotation matrix,
$\hat{\Lambda}_{[k]}{}^{(i)} \hat{\Lambda}^{[k](j)} = \delta_{ij}$.
Therefore $\hat{\Omega}^{(i)(j)}$ is antisymmetric and should be interpreted
as an angular velocity tensor. The action thus has the canonical structure
momenta times velocities minus Hamiltonian $H_M$. The Poisson brackets for the matter part read
\begin{gather}
\{ \hat{z}^i, \hat{p}_j \} = \delta_{ij} \,, \qquad
\{ \hat{\Lambda}^{[i](j)}, \hat{S}_{(k)(l)} \} = \hat{\Lambda}^{[i](k)} \delta_{lj} - \hat{\Lambda}^{[i](l)} \delta_{kj} \,, \label{PBmatcan} \\
\{ \hat{S}_{(i)(j)}, \hat{S}_{(k)(l)} \} = \delta_{ik} \hat{S}_{(j)(l)} - \delta_{jk} \hat{S}_{(i)(l)}
	- \delta_{il} \hat{S}_{(j)(k)} + \delta_{jl} \hat{S}_{(i)(k)} \,, \label{PBScan}
\end{gather}
all other zero, similar to (\ref{nrPB1}, \ref{nrPB2}). It is important that all extrinsic
curvature terms are eliminated from (\ref{Lmat31}, \ref{Hcov}, \ref{Hicov}) by the redefinition of
the linear momentum (\ref{predef}). Terms of this type are the reason for
potential problems with derivative-coupled theories \cite{Isenberg:Nester:1980},
so it is good that they disappear. This is similar to the Dirac field case,
which can be made a nonderivative-coupled theory by a redefinition of the
Dirac field. Further the $\dot{p}_i$-term in (\ref{Lmat31}) was removed by
the redefinition of the position (\ref{zredef}).

If we consider test spinning bodies in an external field, then one immediately
gets the fully reduced Hamiltonian in the time gauge by inserting the metric (i.e., $\gamma_{ij}$, $N$, and $N^i$)
as well as a suitable triad $e_{(k)i}$ (subject only to $e_{(k)i} {e^{(k)}}_{j} = \gamma_{ij}$)
into (\ref{Htest}). Canonical formulations of test spinning bodies were already obtained in
\cite{Kunzle:1972} by a direct construction of the symplectic structure
and also very recently in \cite{Barausse:Racine:Buonanno:2009}
using a Dirac bracket approach. In the latter paper the Hamiltonian was explicitly
obtained for the Kerr metric.
In the next {\secname} we will also be able to put the field part into canonical form.

Given the fact that, at least in the time gauge (\ref{timegauge}), the supplementary
condition (\ref{canSSC}) leads to a canonical spin and position variable, it seems to be simpler
to immediately start with an action implementing (\ref{canSSC}), thus skipping the need for
variable redefinitions. However, one can not be sure in advance that
(\ref{canSSC}) leads to canonical variables. Further, it should be noted that only the \emph{structure} of
the action was simplified by above redefinitions. The redefinitions still
have to be applied to (\ref{Hcov}, \ref{Hicov}), making these expressions more complicated, see (\ref{cansource1}--\ref{cansource3}).
Thus one has a conservation of trouble here and starting directly with (\ref{canSSC}) does
not seem to simplify the calculation. In fact, it could be subtle to correctly implement
the noncovariant condition (\ref{canSSC}) into the action.
However, this succeeded for test spinning objects in \cite{Barausse:Racine:Buonanno:2009}.

\mysection{Full Gauge Reduction}
The discussion of the field part is not as simple as for nonspinning objects.
First, we need the tetrad form of the ADM formalism as derived in
\cite{Deser:Isham:1976}. Second, the matter action depends on the partial
time derivative of the tetrad, which necessitates matter corrections to the
canonical field momentum. Indeed, the canonical momentum conjugate to $e_{(k)j}$ is given by
\begin{equation}\label{pecan}
\bar{\pi}^{(k)j} = 8\pi \frac{\partial ( \mathcal{L}_G + \mathcal{L}_M )}{\partial e_{(k)j,0}}
	= {e^{(k)}}_{i} \, \pi^{ij} + 8\pi {e^{(k)}}_i \, \hat{A}^{ij} \hat{\delta} \,,
\end{equation}
where $\mathcal{L}_M$ is the density version of (\ref{Ltest}), obtained by introducing
$\hat{\delta} = \delta(x^i - \hat{z}^i)$ in certain terms, and $\pi^{ij}$ is still given by (\ref{pfield}).
Remember that (\ref{Hcov}, \ref{Hicov}) do not contain the extrinsic curvature after
redefining the matter variables.
Legendre transformation leads to
\begin{gather}
W = \frac{1}{8\pi} \int \dd^4 x \, \bar{\pi}^{(k)j} e_{(k)j,0}
	+ \int \dd t \, \bigg[ \hat{p}_i \dot{\hat{z}}^i + \frac{1}{2} \hat{S}_{(i)(j)} \hat{\Omega}^{(i)(j)} - H \bigg] \,, \\
H = \int \dd^3 x \left( N \mathcal{H} - N^i \mathcal{H}_i + \lambda_{ij} \pi^{[ij]} \right) + E[\gamma_{ij}] \,, \label{Htriad}
\end{gather}
where $\mathcal{H} \equiv \mathcal{H}^{\text{field}} + \mathcal{H}^{\text{matter}}$ and
$\mathcal{H}_i \equiv \mathcal{H}^{\text{field}}_i + \mathcal{H}^{\text{matter}}_i$ with
(\ref{HHifield}) and (\ref{Hcov}, \ref{Hicov}).
In tetrad gravity one has the additional constraint $\pi^{[ij]} = 0$,
or $\bar{\pi}^{[ij]} = 8\pi \hat{A}^{[ij]} \hat{\delta}$, which was added to the Hamiltonian $H$
via a Lagrange multiplier $\lambda_{ij} = - \lambda_{ji}$.

\mysubsection{Spatial Symmetric Gauge}
The constraint $\pi^{[ij]} = 0$ is eliminated by a further partial gauge fixing now.
The spatial symmetric gauge for the triad $e_{(i)j} = e_{ij} = e_{ji}$ is
imposed, which was suggested by Kibble for a canonical formulation
of the Dirac field coupled to gravity \cite{Kibble:1963}
(however, Kibble was using the Schwinger canonical formalism \cite{Schwinger:1963:1}).
In this gauge, the triad is the symmetric matrix square-root
of the positive definite induced metric, $e_{ij} e_{jk} = \gamma_{ik}$, or
\begin{equation}\label{egauge}
	(e_{ij}) = \sqrt{(\gamma_{ij})} \,.
\end{equation}
Thus the triad is fully given in terms of the metric, which is now the variable to be varied.
We may therefore define an object $B^{kl}_{ij}$ as
\begin{equation}\label{Bedef}
2 B^{kl}_{ij} = e_{mi} \frac{\partial e_{mj}}{\partial \gamma_{kl}} - e_{mj} \frac{\partial e_{mi}}{\partial \gamma_{kl}} \,,
\end{equation}
which enables us to write
\begin{equation}\label{eder}
{e^{(k)}}_i e_{(k)j,\mu} = B^{kl}_{ij} \gamma_{kl,\mu} + \frac{1}{2} \gamma_{ij,\mu} \,.
\end{equation}
The action obviously takes on the form
\begin{gather}
W = \frac{1}{16\pi} \int \dd^4 x \, \hat{\pi}^{ij} \gamma_{ij,0}
	+ \int \dd t \, \bigg[ \hat{p}_i \dot{\hat{z}}^i + \frac{1}{2} \hat{S}_{(i)(j)} \hat{\Omega}^{(i)(j)} - H \bigg] \,, \\
H = \int \dd^3 x \left( N \mathcal{H} - N^i \mathcal{H}_i \right) + E[\gamma_{ij}] \,,
\end{gather}
with the new canonical field momentum conjugate to $\gamma_{ij}$ given by
\begin{equation}\label{NWpi}
\hat{\pi}^{ij} = \pi^{ij} + 8\pi \hat{A}^{(ij)} \hat{\delta} + 16\pi B^{ij}_{kl} \hat{A}^{[kl]} \hat{\delta} \,.
\end{equation}
We have thus reduced the tetrad form of the ADM formalism to
its metric form, still coupled to spinning objects.

\mysubsection{ADM Transverse-Traceless Gauge}
Finally, the gauge fixing for the induced metric follows along the same lines as
for nonspinning objects in \Sec{matcoup}. We apply the gauge conditions
\begin{equation}\label{ADMTTcond2}
\partial_j ( \gamma_{ij} - \tfrac{1}{3} \gamma_{kk} \delta_{ij} ) = 0 \,, \qquad
\hat{\pi}^{ii} = 0 \,.
\end{equation}
However, notice that the ADM transverse-traceless condition
for the canonical field momentum $\hat{\pi}^{ii} = 0$ differs from the original one,
$\pi^{ii} = 0$. Correspondingly we now have the decomposition
\begin{equation}
\hat{\pi}^{ij} = \hat{\tilde{\pi}}^{ij} + \hat{\pi}^{ij\text{TT}} \,, \label{picandecomp} \qquad
\hat{\tilde{\pi}}^{ij} = \hat{\tilde{\pi}}^i{}_{, j} + \hat{\tilde{\pi}}^j{}_{, i}
	- \frac{1}{2} \delta_{ij} \hat{\tilde{\pi}}^k{}_{, k}
	- \frac{1}{2} \Delta^{-1} \hat{\tilde{\pi}}^k{}_{, ijk} \,,
\end{equation}
instead of (\ref{pidecomp}). The decomposition for the metric (\ref{gdecomp})
% \begin{align}
% 	\gamma_{ij} = \left( 1 + \frac{\phi}{8} \right)^4 \delta_{ij} + h^{\text{TT}}_{ij} \,,
% \end{align}
is still valid.
The ADM Hamiltonian then results from solving the field constraints $\mathcal{H} = 0 = \mathcal{H}_i$
together with the gauge conditions as
\begin{equation}\label{HADMspin}
H_{\text{ADM}} = E [\hat{z}^i, \hat{p}_i, \hat{S}_{(i)(j)}, h^{\text{TT}}_{ij}, \hat{\pi}^{ij\text{TT}} ]
	= - \frac{1}{16\pi} \int \dd^3 x \, \Delta \phi \,,
\end{equation}
and the fully reduced Poisson brackets of the field read
\begin{equation}\label{PBfield}
\{h^{\text{TT}}_{ij}({\bf x}), \hat{\pi}^{kl\text{TT}}({\bf x}')\}
	= 16\pi \delta^{\text{TT}kl}_{ij}\delta({\bf x} - {\bf x}') \,,
\end{equation}
all other zero. The Poisson brackets (\ref{PBmatcan}, \ref{PBScan}) of course still hold.
The fully reduced action finally reads
\begin{equation}\label{ADMspinaction}
W = \frac{1}{16\pi} \int \dd^4 x \, \hat{\pi}^{ij\text{TT}} h^{\text{TT}}_{ij,0}
	+ \int \dd t \, \bigg[ \hat{p}_i \dot{\hat{z}}^i + \frac{1}{2} \hat{S}_{(i)(j)} \hat{\Omega}^{(i)(j)}
		- H_{\text{ADM}} \bigg] \,.
\end{equation}
This is the extension of the nonspinning case in (\ref{PMaction}).
The new spin interactions enter via the ADM Hamiltonian $H_{\text{ADM}}$ after
solving the constraints, which now have spin corrections in its source terms,
(\ref{Hcov}, \ref{Hicov}).

\mychapter{Symmetry Generator Approach\label{symmetry}}
As we have seen in \Sec{cangrav} and also in the last {\secname}, after all
constraints as well as supplementary and gauge conditions have been eliminated,
the Hamiltonian is given by the ADM energy depending on the fully reduced
canonical variables. However, while it is not problematic to calculate the
ADM energy at least to some order in a perturbative way, it will then
depend on the variables appearing in the stress-energy tensor (\ref{PDset}) and
equation of motion (\ref{eom}), for which the canonical structure is not known.
If one could someway find the transformation between these variables and fully
reduced canonical variables with usual Poisson brackets, then the ADM energy
can be expressed in terms of canonical variables and turns into the ADM Hamiltonian.
In this {\chapname} we try to construct this variable transformation order-by-order
in some perturbation scheme by looking at certain consistency conditions.
It is expected that if one proceeds to higher and higher orders,
then one also needs to devise more and more consistency conditions.
However, in the post-Newtonian approximation one may reach an order high
enough for all currently relevant applications
by just relying on a specific form of total linear
and angular momentum expressed in terms of canonical variables
\cite{Steinhoff:Schafer:Hergt:2008,Steinhoff:Wang:2009}.
Notice that this approach is not as powerful as the action approach
\cite{Steinhoff:Schafer:2009:2} discussed in the last {\chapname},
however, it succeeded earlier and is still valuable at higher
orders in spin as well as for a check of the action approach
at linear order in spin.

\mysection{Symmetries and Conserved Quantities}
Now the symmetries and corresponding conserved quantities for asymptotically flat
spacetimes are reviewed. These conserved quantities generate their
symmetries on phase space. For total linear and angular momentum this
leads to a very specific form when expressed in terms of canonical variables.

\mysubsection{Global Rotations and Translations\label{globalEucl}}
It is intuitively clear that an asymptotically flat spacetime can be transformed
into a physically equivalent one by a 3-dimensional rotation and/or translation
of each 3-dimensional hypersurface, i.e., of the whole spacetime.
This means that asymptotically flat spacetimes posses a global symmetry\footnote{
A global symmetry depends on parameters which may not vary over spacetime.}
under rotations and translations, i.e., under the 3-dimensional Euclidean group.
In fact, one even has a global symmetry under the Poincar\'e group, which will be
discussed in \Sec{GPoincare}. How the symmetry under the Euclidean group is represented
on the coordinates crucially depends on the chosen coordinate system
even in flat space. If the coordinate system resembles to a Cartesian one in the
asymptotics, then a good guess for the symmetry transformation is
% \begin{equation}
$x^i \rightarrow \Lambda_{ij} ( x^j + a^j )$,
% \end{equation}
with $x^i$ the coordinates of the 3-dimensional hypersurfaces, $a^i$ a constant
vector describing a translation, and a rotation matrix $\Lambda_{ij}$.
The rotation matrix is pa\-ra\-me\-trized by a constant antisymmetric matrix
$\omega^{ij} = - \omega^{ji}$ in the form $\Lambda = e^{\omega}$.
A field, e.g., the induced metric $\gamma_{ij}$, then transforms as
\begin{equation}\label{fieldt}
\gamma_{ij}(\vct{x}) \rightarrow \Lambda_{ik} \Lambda_{jl} \gamma_{kl}(\Lambda^{-1} \vct{x} - \vct{a}) \,,
\end{equation}
where the vector $\vct{a}$ has components $a^i$.
However, for this transformation to be a global symmetry and not just a
particular gauge transformation, the gauge conditions must be invariant
under this representation of the Euclidean group. This is indeed fulfilled for
the ADM gauge conditions (\ref{ADMTTcond}) or (\ref{ADMTTcond2})
(remember that $a^i$ and $\Lambda_{ij}$ are \emph{constant}).
Further, the local basis shall rotate the same way as the coordinate basis, i.e.,
% \begin{equation}
$e_{(i)j}(\vct{x}) \rightarrow \Lambda_{ik} \Lambda_{jl} e_{(k)l}(\Lambda^{-1} \vct{x} - \vct{a})$.
% \end{equation}
We assume here that the tetrad was reduced to a triad with the help of the time gauge (\ref{timegauge}),
as in the action approach.
The triad gauge shall be compatible with this transformation property,
which is the case for (\ref{egauge}).

Looking at infinitesimal transformations, i.e.,
$a^i$ and $\omega^{ij}$ shall be small, it holds
\begin{equation}\label{eucl}
	x^i \rightarrow x^i + a^i + \omega^{ij} x^j \,,
\end{equation}
or for a tensor field (\ref{fieldt})
\begin{equation}
\gamma_{ij} \rightarrow \gamma_{ij} - a^k \partial_k \gamma_{ij} - \omega^{kl} x^l \partial_k \gamma_{ij}
	+ \omega^{ik} \gamma_{kj} + \omega^{jk} \gamma_{ik} \,.
\end{equation}
This is just the Lie-shift given by the infinitesimal coordinate transformation (\ref{eucl}),
i.e., $\gamma_{ij} \rightarrow \gamma_{ij} - \mathcal{L}_{\delta x^k} \gamma_{ij}$.
Similarly, the canonical variables transform as
\begin{gather}
\hat{z}_a^i \rightarrow \hat{z}_a^i + a^i + \omega^{ij} \hat{z}_a^j \,, \label{ztrans} \qquad
\hat{p}_{a i} \rightarrow \hat{p}_{a i} + \omega^{ij} \hat{p}_{a j} \,, \\
\hat{\Lambda}^{[i](j)}_a \rightarrow \hat{\Lambda}^{[i](j)}_a + \omega^{jk} \hat{\Lambda}^{[i](k)}_a \,, \qquad
\hat{S}_{a(i)(j)} \rightarrow \hat{S}_{a(i)(j)} + \omega^{im} \hat{S}_{a(m)(j)} + \omega^{jm} \hat{S}_{a(i)(m)} \,, \label{Ltrans} \\
h^{\text{TT}}_{i j} \rightarrow h^{\text{TT}}_{i j}
	- a^k \partial_k h^{\text{TT}}_{i j} - \omega^{kl} x^l \partial_k h^{\text{TT}}_{i j}
	+ \omega^{ik} h^{\text{TT}}_{k j} + \omega^{jk} h^{\text{TT}}_{i k} \,, \label{htttrans} \\
\hat{\pi}^{i j \text{TT}} \rightarrow \hat{\pi}^{i j \text{TT}}
	- a^k \partial_k \hat{\pi}^{i j \text{TT}} - \omega^{kl} x^l \partial_k \hat{\pi}^{i j \text{TT}}
	+ \omega^{ik} \hat{\pi}^{k j \text{TT}} + \omega^{jk} \hat{\pi}^{i k \text{TT}} \,. \label{pitttrans}
\end{gather}
A label index was attached to the matter variables now.
In (\ref{Ltrans}) the transformation property of the local basis was used.
Notice that the body-fixed basis in (\ref{Ltrans}) stays unchanged.
For (\ref{htttrans}) and (\ref{pitttrans}) the transverse-traceless projection
was commuted with the infinitesimal coordinate change.

\mysubsection{Symmetry Generators}
Now we try to construct the generators of infinitesimal rotations
and translations, $P_i$ and $J_{ji}$. These are of course nothing else than 3-dimensional
total linear and angular momentum. With the help of these generators
the transformation rule for an arbitrary phase space function $A$ must read
\begin{equation}
A \rightarrow A + \tfrac{1}{2} \omega^{ij} \{ A , J_{ji} \} + a^i \{ A , P_i \} \label{trans} \,.
\end{equation}
It is sufficient to guarantee this transformation rule for all canonical variables.
Comparing (\ref{trans}) with (\ref{ztrans}--\ref{pitttrans}), using the standard Poisson brackets
(\ref{PBmatcan}, \ref{PBScan}) for each object as well as (\ref{PBfield}), one can indeed
construct $P_i$ and $J_{ij}$. It is immediately clear that $P_i$ and $J_{ij}$
are a sum of matter and field parts,
\begin{equation}\label{PJsum}
P_i = P_i^{\text{matter}} + P_i^{\text{field}} \,, \qquad
J_{ij} = J_{ij}^{\text{matter}} + J_{ij}^{\text{field}} \,.
\end{equation}
In order to get $P_i$, one sets $\omega^{ij} = 0$ and $a^i$ is taken to be arbitrary.
Then among the matter variables only $\hat{z}_a^i$ is transformed.
Comparing (\ref{trans}) with (\ref{ztrans}) one obtains
$\delta_{ij} = \{ \hat{z}_a^i , P_j \} = \frac{\partial P_j}{\partial \hat{p}_{a i}}$
for each particle, and thus
\begin{equation}
P_i^{\text{matter}} = \sum_a \hat{p}_{ai} \,. \label{Pmatter}
\end{equation}
Similarly, for the field part one gets $- \partial_k h^{\text{TT}}_{i j} = \{ h^{\text{TT}}_{i j} , P_k \}$
as well as $- \partial_k \hat{\pi}^{i j \text{TT}} = \{ \hat{\pi}^{i j \text{TT}} , P_k \}$, which leads to
\begin{equation}
P_i^{\text{field}} = - \frac{1}{16\pi} \int \dd^3x \, \hat{\pi}^{kl\text{TT}} h^{\text{TT}}_{kl,i} \,. \label{Pfield}
\end{equation}
The derivation of $J_{ij}$ is analogous, with the result
\begin{align}
J_{ij}^{\text{matter}} &= \sum_a ( \hat{z}_a^i \hat{p}_{aj} - \hat{z}_a^j \hat{p}_{ai} ) + \sum_a \hat{S}_{a(i)(j)} \,, \label{Jmatter} \\
\begin{split}
J_{ij}^{\text{field}} &=
	- \frac{1}{16\pi} \int \dd^3x \, ( x^i \hat{\pi}^{kl\text{TT}} h^{\text{TT}}_{kl,j} - x^j \hat{\pi}^{kl\text{TT}} h^{\text{TT}}_{kl,i} ) \nl
	- \frac{1}{16\pi} \int \dd^3x \, 2 ( \hat{\pi}^{ik\text{TT}} h^{\text{TT}}_{kj}
		- \hat{\pi}^{jk\text{TT}} h^{\text{TT}}_{ki} ) \,. \label{Jfield}
\end{split}
\end{align}
The ADM Hamiltonian $H_{\text{ADM}}$ is by construction manifestly invariant under global rotations
and translations (at least in the considered gauges). Comparing $H_{\text{ADM}} \rightarrow H_{\text{ADM}}$
with (\ref{trans}) one sees that total linear and angular momentum have
vanishing Poisson brackets with the ADM Hamiltonian and are thus conserved.

Yet another symmetry specific to objects with spin is given
by constant rotations of the body-fixed frame,
\begin{equation}\label{bodysym}
\hat{\Lambda}^{[i](j)}_a \rightarrow \hat{\Lambda}^{[i](j)}_a + \omega^{[i][k]}_a \hat{\Lambda}^{[k](j)}_a \,, 
\end{equation}
parametrized by a constant antisymmetric matrix $\omega^{[i][j]}_a = - \omega^{[j][i]}_a$ for each object.
The corresponding generators read
% \begin{equation}\label{Jbody}
$J_{a[i][j]}^{\text{body}} = \hat{\Lambda}^{[i](k)}_a \hat{\Lambda}^{[j](l)}_a \hat{S}_{a(k)(l)}$ %\,,
% \end{equation}
and are also conserved quantities, $J_{a[i][j]}^{\text{body}} = \text{const}$. A corollary of this is that
\begin{equation}\label{JbodySq}
J_{a[i][j]}^{\text{body}} J_{a[i][j]}^{\text{body}} = \hat{S}_{a(i)(j)} \hat{S}_{a(i)(j)} = \text{const} \,.
\end{equation}

As the ADM Hamiltonian $H_{\text{ADM}}$ is invariant under the transformations
(\ref{ztrans}--\ref{pitttrans}) and (\ref{bodysym}), these transformations are also a symmetry
of the action (\ref{ADMspinaction}). The corresponding conserved quantities $P_i$,
$J_{ij}$, and $J_{a[i][j]}^{\text{body}}$ can then be obtained by standard Noether
arguments \cite{Noether:1918} and come out identical to above results.

\mysubsection{Global Poincar\'e Invariance\label{GPoincare}}
The global symmetry under the Euclidean group discussed in the last {\secname} is only a part of the
bigger global symmetry under the Poincar\'e group. Besides total linear
and angular momentum, also the boost vector $J^{i0}$ and the energy $E=H_{\text{ADM}}$ of the system
generate a symmetry of the action and are conserved quantities for asymptotically flat spacetimes.
However, the infinitesimal transformations generated by $H_{\text{ADM}}$ and $J^{i0}$, similar to
(\ref{trans}), are in general highly nonlinear in the considered gauges and
may not be written down immediately, as opposed to (\ref{ztrans}--\ref{pitttrans}).
Further, $J^{i0}$ explicitly depends on time, see (\ref{boost}).
$H_{\text{ADM}}$ and $J^{i0}$ can be calculated by surface integrals at spatial
infinity, see, e.g., \cite{Arnowitt:Deser:Misner:1961:3,Regge:Teitelboim:1974}. For the total energy $E=H_{\text{ADM}}$
this was already found in (\ref{Esurf}) and the boost vector $J^{i0}$
is given by (\ref{boost}) with
\begin{equation}
	G^i = \frac{1}{16\pi}\oint \dd^2 s_k \left[ x^i ( \gamma_{kl,l} - \gamma_{ll,k} )
		- \gamma_{ik} + \delta_{ik} \gamma_{ll} \right] \,.
\end{equation}
% where the spatial coordinates are denoted $x^i$.
Similarly, for 3-dimensional total linear and angular momentum it holds
\begin{equation}
	P_i = - \frac{1}{8\pi}\oint \dd^2 s_k \pi^{ik} \,, \qquad
	J_{ij} = - \frac{1}{8\pi}\oint \dd^2 s_k ( x^i \pi^{jk} - x^j \pi^{ik}) \,.
\end{equation}
When these quantities are expressed in terms of canonical variables (after
gauge fixing), they fulfill the Poincar\'e algebra (\ref{poinc1}, \ref{poinc2}). Notice that all Poisson
brackets in (\ref{poinc1}, \ref{poinc2}) involving $P_i$ and $J_{ij}$ just reflect the transformation property (\ref{trans}).
Similar to the special relativistic case in \Sec{spinSR}, one can define different total spins and centers
for a gravitating system in asymptotically flat spacetimes. In particular,
a center and total spin of the system with standard Poisson brackets can be constructed
(this was exploited recently in \cite{Rothe:Schafer:2010}).

\mysubsection{Symmetry Generators from Integral Formulas}
For simplicity we assume that $\gamma_{ij}$ does not need a redefinition
in order to receive a canonical meaning. However, this might be necessary
at higher orders in spin. For the canonical field momentum $\hat{\pi}^{ij}$
we allow spin corrections by the ansatz
\begin{equation}\label{pican}
	\hat{\pi}^{ij} = \pi^{ij} + 16 \pi \sum_a \pi^{ij}_a \hat{\delta}_a \,,
\end{equation}
where $\pi^{ij}_a$ contains the yet undetermined spin corrections. 
The gauge condition $\hat{\pi}^{ii} = 0$ with the subsequent decomposition (\ref{picandecomp}) is assumed to hold.
The surface integrals from the last {\secname} can be transformed into volume
integrals using the Gauss theorem. With the decomposition (\ref{gdecomp}) it follows
\begin{equation}
	E = - \frac{1}{16\pi} \int \dd^3x \, \Delta \phi \,, \qquad
	G^i = - \frac{1}{16\pi} \int \dd^3x \, x^i \Delta \phi \,. \label{Gint}
\end{equation}
However, it is not possible to express $E$ and $G^i$ in terms of the canonical variables
without solving the nonlinear constraint equations for $\phi$. Similarly one gets
\begin{equation}\label{PJint}
	P_i = - \frac{1}{8\pi} \int \dd^3x \, \hat{\pi}^{ik}{}_{,k} \,, \qquad
	J_{ij} = - \frac{1}{8\pi} \int \dd^3x \, ( x^i \hat{\pi}^{jk}{}_{,k} - x^j \hat{\pi}^{ik}{}_{,k}) \,.
\end{equation}
Here one can exploit the momentum constraint
$\mathcal{H}_i \equiv \mathcal{H}^{\text{field}}_i + \mathcal{H}^{\text{matter}}_i = 0$
to further evaluate $P_i$ and $J_{ij}$ without needing to actually solve the constraints.
Using (\ref{HHifield}, \ref{gdecomp}, \ref{picandecomp}) the momentum constraint can \emph{exactly} be written as
\begin{equation}\label{cmom}
\hat{\pi}^{ik}{}_{,k} =
	- 8\pi ( \mathcal{H}^{\text{matter}}_{i}
		+ \mathcal{H}^{\pi \text{matter}}_i )
	+ \frac{1}{2} \hat{\pi}^{jk\text{TT}} h_{jk,i}^{\text{TT}}
	- ( \hat{\pi}^{jk\text{TT}} h_{ki}^{\text{TT}} )_{,j}
	- \Delta \left( \hat{V}^k h_{ki}^{\text{TT}} \right)
	+ \hat{B}^{ij}{}_{,j} \,,
\end{equation}
with the definitions
\begin{gather}
\mathcal{H}^{\pi \text{matter}}_i = \sum_a \left[
	\pi^{jk}_a \gamma_{jk,i} \hat{\delta}_a
	- 2 ( \gamma_{ik} \pi^{kj}_a \hat{\delta}_a )_{,j} \right] \,, \label{HSpi} \\
\hat{B}^{ij} = \left[ 1 - \left( 1 + \tfrac{1}{8} \phi \right)^4 \right] ( \hat{\tilde{\pi}}^{ij}
		+ \hat{\pi}^{ij\text{TT}} )
	+ \hat{V}^k ( h_{ki,j}^{\text{TT}} + h_{kj,i}^{\text{TT}} - h_{ij,k}^{\text{TT}} )
	- \frac{1}{3} \hat{V}^k{}_{,k} h_{ij}^{\text{TT}} \,, \label{Bijcan}
\end{gather}
and the alternative vector potential
\begin{equation}\label{Vcan}
\hat{V}^i = \left( \delta_{ij} - \frac{1}{4} \partial_i \partial_j \Delta^{-1} \right) \hat{\tilde{\pi}}^j \,,
\end{equation}
for which it holds
\begin{equation}\label{picanV}
\hat{\tilde{\pi}}^{ij} = \hat{V}^i{}_{, j} + \hat{V}^j{}_{, i} - \frac{2}{3} \delta_{ij} \hat{V}^k{}_{, k} \,.
\end{equation}
One can calculate $\mathcal{H}^{\text{matter}}_i$ using (\ref{Himatter}).
Notice that $\hat{B}^{ij} = \hat{B}^{ji}$ and $\hat{B}^{ii} = 0$.
Further the last two terms in (\ref{cmom}) do not contribute to (\ref{PJint}).
Obviously, (\ref{PJint}) are a sum of matter and field parts, (\ref{PJsum}).
The field parts are identical to (\ref{Pfield}) and (\ref{Jfield}).
However, the matter parts now read
\begin{gather}
P_i^{\text{matter}} = \int \dd^3x \, ( \mathcal{H}^{\text{matter}}_i + \mathcal{H}^{\pi \text{matter}}_i ) \,, \label{PJmatter1} \\
J_{ij}^{\text{matter}} = \int \dd^3x \, ( x^i \mathcal{H}^{\text{matter}}_j
		+ x^i \mathcal{H}^{\pi \text{matter}}_j
		- x^j \mathcal{H}^{\text{matter}}_i - x^j \mathcal{H}^{\pi \text{matter}}_i ) \,. \label{PJmatter2}
\end{gather}
For consistency, these must turn into (\ref{Pmatter}) and (\ref{Jmatter}) when expressed in terms
of canonical variables.

\mysection{Construction of Canonical Variables}
In this {\secname} we will formulate the important consistency conditions
and apply them order-by-order in the post-Newtonian approximation
to find canonical variables.

\mysubsection{Consistency Conditions}
In \Sec{sscsec} it was seen that the spin length $S_a$ given by $2 S^2_a = S_{a\mu\nu} S^{\mu\nu}_a$
is a conserved quantity in the covariant spin supplementary condition.
This can also be derived from the action (\ref{pdaction}) using the symmetry
under constant 4-dimensional Lorentz transformations of the body-fixed frame,
see also \cite{Hanson:Regge:1974}.
This conserved quantity must be identical to the one in (\ref{JbodySq}), as both
were derived from the same symmetry (though only the 3-dimensional rotation
part is relevant after the supplementary conditions were eliminated).
Thus it must hold
\begin{equation}\label{spinSq}
S_{a\mu\nu} S^{\mu\nu}_a = \hat{S}_{a(i)(j)} \hat{S}_{a(i)(j)} \,,
\end{equation}
providing a relation between covariant spin $S_{a\mu\nu}$ and
canonical spin $\hat{S}_{a(i)(j)}$.
This is one important consistency condition we will impose.

Further, one can calculate $\mathcal{H}^{\text{matter}}_{i}$ and thus (\ref{PJmatter1}, \ref{PJmatter2})
in terms of (noncanonical) variables in the covariant supplementary condition
with the help of (\ref{Himatter}) and (\ref{PDset}),
\begin{equation}
\mathcal{H}^{\text{matter}}_i = \sum_a \left[ (p_{ai} + K_{ij} nS^j_a ) \delta_a
	+ \bigg( \frac{1}{2} \gamma^{jk} S_{aik} \delta_a
		+ \gamma^{jk} \frac{p_{a(i} nS_{ak)}}{np_a} \delta_a \bigg)_{;j} \right] \,, \label{Hicov2}
\end{equation}
see (\ref{Hicov}).
Then (\ref{PJmatter1}, \ref{PJmatter2}) must coincide with (\ref{Pmatter}, \ref{Jmatter}),
leading to conditions on the transformation between canonical variables and
variables in the covariant supplementary condition.
We write this as a condition on $\mathcal{H}^{\text{matter}}_{i}$ in
the form
\begin{equation}
	\mathcal{H}^{\text{matter}}_{i} = \sum_a \bigg[ (\hat{p}_{ai} - \pi^{jk}_a \gamma_{jk,i}) \hat{\delta}_a
		+ \frac{1}{2} (s_a^{ij} \hat{\delta}_a)_{,j} \bigg] \,, \label{defP}
\end{equation}
where the symmetric part of $s_a^{ij}$ is not constrained, but it has to hold
\begin{equation}
	s_a^{[ij]} = \hat{S}_{a(i)(j)}
		+ 2 \pi_a^{jk} h^{\text{TT}}_{ki}
		- 2 \pi_a^{ik} h^{\text{TT}}_{kj} \,. \label{Scons}
\end{equation}
This condition on $\mathcal{H}^{\text{matter}}_i$ is the most general
one\footnote{In the pole-dipole approximation at most one partial derivative can appear in $\mathcal{H}^{\text{matter}}_i$.
Further it was assumed that the variables from different objects do not mix
(e.g., as $\hat{p}_1 \hat{\delta}_2$) at this stage.}
that guarantees that (\ref{PJmatter1}, \ref{PJmatter2}) coincide with (\ref{Pmatter}, \ref{Jmatter}).

Above conditions are sufficient for the post-Newtonian order considered here.
Another condition that could be useful at even higher orders
(in particular also higher orders in spin) would be the fulfillment of the Poincar\'e algebra.
However, all Poisson brackets in (\ref{poinc1}, \ref{poinc2}) involving $P_i$ and $J_{ij}$
are fulfilled by construction due to the transformation property (\ref{trans})
if above conditions hold, thus giving nothing new.
% Further, $E$ and $G^i$ are quite complicated and inserting a general ansatz for a variable
% transformation into them \dots
In \cite{Steinhoff:Schafer:Hergt:2008} it was considered whether the construction of the constraint algebra
(\ref{constalg1}--\ref{constalg3}), which is related to diffeomorphism invariance
and thus more fundamental than global Poincar\'e invariance,
could be used to construct canonical variables. However, this approach
seems to be unmanageable.

\mysubsection{Canonical Variables}
First we evaluate the condition on the spin length given by (\ref{spinSq}).
We will first construct a specific transformation between $S_{aij}$
and $\hat{S}_{a(i)(j)}$ and then discuss its uniqueness.
Inspired by the flat space case (\ref{covToNW}), we first apply
the transformation
\begin{equation}
	S_{a ij} = \hat{S}_{a ij} - \frac{p_{a i} nS_{a j}}{m_a-np_a} + \frac{p_{a j} nS_{a i}}{m_a-np_a} \,, \qquad
	nS_{a i} = - \frac{p_{a k} \gamma^{kj} \hat{S}_{a ji}}{m_a} \,, \label{NWspin}
\end{equation}
to the conserved quantity
% \begin{align}
$S_{a\mu\nu} S^{\mu\nu}_a = \gamma^{ki} \gamma^{lj} S_{a kl} S_{a ij}
	- 2 \gamma^{ij} nS_{ai} nS_{aj}$,
% \end{align}
with the result
% \begin{align}
$S_{a\mu\nu} S^{\mu\nu}_a = \gamma^{ki} \gamma^{lj} \hat{S}_{a kl} \hat{S}_{a ij}$.
% \end{align}
With the help of an arbitrary triad $e_{(i)j}$ this can be written in a
local basis as
% \begin{align}
$S_{a\mu\nu} S^{\mu\nu}_a = \hat{S}_{a(i)(j)} \hat{S}_{a(i)(j)}$,
% \end{align}
so we have found a possible transformation allowed by (\ref{spinSq}).
The ambiguities that are left can best be discussed in terms of the
spin vector $\hat{S}_{a(i)}$.
As we are still considering the linear order in spin, any further
transformation of $\hat{S}_{a(i)}$ must be linear in spin
and must leave the expression $\hat{S}_{a(i)} \hat{S}_{a(i)}$ \emph{invariant}
(notice $\hat{S}_{a(i)(j)} \hat{S}_{a(i)(j)} = 2 \hat{S}_{a(i)} \hat{S}_{a(i)}$).
Therefore only a rotation of the spin vector as a further transformation
is possible, which can be absorbed into the yet arbitrary triad $e_{(i)j}$.

A comparison of (\ref{Hicov2}) with (\ref{defP}) leads to
\begin{equation}
\hat{p}_{ai} = p_{ai} + K_{ij} nS^j_a + \pi^{jk}_{a} \gamma_{jk,i}
	- \bigg( \frac{1}{2} S_{akj}
		+ \frac{p_{a(k} nS_{aj)}}{np_a}
	\bigg) \Gamma^{kj}{}_i \,, \label{NWmom}
\end{equation}
without any ambiguity.
Now (\ref{Hicov2}) is of the form (\ref{defP}),
so (\ref{Scons}) is the only condition that is left.
In order to evaluate (\ref{Scons}) we first need to read off $s_a^{ij}$.
For the redefinition of the position variable we use
\begin{equation}\label{NWpos}
	z^i_a = \hat{z}^i_a - \frac{nS^i_a}{m_a - np_a} + z^i_{\Delta a} \,,
\end{equation}
where $z^i_{\Delta a}$ is a yet unknown correction to the flat space case (\ref{covToNW}).
Comparing (\ref{Hicov2}) expressed in terms of the new variables with (\ref{defP}) leads to
\begin{equation}\label{scons}
s_a^{ij} = \gamma^{jk} \hat{S}_{aik}
	+ \gamma^{jk} \gamma^{lp} \frac{2 \hat{p}_{al} \hat{p}_{a(i} \hat{S}_{ak)p}}{n\hat{p}_a (m_a - n\hat{p}_a)}
	- 2 \hat{p}_{ai} z^j_{\Delta a} \,,
\end{equation}
with the definition $n\hat{p}_a = - \sqrt{m_a^2 + \gamma^{ij} \hat{p}_{ai} \hat{p}_{aj}}$.
The only ambiguities in the transition to canonical variables are now
given by $\pi_a^{ij}$, $z^i_{\Delta a}$, and the triad $e_{(i)j}$.
We try to fix these ambiguities by considering (\ref{Scons}) with (\ref{scons})
order-by-order in the post-Newtonian approximation, which is introduced in
the next {\secname}.

From the action approach we know that the ambiguity of $e_{(i)j}$
should just be a gauge freedom. Thus different choices for $e_{(i)j}$
should be canonically equivalent. Indeed, it was shown in
\cite{Damour:Jaranowski:Schafer:2008:1} that a spin rotation is
just a canonical transformation at linear order in spin.
However, a canonical transformation may change all variables,
but $\hat{p}_{ai}$ as well as $h^{\text{TT}}_{ij}$ can not be changed any more.
Thus the canonical representation was already partly fixed
and we must therefore still keep $e_{(i)j}$ as general as allowed
by the restriction on the triad gauge made in \Sec{globalEucl}.

\mysubsection{Post-Newtonian Approximation\label{count}}
The idea behind the post-Newtonian approximation is that for slowly moving
bodies and weak gravitational forces the Newtonian physics is recovered as a first approximation.
For two objects this means that their relative velocity $v$ shall be
small compared to the speed of light $c$. In Newtonian physics
the time average of kinetic and potential energy is of the same order
if the virial theorem applies, which is the case for bound systems.
Then one has
\begin{equation}\label{PNcount}
\frac{v^2}{c^2} \sim \frac{G M}{c^2 r} \ll 1 \,,
\end{equation}
where $M$ is the total mass of the system and $r$ the typical distance
of the objects. An expansion in the dimensionless quantities
(\ref{PNcount}) obviously is also an expansion in $c^{-2}$.
We will therefore think of the post-Newtonian expansion as
an expansion in $c^{-2}$. However, this is a rather formal point of view
as it depends on the choice of units
whether $c^{-2}$ is actually a small number (e.g., in our units it is equal to one).
As seen later, there may be half post-Newtonian orders corresponding
to $c^{-1}$.

As post-Newtonian orders are formally counted in terms of the velocity
of light $c$ originally present in the equations, i.e., before setting $c = 1 = G$,
one should introduce $G$ and $c$ back into all expressions. However, this would undo
the advantages achieved by setting $c = 1 = G$. Instead, we will
assign an order in powers of $c^{-1}$ directly to our variables. When setting $c = 1 = G$
only one unit is needed, which we choose to be the unit of spatial distances, e.g., meters.
Then the values of all masses $m_a$ must be given in meters, which is obtained by multiplying
their values in kilograms by $G/c^2$. Therefore we just count the
masses to be of the order $c^{-2}$, as this is the power of the speed of light
that would be introduced into the expressions if we restore the
original units. Similar arguments apply to the other matter variables
and we have the counting rules
\begin{equation}\label{matcount}
\hat{z}_a = \Order{\left(c^{0}\right)} \,, \qquad
m_a = \Order{\left(c^{-2}\right)} \,, \qquad
\hat{p}_a = \Order{\left(c^{-3}\right)} \,.
\end{equation}
Notice that an energy receives a counting of $c^{-4}$, which gives the
absolute order of the Newtonian Hamiltonian (being an energy) within
these counting rules. However, one obtains \emph{different} counting
rules for the matter variables if one uses kilograms instead of meters
to replace all units when setting $c = 1 = G$. This convention is also
often used and leads to different \emph{absolute} orders in $c^{-1}$, e.g.,
a mass now receives a counting of $c^{0}$ and the Newtonian Hamiltonian
is at the absolute order $c^{-2}$. But relative orders
are always the same, so only a counting relative to the Newtonian order
(or to the leading order if the Newtonian order vanishes) finally makes sense
when using such counting rules.
% If one would actually count the powers
% of $c^{-1}$ as originally present in the equations (instead of using counting rules),
% one would obtain the correct absolute orders.
The correct absolute Newtonian order is $c^{0}$,
as it must prevail when $c^{-1} \rightarrow 0$.

The formal counting may nicely be applied to more complicated situations,
e.g., when spins are present. For dimensional reasons only we are thus counting
the spins of the order $c^{-3}$.
This has some computational advantages, e.g., similarities to calculations
for nonspinning objects are more manifest, see \Sec{HLO}.
Here post-Newtonian orders should always be understood in the formal sense
if not otherwise stated.
However, the spin of a (Kerr) black hole is given by $G m^2 a / c$, where
$m$ is the mass of the black hole and $a = 0 \dots 1$ is the dimensionless Kerr parameter.
The maximal spin of an object is defined as $G m^2 / c$ (which \emph{is}
the maximal spin of a black hole, $a=1$), and additionally has to be counted as $c^{-1}$.
If the spins are maximal, one therefore has to add half a post-Newtonian
order relative to the formal counting for each spin variable
appearing in a specific expression.

If the spins are not maximal, one has to be careful when classifying spin effects
into post-Newtonian orders.
For example, if the spin is $\frac{1}{100}$ of the maximal one and the orbital velocity is $\frac{1}{100}$ of
the speed of light, then each spin variable corresponds to one extra order in $v/c$ relative to the maximal spin case,
or half a post-Newtonian order. At a later time during the inspiral the spin length
has not changed much\footnote{In the approximation considered here the spin length is even exactly constant.},
however, the orbital velocity might have increased, e.g., to $\frac{1}{10}$ of the speed of light.
Then each spin variable even corresponds to two additional orders in the velocity or one post-Newtonian order relative to the maximal spin case.
To conclude, while the spin length does essentially stay constant during the inspiral,
the orbital velocity will increase and one expects that spin effects slightly shift
to higher post-Newtonian orders during inspiral. Therefore, assigning a post-Newtonian order to spin
contributions in the Hamiltonian seems to make no sense in general, except for maximal spins or within the formal counting.
However, this discussion is only superficial, the relevance of spin effects also
crucially depends on the orientation of the spins and the mass ratio of the objects.
Due to these problems we will prefer to classify spin effects by leading order,
next-to-leading order, etc.\ when possible.

Counting rules for other quantities may be derived from (\ref{matcount}).
For example, $\phi$ results from solving the constraints, and one may easily
see that its leading order must be identical to the leading order of the matter
source of the Hamilton constraint $\mathcal{H}^{\text{matter}}$, which is $c^{-2}$
(this will become obvious in \Sec{expansion}). Similarly one gets counting rules for
the other field variables by considering the matter source of the field equations.
Without going into detail, we state here that
\begin{equation}\label{frules}
\phi = \Order{\left(c^{-2}\right)} \,, \qquad
h^{\text{TT}}_{ij} = \Order{\left(c^{-4}\right)} \,, \qquad
\tilde{\pi}^{ij} = \Order{\left(c^{-3}\right)} \,, \qquad
\pi^{ij\text{TT}} = \Order{\left(c^{-5}\right)} \,,
\end{equation}
are the correct counting rules for the fields.
In general the fields include different post-Newtonian orders, (\ref{frules})
only gives the leading orders. The Taylor expansion of the fields in terms
of $c^{-1}$ is written as, e.g.,
\begin{equation}
\phi = \phi_{(2)} + \phi_{(4)} + \phi_{(6)} + \Order{(c^{-7})} \,,
\end{equation}
where a number in round brackets denotes the absolute order in $c^{-1}$ within
the counting given by (\ref{matcount}) (this should not be confused with
indices in the local basis). The vanishing of the odd
orders $\phi_{(3)}$ and $\phi_{(5)}$ is explained by the vanishing of the
corresponding orders in the source terms.
% ; however, see also \Sec{matteronly}.

\mysubsection{Final Fixation of the Canonical Variables}
First we try to find a way to parametrize the ambiguity in the triad
when the induced metric is kept fixed. If one considers the perturbative
expansion of 
% \begin{equation}
$e^{i(k)} e^{j(k)} = \gamma^{ij}$ %\,,
% \end{equation}
under the assumption that the leading order is given by $e^{i(k)}_{(0)}=\delta_{ik}$,
then one sees that the symmetric part of $e^{i(k)}$ is uniquely fixed
at each order, while the antisymmetric part $\hat{e}^{ij} \equiv \frac{1}{2} (e^{i(j)} - e^{j(i)})$
is arbitrary. Therefore $\hat{e}^{ij}$ parametrizes the rotational degrees of freedom left in
the definition of the local basis and thus the ambiguity of the canonical spin variable.
In particular, the leading post-Newtonian orders read
\begin{equation}
	e^{i(j)}_{(2)} = \hat{e}^{ij}_{(2)} - \frac{1}{4} \delta_{ij} \phi_{(2)} \,, \quad
	e^{i(j)}_{(4)} = \hat{e}^{ij}_{(4)} - \frac{1}{2} \hat{e}^{ik}_{(2)} \hat{e}^{jk}_{(2)}
		- \frac{1}{4} \delta_{ij} \phi_{(4)} + \frac{3}{64} \delta_{ij} \phi_{(2)}^2 - \frac{1}{2} h^{\text{TT}}_{ij} \,.
\end{equation}
Notice that $\hat{e}^{ij}$ is needed only on the worldlines.
In the following we use the abbreviation $\hat{e}^{ij}_a \equiv \hat{e}^{ij}(\hat{z}_a^k)$.

Next we make an ansatz for $\pi_a^{ij}$, $z^i_{\Delta a}$,
and $\hat{e}^{ij}_a$ at each post-Newtonian order.
For this purpose it is important that $\pi_a^{ij}$ has the
dimension length squared, $z^i_{\Delta a}$ the dimension length,
and $\hat{e}^{ij}_a$ is dimensionless.
Further, $\pi_a^{ij}$ and $z^i_{\Delta a}$ must be linear in spin, while
$\hat{e}^{ij}_a$ must be independent of the spins.
The fields $h^{\text{TT}}_{i j}$ and $\hat{\pi}^{i j \text{TT}}$ are
always taken at the position $\hat{z}_a^i$ in such an ansatz
and $\hat{z}_a^i$ should not appear directly.
Of course one also has to take into account that
$\pi_a^{ij}$ must be symmetric and $\hat{e}^{ij}_a$ antisymmetric.
Considering possible ans\"atze under these restrictions we
infer that the leading orders are at least
% \begin{equation}
$\pi_a^{ij} = \Order{\left(c^{-5}\right)}$, %\,, \qquad
$z^i_{\Delta a} = \Order{\left(c^{-2}\right)}$, and %\,, \qquad
$\hat{e}^{ij}_a= \Order{\left(c^{-6}\right)}$. %\,.
% \end{equation}
From (\ref{scons}) the first orders of $s_{a}^{[ij]}$ then follow as
\begin{equation}
s_{a(3)}^{[ij]} = \hat{S}_{a(i)(j)} \,, \quad
s_{a(5)}^{[ij]} = \hat{p}_{aj} z^i_{\Delta a(2)} - \hat{p}_{ai} z^j_{\Delta a(2)} \,, \quad
s_{a(7)}^{[ij]} = \hat{p}_{aj} z^i_{\Delta a(4)} - \hat{p}_{ai} z^j_{\Delta a(4)} \,.
\end{equation}
Evaluating (\ref{Scons}) one concludes that
% \begin{equation}
$z^i_{\Delta a(2)} = 0$ and %\,, \qquad
$z^i_{\Delta a(4)} = 0$. %\,.
% \end{equation}
Thus we have anticipated the correct redefinition of the position (\ref{NWpos})
to this order.

For $s^{[ij]}_{a (9)}$ one has
\begin{equation}
	s^{[ij]}_{a (9)} = \hat{e}^{ik}_{a(6)} \hat{S}_{a(k)(j)} - \hat{p}_{ai} z^j_{\Delta a(6)}
		+ \frac{1}{4m_a^2} \hat{p}_{a k} h_{l j}^{\text{TT}} ( \hat{p}_{a i} \hat{S}_{a (l) (k)}
			+ \hat{p}_{a l} \hat{S}_{a (i) (k)} )
		- (i \leftrightarrow j) \,,
\end{equation}
where $(i \leftrightarrow j)$ denotes an exchange of the indices $i$ and $j$.
The most general solution of (\ref{Scons}) under above restrictions is
\begin{gather}
\pi^{ij}_{a(5)} = \frac{1-C}{8m_a^2} ( \hat{p}_{a i} \hat{p}_{a k} \hat{S}_{a (k) (j)} + \hat{p}_{a j} \hat{p}_{a k} \hat{S}_{a (k) (i)} ) \,, \\
\hat{e}^{ij}_{a(6)} = \frac{C}{4m_a^2} \hat{p}_{ak} ( \hat{p}_{ai} h_{j k}^{\text{TT}} - \hat{p}_{aj} h_{i k}^{\text{TT}} ) \,, \quad
z^i_{\Delta a(6)} = \frac{C}{4m_a^2} \hat{p}_{aj} ( \hat{S}_{a(k)(i)} h_{jk}^{\text{TT}} + \hat{S}_{a(k)(j)} h_{ik}^{\text{TT}} ) \,,
\end{gather}
at this order and now depends on an arbitrary constant $C$.

However, we can remove the ambiguity $C$ by a canonical transformation with an infinitesimal generator
\begin{equation}
	g = \frac{C}{4m_a^2} \hat{p}_{a i} \hat{p}_{a k} \hat{S}_{a (k) (j)} \int d^3x \, h^{\text{TT}}_{i j} \hat{\delta}_a \,.
\end{equation}
An arbitrary phase space function $A$ then transforms as
$A \rightarrow A + \{ A , g \}$ to the required order.
Applied to the canonical variables one obtains
\begin{gather}
	h^{\text{TT}}_{i j} \rightarrow h^{\text{TT}}_{i j} \,, \qquad
	\hat{\pi}^{i j \text{TT}} \rightarrow \hat{\pi}^{i j \text{TT}}
		- \delta^{\text{TT}ij}_{kl} \sum_a \frac{4\pi C}{m_a^2} \hat{p}_{a k} \hat{p}_{a m} \hat{S}_{a (m) (l)} \hat{\delta}_a \,, \\
	\hat{S}_{a(i)(j)} \rightarrow \hat{S}_{a(i)(j)} - \hat{e}^{ik}_{a(6)} \hat{S}_{a(k)(j)} - \hat{e}^{jk}_{a(6)} \hat{S}_{a(i)(k)} \,, \\
	\hat{z}^i_a \rightarrow \hat{z}^i_a - z^i_{\Delta a(6)} \,, \qquad
	\hat{p}_{a i} \rightarrow \hat{p}_{a i} - \frac{C}{4m_a^2} \hat{p}_{a l} \hat{p}_{a j} \hat{S}_{a (j) (k)} h^{\text{TT}}_{kl,i} \,.
\end{gather}
This indeed removes all terms depending on $C$ from the source expressions
$\mathcal{H}^{\text{matter}}$ and $\mathcal{H}^{\text{matter}}_{i}$ at the considered order.
We can therefore choose $C=0$, which leads to agreement with the action approach.
The triad then is in the spatial symmetric gauge $\hat{e}^{ij} = 0$ and
all variable transformations are the same as in the action approach
at the considered post-Newtonian order. In particular, using (\ref{eder}) and
(\ref{NWpi}) in (\ref{predef}) leads to (\ref{NWmom}).
Further, in the action approach we found that $z_{\Delta a}^i = 0$ and
\begin{equation}\label{defpiaA}
\pi_a^{ij} = \frac{1}{2} \hat{A}^{(ij)}_a + B^{ij}_{kl} \hat{A}^{[kl]}_a \,,
\end{equation}
or more explicitly using (\ref{defAhat})
\begin{equation}\label{defpia}
\pi_a^{ij} = \gamma^{ik} \gamma^{jl} \frac{m_a \hat{p}_{a(k} nS_{al)}}{2 n\hat{p}_a (m_a - n\hat{p}_a)}
	+ \frac{1}{2} B^{ij}_{kl} \gamma^{km} \gamma^{ln} \hat{S}_{amn} \,.
\end{equation}
Using $B^{ij}_{kl} = \Order{(c^{-4})}$, cf.\ \eqname\ (3.37) in \cite{Steinhoff:Wang:2009},
the post-Newtonian expansion of (\ref{defpia})
agrees with the findings in this {\secname}.
The check of the action approach given here is valid to the formal 3.5 post-Newtonian order.

\mychapter{Higher Orders in Spin\label{higher}}
Higher orders in spin require higher multipole moments, e.g.,
a black hole has a nonzero quadrupole at the quadratic level in
spin \cite{Thorne:1980}. We will constrain to quadrupole
and quadratic order in spin in this {\chapname}.
Besides spin-induced quadrupole deformations discussed here,
also tidal deformations induced through the gravitational field of
other objects have been treated in the literature, see, e.g.,
\cite{Hartle:1974,*Taylor:Poisson:2008,Damour:Nagar:2009}.

\mysection{Quadrupole Approximation\label{QuadSec}}
The extension of the pole-dipole approximation to higher multipoles
was already essentially completed some time ago
\cite{Dixon:1964,Madore:1969,*Dixon:1970:1,*Dixon:1970:2,Dixon:1973,Dixon:1974,*Dixon:1979},
see also \cite{Schattner:Lawitzki:1984,Ilhan:2009}, most notably by Dixon.
It should be stressed that Dixon's method incorporates
Mathisson's pioneering ideas \cite{Dixon:2008}.

\mysubsection{Quadrupole Approximation from Tulczyjew's Method}
A more direct application of Mathisson's ideas to the quadrupole order
was given in \cite{Steinhoff:Puetzfeld:2009} with the help
of W.\,M.\ Tulczyjew's method \cite{Tulczyjew:1959}, see also \cite{Trautman:2002}.
There the quadrupole moment $t^{\mu\nu\alpha\beta}$ was kept in
(\ref{Texp}) when evaluating (\ref{Tvar}). In addition to $p^{\mu}$ and
$S^{\mu\nu}$ now various quadrupole moments appear. It is suitable to introduce
a reduced quadrupole moment $J^{\mu\nu\alpha\beta}$ with symmetries
\begin{equation}
J^{\nu\rho\beta\alpha} = J^{[\nu\rho][\beta\alpha]}
	= J^{\beta\alpha\nu\rho} \,, \qquad
J^{\nu[\rho\beta\alpha]} = 0 \quad \Leftrightarrow \quad
	J^{\nu\rho\beta\alpha} + J^{\nu\beta\alpha\rho}
	+ J^{\nu\alpha\rho\beta} = 0 \,.
\end{equation}
Thus $J^{\rho\beta\alpha\nu}$ has the same (algebraic) symmetries as
the Riemann tensor. This quadrupole moment is able to incorporate all
quadrupole contributions from $t^{\mu\nu\alpha\beta}$ that remain after (\ref{Tvar}) was evaluated.
The equations of motion then take on the simple form
\begin{equation}\label{QEOM}
\frac{\dD S^{\mu\nu}}{\dd \tau}
	= 2 p^{[\mu} u^{\nu]} + \frac{4}{3} \Riem^{(4)}_{\alpha\beta\rho}{}^{[\mu} J^{\nu]\rho\beta\alpha} \,, \quad
\frac{\dD p_{\mu} }{\dd \tau} =
	- \frac{1}{2} \Riem^{(4)}_{\mu\rho\beta\alpha} u^{\rho} S^{\beta\alpha}
	- \frac{1}{6} \Riem^{(4)}_{\nu\rho\beta\alpha || \mu} J^{\nu\rho\beta\alpha} \,,
\end{equation}
and agree with Dixon \cite{Dixon:1974,*Dixon:1979}.
The reduced moment $J^{\mu\nu\alpha\beta}$ is also optimal to give
a simplified expression for the stress-energy tensor, reading
\begin{equation}\label{Qset}
\sqrt{-g} T^{\mu\nu} = \! \int \! \dd \tau \bigg[
	u^{(\mu} p^{\nu)} \delta_{(4)}
	+ \frac{1}{3} \Riem^{(4)}_{\alpha\beta\rho}{}^{(\mu} J^{\nu)\rho\beta\alpha} \delta_{(4)}
	+ \left( u^{(\mu} S^{\nu)\alpha} \delta_{(4)} \right)_{|| \alpha}
	- \frac{2}{3} \left( J^{\mu\alpha\beta\nu} \delta_{(4)} \right)_{|| (\alpha\beta)} \!
\bigg] \!.
\end{equation}
In this form the stress-energy tensor was first given in \cite{Steinhoff:Puetzfeld:2009}.
This stress-energy tensor, together with the ansatz for $J^{\mu\nu\alpha\beta}$
at the quadratic level in spin given below, can be applied to the
derivation of the next-to-leading order radiation field, see
\cite{Blanchet:Buonanno:Faye:2006,*Blanchet:Buonanno:Faye:2006:err}
for the spin-orbit case (the leading order is given in \cite{Kidder:1995}).
Besides this formula for the stress-energy tensor, a further interesting
result in \cite{Steinhoff:Puetzfeld:2009} is the relation between the
$t^{\mu\nu\dots}$ moments and Dixon's reduced moments
$p^{\mu}$, $S^{\mu\nu}$, and $J^{\mu\nu\alpha\beta}$. This relation
could be used to study alternatives to Dixon's integral formulas
for the multipole moments or to discuss the relation between
moments belonging to different representative worldlines
(for the latter see section VII.C in \cite{Steinhoff:Puetzfeld:2009}).

The spin supplementary condition $S^{\mu\nu} f_{\nu} = 0$ is preserved in time if
\begin{equation}\label{PUrelQuad}
p^{\mu} = \frac{1}{- f_{\alpha} u^{\alpha}} \left( - f_{\nu} p^{\nu} u^{\mu} + S^{\mu\nu} \frac{\dD f_{\nu}}{\dd \tau}
	+ \frac{4}{3} f_{\nu} \Riem^{(4)}_{\alpha\beta\rho}{}^{[\mu} J^{\nu]\rho\beta\alpha} \right) \,,
\end{equation}
which should give a relation between $p^{\mu}$ and $u^{\mu}$.
This extends (\ref{PUrel}) to the quadrupole
approximation. The extension of (\ref{PUnossc}) reads
\begin{equation}\label{PUnosscQuad}
p^{\mu} = - u_{\nu} p^{\nu} u^{\mu} - \frac{\dD ( S^{\mu\nu} )}{\dd \tau} u_{\nu}
	+ \frac{4}{3} u_{\nu} \Riem^{(4)}_{\alpha\beta\rho}{}^{[\mu} J^{\nu]\rho\beta\alpha} \,.
\end{equation}

\mysubsection{Decomposition of the Quadrupole}
In order to parametrize the quadrupole deformation due to spin we try to find
the most general covariant ansatz for $J^{\mu\nu\alpha\beta}$ quadratic in
the spin tensor that is relevant for the post-Newtonian order in question.
It is suitable to consider the orthogonal decomposition of $J^{\mu\nu\alpha\beta}$
with respect to the vector $f_{\mu}$ to which the spin is orthogonal, $S^{\mu\nu} f_{\nu} = 0$.
This decomposition reads
\begin{equation}\label{Jdecomp}
J^{\nu\rho\beta\alpha} = Q^{\nu\rho\beta\alpha}
	- \frac{1}{\sqrt{-f_{\nu}f^{\nu}}} ( f^{[\nu} Q^{\rho]\beta\alpha}
		+ f^{[\alpha} Q^{\beta]\rho\nu} )
	- \frac{3}{-f_{\nu}f^{\nu}} f^{[\nu} Q^{\rho][\beta} f^{\alpha]} \,,
\end{equation}
where $Q^{\nu\rho\beta\alpha}$, $Q^{\rho\beta\alpha}$, and $Q^{\rho\beta}$ are
called stress, flow, and mass quadrupole here and
are orthogonal to $f_{\mu}$ in each index, see also \cite{Ehlers:Rudolph:1977}.
They further have the symmetries
\begin{gather}
Q^{\nu\rho\beta\alpha} = Q^{[\nu\rho][\beta\alpha]} = Q^{\beta\alpha\nu\rho} \,, \qquad
Q^{\nu[\rho\beta\alpha]} = 0 \quad \Leftrightarrow \quad
	Q^{\nu\rho\beta\alpha} + Q^{\nu\beta\alpha\rho}
	+ Q^{\nu\alpha\rho\beta} = 0 \,, \\
Q^{\rho\beta\alpha} = Q^{\rho[\beta\alpha]} \,, \qquad
Q^{[\rho\beta\alpha]} = 0 \quad \Leftrightarrow \quad
	Q^{\rho\beta\alpha} + Q^{\beta\alpha\rho}
	+ Q^{\alpha\rho\beta} = 0 \,, \qquad
Q^{\rho\beta} = Q^{(\rho\beta)} \,.
\end{gather}

In a local basis with $f^{\mu}$ giving the time direction these moments only
have spatial components (due to the orthogonality of these moments to $f^{\mu}$).
One may therefore decompose these moments further in the local basis into
parts transforming under irreducible representations of the \emph{3-dimensional}
rotation group SO(3). For the mass quadrupole this SO(3)-decomposition reads,
written in the coordinate frame,
\begin{equation}\label{Qdecomp}
Q_{\mu\nu} = Q_{\mu\nu}^{\text{STF}} + \frac{1}{3} P_{\mu\nu} {Q^{\rho}}_{\rho} \,,
\end{equation}
with the orthogonal projector $P^{\mu\nu} = g^{\mu\nu} - \frac{1}{f_{\rho}f^{\rho}} f^{\mu} f^{\nu}$ (notice $P^{\mu\nu} g_{\mu\nu} = 3$). Here
$Q_{\mu\nu}^{\text{STF}}$ is symmetric and trace-free (STF) in the local frame.
In the coordinate frame the trace-free property reads $Q_{\mu\nu}^{\text{STF}} g^{\mu\nu} = 0$
and of course it holds $Q_{\mu\nu}^{\text{STF}} f^{\nu} = 0$.
Obviously $Q^{\mu\nu}$ has six independent components, five contained in the
symmetric trace-free part and one in the scalar part ${Q^{\rho}}_{\rho}$. The same
holds for $Q^{\mu\nu\alpha\beta}$ with a more complicated decomposition into
symmetric trace-free and scalar parts, whereas $Q^{\mu\nu\alpha}$ has even
eight independent components corresponding to a symmetric trace-free
and a vector part \cite{Ehlers:Rudolph:1977}.

\mysubsection{Ansatz for the Mass Quadrupole}
We will now constrain to $f_{\mu} = p_{\mu}$ and to the Newtonian limit, the
latter to identify the dominant contributions.
Then only the mass multipoles are important for the dynamics.
Though flow and stress multipoles do in general not vanish in the Newtonian
limit \cite{Dixon:1973}, they give no contribution to the gravitational field
and can be neglected. Therefore the decomposition of the quadrupole moment (\ref{Jdecomp}) just reads
\begin{equation}\label{Jansatz}
	J^{\nu\rho\beta\alpha} = - \frac{3}{m_p^2} p^{[\nu} Q^{\rho][\beta} p^{\alpha]} \,,
\end{equation}
where the dynamical mass defined by $p_{\mu} p^{\mu} = -m_p^2$ is now denoted as $m_p$.
Also the trace part of the mass quadrupole gives no contribution to the
gravitational field outside the body and one can thus assume ${Q^{\rho}}_{\rho} = 0$.
The mass quadrupole induced by spin is then given by the ansatz
\begin{equation}\label{Qansatz}
Q_{\mu\nu} = Q_{\mu\nu}^{\text{STF}} = \frac{C_Q}{m_p} \left( S_{\mu\rho} {S_{\nu}}^{\rho}
		- \frac{1}{3} P_{\mu\nu} S^{\rho\sigma} S_{\rho\sigma} \right) \,,
\end{equation}
and is parametrized only by $C_Q$ in the Newtonian limit and quadratic level in spin, see
also \cite{Poisson:1998}. For black holes one has $C_Q = 1$ \cite{Thorne:1980} while for
neutron star models $C_Q$ depends on the equation of state \cite{Laarakkers:Poisson:1999}.

Though the ansatz for the quadrupole (\ref{Jansatz}, \ref{Qansatz}) was given in the
Newtonian limit only, it was written in a manifestly covariant way and we will
now consider its implications in full general relativity. However, we stay at the
quadratic level in spin. It will be shown in the next {\secname} by relying on investigations in \cite{Porto:Rothstein:2008:2}
that this ansatz indeed holds to next-to-leading order in the post-Newtonian approximation. It is easy to see
from (\ref{QEOM}) that the spin length $S$ given by $2 S^2 = S^{\mu\nu} S_{\mu\nu}$
is conserved for our quadrupole ansatz and spin supplementary condition $S^{\mu\nu} p_{\mu} = 0$.
But the mass $m_p$ is not conserved. However, the new mass-like parameter $m$
defined by
\begin{equation}\label{Qmass}
m = m_p - \frac{1}{6} \Riem^{(4)}_{\nu\rho\beta\alpha} J^{\nu\rho\beta\alpha} \,,
\end{equation}
is conserved for our ansatz quadratic in spin. Finally (\ref{PUrelQuad}) can be written as
\begin{equation}\label{pmassQ}
p^{\mu} = m u^{\mu} - \frac{1}{2 m} \Riem_{\rho\nu\alpha\beta}^{(4)} S^{\mu\rho} S^{\alpha\beta} u^{\nu}
	+ \frac{1}{2} \Riem_{\delta\alpha\beta\nu}^{(4)} Q^{\alpha\beta} u^{\nu}
		( 2 g^{\delta\mu} + u^{\delta} u^{\mu} ) \,,
\end{equation}
and gives a relation between $p^{\mu}$ and $u^{\mu}$.

\mysection{Action Approach\label{QuadAction}}
It is shown in this {\secname} that allowing nonminimal couplings
in the action approach from \Sec{mincoup} corresponds to certain
higher multipole corrections, see also \cite{Bailey:Israel:1975}.
The couplings in the action needed for spin-induced quadrupole deformations
at next-to-leading order in the post-Newtonian approximation can be found
in \cite{Porto:Rothstein:2008:2}.

\mysubsection{Nonminimal Couplings\label{nonmincoup}}
We now generalize the ansatz (\ref{PDaction}) for the action to nonminimal couplings.
More precisely, the Lagrangian is allowed to contain the Riemann curvature tensor,
\begin{equation}\label{WMquad}
	W_M[e_{I\mu}, z^{\mu}, \Lambda^{AI}] = \int \dd \tau \,
		L_M(u^{\mu}, \Omega^{\mu\nu}, g^{\mu\nu}(z^{\rho}),
			g_{\mu\nu}(z^{\rho}), \Riem^{(4)}_{\mu\nu\alpha\beta}(z^{\rho})) \,.
\end{equation}
The Euler-Lagrange equations of this action follow as in \Sec{mincoup} in a straightforward way.
It is easy to see that (\ref{SEOM}) stays unchanged,
\begin{equation}\label{SEOM3}
\frac{\dD}{\dd \tau} \left[ \frac{\partial L_M}{\partial \Omega^{\mu\nu}} \right]
	= \frac{\partial L_M}{\partial \Omega^{\mu\rho}} {\Omega^{\rho}}_{\nu}
		- \frac{\partial L_M}{\partial \Omega^{\nu\rho}} {\Omega^{\rho}}_{\mu} \,.
\end{equation}
However, the important relation (\ref{RPI2}) now reads
\begin{equation}
0 = \frac{\partial L_M}{\partial u^{\alpha}} u^{\beta}
	+ 2 \frac{\partial L_M}{\partial \Omega^{\alpha\nu}} \Omega^{\beta\nu}
	+ 2 \frac{\partial L_M}{\partial g^{\alpha\nu}} g^{\beta\nu}
	- 2 \frac{\partial L_M}{\partial g_{\beta\nu}} g_{\alpha\nu}
	- 4 \frac{\partial L_M}{\partial \Riem^{(4)}_{\beta\nu\rho\delta}} \Riem^{(4)}_{\alpha\nu\rho\delta} \,.
\end{equation}
Using this identity and the definitions
\begin{equation}
p_{\mu} = \frac{\partial L_M}{\partial u^{\mu}} \,, \qquad
S_{\mu\nu} = 2 \frac{\partial L_M}{\partial \Omega^{\mu\nu}} \,, \qquad
J^{\mu\nu\alpha\beta} = - 6 \frac{\partial L_M}{\partial \Riem^{(4)}_{\mu\nu\alpha\beta}} \,, \label{JDefA}
\end{equation}
the Euler-Lagrange equations for the matter variables turn into (\ref{QEOM}),
whereas for the field variables one obtains the
Einstein equations with the stress-energy tensor (\ref{Qset}).
Higher multipoles are covered by including symmetrized covariant derivatives
of the curvature tensor in the Lagrangian \cite{Bailey:Israel:1975}.

An action invariant under general coordinate transformations always leads to a
stress-energy tensor fulfilling (\ref{Tvar}). As well known, this can be shown
from the Noether identity \cite{Noether:1918} following from general covariance,
see also \eqname\ (18.23) in \cite{DeWitt:1964}. However, this does not mean that an
action approach as envisaged here always leads to the most general
equations of motion allowed by (\ref{Tvar}).
Dixon's derivation essentially only evaluated (\ref{Tvar}) and thus covers a much more general situation,
though the Euler-Lagrange equations obtained in this {\secname} are identical
to the equations of motion found by Dixon (but it is not a priori clear
that this will be the case).
In particular, the multipole moments will always be implicitly fixed by the other variables on which the
Lagrangian was chosen to depend on, cf.\ (\ref{JDefA}).
This means that the quadrupole is not a dynamical variable within our
action approach and our action does not cover, e.g., quadrupole oscillation
modes or tidal resonances, see, e.g., \cite{Kokkotas:Schmidt:1999,*Alexander:1987}.
Notice that (\ref{Tvar}) puts
no constraints on equations of motion related to dynamical quadrupole degrees of freedom.
However, a dynamical quadrupole requires further dynamical variables in the action
principle. For a good effective description of extended objects via
an action one thus needs some intuition on the relevant
degrees of freedom that should enter into an ansatz for the effective action.
The action (\ref{WMquad}) includes translational and rotational
degrees of freedom of the object, which is expected to be a
good choice for the inspiral phase.

An advantage of the action approach is that it is easier to find
conserved quantities or constant parameters. In particular, the conservation
of the spin length immediately follows from the symmetry under Lorentz transformations
of the body-fixed frame. Further all parameters in the action, e.g., a mass-like parameter,
are constant simply by assumption. It is much more difficult to find
such constant quantities if one only considers Dixon's results together
with a specific ansatz for the quadrupole moment,
see, e.g., \Eq{Qmass} or the discussion in
reference \cite{Steinhoff:Puetzfeld:2009}.

It is important that one may eliminate the Ricci tensor (and scalar) from the matter
Lagrangian $L_M$ by a suitable redefinition of the metric \cite{Goldberger:Rothstein:2006}.
This was already found in \cite{Damour:EspositoFarese:1998}
within a slightly different situation and is based on the observation that in a perturbative
context the use of lower order equations of motion in the perturbation part of the
action corresponds to a redefinition of variables, see \cite{Schafer:1984}.
If we take some additions to the point-mass Lagrangian as a perturbation,
then we may eliminate the Ricci tensor by using the Einstein field equations
with the point-mass stress-energy tensor as a source, corresponding to an
irrelevant redefinition of the metric. However, the point-mass stress-energy
tensor then gives rise to singular self-interactions in the matter Lagrangian $L_M$,
which are formally neglected \cite{Damour:EspositoFarese:1998}. The
conclusion is that one may use the \emph{vacuum} field equations
$\Riem^{(4)}_{\mu\nu} = 0$ in the matter Lagrangian $L_M$.
(This will also be used in a slightly modified way in \Sec{S2LOA}.)
The matter Lagrangian can therefore be restricted to depend
on the completely trace-free Weyl tensor $C^{(4)}_{\mu\nu\alpha\beta}$,
\begin{equation}\label{Weyl}
C^{(4)}_{\mu\alpha\nu\beta} = \Riem^{(4)}_{\mu\alpha\nu\beta}
		+ g_{\alpha[\nu} \Riem^{(4)}_{\beta]\mu}
		- g_{\mu[\nu} \Riem^{(4)}_{\beta]\alpha}
		+ \frac{1}{3} g_{\mu[\nu} g_{\beta]\alpha} \Riem^{(4)} \,,
\end{equation}
instead of $\Riem^{(4)}_{\mu\nu\alpha\beta}$. This would give rise to
corresponding modified multipole moments defined analogous to (\ref{JDefA}).
Further, the Weyl tensor can be split into electric $E_{\mu\nu}^{(4)}$
and magnetic $B_{\mu\nu}^{(4)}$ parts,
\begin{equation}\label{EMWeyl}
E_{\mu\nu}^{(4)} = C^{(4)}_{\mu\alpha\nu\beta} u^{\alpha} u^{\beta} \,, \qquad
B_{\mu\nu}^{(4)} = \frac{1}{2} \epsilon_{\mu\rho\alpha\beta}^{(4)} C^{(4)}_{\nu\sigma}{}^{\alpha\beta} u^{\rho} u^{\sigma} \,,
\end{equation}
with $\epsilon_{\mu\alpha\beta\rho}^{(4)}$ the 4-dimensional Levi-Civita symbol,
leading to definitions for corresponding electric and magnetic multipoles
as partial derivatives of $L_M$. It could be interesting to consider
the impact on the equations of motion and the stress-energy tensor
from letting $L_M$ depend on $C^{(4)}_{\mu\nu\alpha\beta}$ or
$E_{\mu\nu}^{(4)}$ and $B_{\mu\nu}^{(4)}$ instead of $\Riem^{(4)}_{\mu\nu\alpha\beta}$.
However, this will not be necessary here.

\mysubsection{Legendre Transforms and Supplementary Conditions\label{ASSC}}
Up to now the Lagrangian $L_M$
is completely arbitrary and the equations of motion fully agree
with Dixon at the quadrupole level. The question is which supplementary
conditions (\ref{Lssc}) and (\ref{covSSC2}) belong to $L_M$, or how $L_M$
must be chosen to fit with specific supplementary conditions.
We only require here that (\ref{Lssc}) and (\ref{covSSC2})
are preserved in time, which leads to
\begin{equation}
\frac{\dD \Lambda^{A I}}{\dd \tau} f_I
	+ \Lambda^{A I} \left( \eta_{IJ} - \frac{f_I f_J}{f_K f^K} \right) \frac{\dD f^J}{\dd \tau} = 0 \,, \qquad
{S^{\mu}}_{\rho} \Omega^{\rho\nu} f_{\nu}
	+ S^{\mu\nu} \frac{\dD f_{\nu}}{\dd \tau} = 0 \,, \label{Acond2}
\end{equation}
where (\ref{SEOM3}) with (\ref{JDefA}) was used.
Both conditions are fulfilled if we have
\begin{equation}\label{Acond}
\Omega^{\mu\nu} f_{\nu} + P^{\mu\rho} \frac{\dD f_{\rho}}{\dd \tau} = 0 \,.
\end{equation}
In this sense the conditions (\ref{Lssc}) and (\ref{covSSC2}) belong together
(however, there may be exceptions). This is the condition our action shall fulfill here.
Notice that (\ref{PUrelQuad}) only guarantees that (\ref{covSSC2}) is preserved
in time so that the second relation in (\ref{Acond2}) holds, but this does not imply (\ref{Acond}).
However, comparing (\ref{JDefA}) with (\ref{PUrelQuad}) can still be useful.
Further, at the quadratic level in spin we need to fulfill (\ref{Acond}) only to
linear order in spin.

It is suitable to define a new function $R_M(u^{\mu}, S_{\mu\nu}, g^{\mu\nu}, g_{\mu\nu}, \Riem^{(4)}_{\mu\nu\alpha\beta})$
via Legendre transformation,
% \begin{gather}
$R_M = L_M - \frac{1}{2} S_{\mu\nu} \Omega^{\mu\nu}$. %\,.
% \end{gather}
It holds
\begin{equation}
p_{\mu} = \frac{\partial R_M}{\partial u^{\mu}} \,, \qquad
\Omega^{\mu\nu} = - 2 \frac{\partial R_M}{\partial S_{\mu\nu}} \,, \qquad
J^{\mu\nu\alpha\beta} = - 6 \frac{\partial R_M}{\partial \Riem^{(4)}_{\mu\nu\alpha\beta}} \,. \label{Rpd}
\end{equation}
An ansatz for $R_M$ then has to fulfill the condition (\ref{Acond})
with (\ref{Rpd}) inserted. This gives a partial differential equation
for $R_M$. It holds $R_M = p_{\mu} u^{\mu}$, which is a consequence
of the reparametrization invariance of the matter action.
Notice that $R_M$ is similar to the Routhian used in
\cite{Yee:Bander:1993,Porto:Rothstein:2008:1,Porto:Rothstein:2008:2}.
For the Routhian the Ricci rotation part in the term $\frac{1}{2} S_{\mu\nu} \Omega^{\mu\nu}$,
cf.\ \Eq{Odef}, is not subtracted from the Lagrangian $L_M$. Therefore
the Routhian is not a covariant function, whereas $R_M$ introduced
here is covariant.

Due to reparametrization invariance a full Legendre transformation
in $u^{\mu}$ and $\Omega^{\mu\nu}$ leads to a vanishing result.
However, as in \Sec{matcoup} we define a function $H_{M\tau}$ which
contains the mass-shell constraint, suitably generalized to the
quadratic-in-spin level, together with a Lagrange multiplier $\lambda$.
It holds
\begin{equation}
u^{\mu} = \frac{\partial H_{M\tau}}{\partial p_{\mu}} \,, \qquad
\Omega^{\mu\nu} = 2 \frac{\partial H_{M\tau}}{\partial S_{\mu\nu}} \,, \qquad
J^{\mu\nu\alpha\beta} = 6 \frac{\partial H_{M\tau}}{\partial \Riem^{(4)}_{\mu\nu\alpha\beta}} \,. \label{Hpd}
\end{equation}
It is also possible to give an ansatz for the mass-shell constraint
and thus for $H_{M\tau}$ directly. This ansatz must be chosen such that
the condition (\ref{Acond}) with (\ref{Hpd}) inserted is fulfilled.
Further, one may simplify the quadratic-in-spin corrections to $H_{M\tau}$
by using the leading order constraint $p^{\mu}p_{\mu} = - m^2$, corresponding
to a redefinition of the Lagrange multiplier \cite{Schafer:1984}.

\mysubsection{Leading Order\label{S2LOA}}
The coupling terms found in \cite{Porto:Rothstein:2008:2}
adapted to our notation and conventions read
\begin{equation}
R_M = \frac{1}{\sqrt{- u_{\sigma} u^{\sigma}}} \left( m u_{\mu} u^{\mu}
	- \frac{1}{2 m} \Riem^{(4)}_{\mu\nu\alpha\beta} S^{\rho\mu} S^{\alpha\beta} u^{\nu} u_{\rho}
	+ \frac{C_{ES^2}}{2 m} E^{(4)}_{\mu\nu} {S^{\mu}}_{\rho} S^{\rho\nu} \right) \,.
\end{equation}
It was found in \cite{Porto:Rothstein:2008:2} that these coupling terms
are the most general ones at quadratic level in spin sufficient for the next-to-leading order in
the post-Newtonian approximation. Besides these terms corresponding
to quadrupole deformation due to spin, one could also treat tidal
deformations, see, e.g., \cite{Hartle:1974,*Taylor:Poisson:2008},
using nonminimal couplings in the action given in
\cite{Damour:Nagar:2009,Goldberger:Rothstein:2006:2}.
With the equivalence of Riemann and Weyl tensors within the matter
action, see \Sec{nonmincoup}, we can write $R_M$ as
\begin{equation}\label{Rquad}
R_M = \frac{1}{\sqrt{- u_{\sigma} u^{\sigma}}} \left( m u_{\mu} u^{\mu}
	- \frac{1}{2 m} \Riem_{\mu\nu\alpha\beta}^{(4)}
		S^{\rho\mu} S^{\alpha\beta} u^{\nu} u_{\rho}
	- \frac{1}{2} \Riem_{\alpha\mu\beta\nu}^{(4)}
		Q^{\alpha\beta} u^{\mu} u^{\nu} \right) \,,
\end{equation}
where $Q_{\mu\nu}$ is given by (\ref{Qansatz}) and $C_{ES^2} = C_Q$.
If we set $f_{\mu} = p_{\mu}$ and thus
\begin{equation}
S^{\mu\nu} p_{\nu} = S^{\mu\nu} \frac{\partial R_M}{\partial u^{\nu}} = 0 \,,
\end{equation}
we obviously reproduce (\ref{Jansatz}) and (\ref{pmassQ})
within the gauge $u_{\sigma} u^{\sigma} = -1$ by plugging (\ref{Rquad}) into (\ref{Rpd}).
Further (\ref{Acond}) is fulfilled\footnote{One could also
consider the most general ansatz for $R_M$ and ask for which
choice of $f_{\mu}$ the condition (\ref{Acond}) is fulfilled.
This would allow one to study the impact of the supplementary conditions
on the dynamics.}
to the considered order in spin
by using (\ref{QEOM}) and (\ref{Rpd}), i.e.,
\begin{equation}
0 = - 2 \frac{\partial R_M}{\partial S_{\mu\nu}} \frac{\partial R_M}{\partial u^{\nu}}
- \frac{1}{2} P^{\mu\sigma} \Riem^{(4)}_{\sigma\rho\beta\alpha} u^{\rho} S^{\beta\alpha}
+ P^{\mu\sigma} \Riem^{(4)}_{\nu\rho\beta\alpha || \sigma}
	\frac{\partial R_M}{\partial \Riem^{(4)}_{\nu\rho\beta\alpha}} \,.
\end{equation}
The last term is of higher order here as this condition must be fulfilled to
linear order in spin only.
Notice that $m$ depends on spin according to
% \begin{equation}\label{mdef}
$m = m_0 + \frac{1}{4 I} S_{\alpha\beta} S^{\alpha\beta}$, %\,,
% \end{equation}
see also (\ref{mtom0}).
(Otherwise the Legendre transformation between $L_M$ and $R_M$ would not be possible.)

An equivalent description in terms of $H_{M\tau}$ reads
\begin{equation}\label{HMquad}
H_{M\tau} = \lambda \left( m^2 + p_{\mu} p^{\mu}
	+ \frac{1}{m^2} \Riem_{\mu\nu\alpha\beta}^{(4)} S^{\rho\mu} S^{\alpha\beta} p^{\nu} p_{\rho}
	- \frac{C_Q}{m^2} \Riem_{\alpha\mu\beta\nu}^{(4)} {S^{\alpha}}_{\rho} S^{\rho\beta} p^{\mu} p^{\nu} \right) \,,
\end{equation}
with the action still given by (\ref{pdaction}). The derivation of a
canonical formalism now follows along the same lines as in \Chap{action}.
First the matter constraints are solved. The only difference to the
linear-in-spin case arises in the mass-shell constraint, which follows from
the variation of $\lambda$. The solution of this constraint reads
\begin{equation}
np \equiv n^{\mu} p_{\mu} = - \sqrt{m^2 + \gamma^{ij} p_{i} p_{j}} + \frac{C_Q}{2 m^2 \sqrt{m^2 + \gamma^{ij} p_{i} p_{j}}} \Riem_{\alpha\mu\beta\nu}^{(4)}
		{S^{\alpha}}_{\rho} S^{\rho\beta} p^{\mu} p^{\nu} \,.
\end{equation}
The last term was not yet split into time and space parts. This
splitting leads to quite many terms, so one should
restrict to some post-Newtonian order. Though all formulas
are sufficient for the next-to-leading order, we will
for simplicity only treat the leading order in this {\secname}.
Then we have
\begin{equation}\label{npQuad}
np = - \sqrt{m^2 + \gamma^{ij} p_{i} p_{j}} - \frac{C_Q}{2 m N} \gamma^{kl} \gamma^{im} \gamma^{jn} S_{ik} S_{jl} ( K_{mn,0} +  N_{;mn} ) \,.
\end{equation}
The only contribution to the action quadratic in spin
then arises from the term $N np$ in the matter Lagrangian, see (\ref{Lpm31}).
Problematic is the partial time derivative of the extrinsic curvature.
In consideration of the definition $2 N K_{ij} = - \gamma_{ij,0} + 2 N_{(i;j)}$
we see that the $K_{mn,0}$-term produces time-derivatives of lapse and shift,
as well as a double time-derivative of $\gamma_{ij}$. This does not fit
well to the derivation of the canonical formalism as given in \Chap{action}.
In order to overcome these problems, we eliminate $K_{mn,0}$ with the help
of the vacuum field equations, cf.\ the discussion in \Sec{nonmincoup}.
Finally one ends up with just a quadratic-in-spin correction $\mathcal{H}^{\text{matter}}_{\text{S}^2}$
to the source of the Hamilton constraint $\mathcal{H}^{\text{matter}}$ of the form
\begin{equation}
\mathcal{H}^{\text{matter}}_{\text{S}^2} = \frac{C_Q}{2 m} \gamma^{kl} \Riem^{ij} S_{ik} S_{jl} \delta \,.
\end{equation}
This source term is quite unusual in the sense that it is not
a specific projection of the stress-energy tensor (\ref{Qset}, \ref{Jansatz}, \ref{Qansatz}),
i.e., $\mathcal{H}^{\text{matter}} \neq \sqrt{\gamma}T_{\mu\nu} n^{\mu}n^{\nu}$.
This is due to the implicit redefinition of variables performed by
using the vacuum field equations in the matter action.
However, the leading order Hamiltonian resulting from this
source term is identical to the well-known one obtained in \Sec{HLO}.
The variable redefinitions from \Sec{vartrans} are still correct at the
leading order. (There are no additional terms that need to be cancelled in the action
and all quadratic spin contributions from the redefinitions in \Sec{vartrans}
are of higher order.)

At the next-to-leading order the calculation gets much more involved. In
particular there are more time derivatives of the extrinsic curvature that must be eliminated
and the variable redefinitions from \Sec{vartrans} need corrections quadratic in spin.
It is also relevant whether the field variables in the variable transformations
are taken at the new or at the old particle position.
Further corrections to the canonical field momentum seem to be necessary, too.
We will therefore study an alternative derivation oriented at the symmetry generator
approach from \Chap{symmetry} in the following.

\mysection{Symmetry Generator Approach}
We now sketch the derivation of the canonical formalism at quadratic level in spin
via the approach from \Chap{symmetry}. However, essentially only the calculation
of the source terms of the constraints as certain projections of the stress-energy
tensor is used here, the determination of canonical variables by looking at
the symmetry generators will only be touched lightly.

\mysubsection{Leading Order}
First we calculate the source of the field constraints as certain projections of
the stress-energy tensor (\ref{Qset}, \ref{Jansatz}, \ref{Qansatz}),
e.g., $\mathcal{H}^{\text{matter}} = \sqrt{\gamma}T_{\mu\nu} n^{\mu}n^{\nu}$.
To leading order we have
\begin{equation}\label{HSLO}
\mathcal{H}^{\text{matter}}_{\text{S}^2} = \sum_a
	\left( \frac{1}{2} \gamma^{ki} \gamma^{lj} Q_{a ij} \delta_a \right)_{;kl} \,,
\end{equation}
and no corrections appear in $\mathcal{H}^{\text{matter}}_i$.
The variable redefinitions found at the linear order in spin
are therefore sufficient here, as they are followed from $\mathcal{H}^{\text{matter}}_i$
in the symmetry generator approach. For $C_{Qa} = 1$ this source term is in
agreement with the source of the Kerr metric in approximate ADM
coordinates found in \cite{Hergt:Schafer:2008:2}.
It further gives the correct Hamiltonian, see \Sec{HLO}.

% \begin{equation}
% \begin{split}
% \mathcal{H}^{\rm matter}_{\text{S}^2} &= \sum_a
% 	\left( \frac{1}{2} \gamma^{ki} \gamma^{lj} \hat{Q}_{a ij} \hat{\delta}_a
% 	+ \frac{1}{m_a} C_{Ta} \hat{\vct{S}}_a^2 \gamma^{kl} \hat{\delta}_a \right)_{; kl} \,.
% \end{split}
% \end{equation}

Obviously the derivation of the leading order in this {\secname} is much simpler
than the one via the action approach. But this does not need to be true
at the next-to-leading order. The problem is that it is not guaranteed
that the variable redefinitions can be uniquely fixed by just the conditions
(\ref{defP}) and (\ref{Scons}).
% Further the extrinsic curvature
% could remain in the source terms and those terms can not be handled
% in a systematic way, as opposed to the action approach.
The action approach is much more systematic and should therefore be
preferred at the next-to-leading order. However, in the next {\secname}
a shortcut to the next-to-leading order Hamiltonian is described, which
combines the approach of the present {\secname} with the
Poincar\'e algebra approach in \cite{Hergt:Schafer:2008}.

\mysubsection{Next-to-Leading Order Static Source Terms\label{S2calc}}
In \cite{Hergt:Schafer:2008} Hergt and Sch\"afer constructed the part
of the next-to-leading order Hamiltonian that depends on $\hat{p}_i$ (i.e., the nonstatic part)
from an ansatz for this Hamiltonian (together with a suitable ansatz for
the source of the constraints). The coefficients in this ansatz
could be uniquely fixed up to a canonical transformation by considering
the Poincar\'e algebra (\ref{poinc1}, \ref{poinc2}). The degrees of freedom corresponding
to the ambiguity in the canonical representation are given by
the coefficients that enter via an ansatz for the center of mass
vector $G_{i}$. However, the static (i.e., $\hat{p}_i=0$) part of
the Hamiltonian is left completely undetermined by the Poincar\'e algebra
approach in \cite{Hergt:Schafer:2008}.

In order to get the complete next-to-leading order Hamiltonian only the
static part of the Hamiltonian is missing, as well
as the corresponding center of mass vector $G_{i}$. The latter is needed
to consistently fix the canonical representation of the nonstatic part
of the Hamiltonian given in \cite{Hergt:Schafer:2008}. Fortunately
the center of mass vector does not depend on $\hat{p}_i$ at the considered order. Therefore
both the static part of the Hamiltonian and the center of mass vector
are determined if we only know the static part of the source of the constraints.
For $p_i=0$ we get from the stress-energy tensor (\ref{Qset}, \ref{Jansatz}, \ref{Qansatz})
\begin{equation}
\mathcal{H}^{\text{matter}}_{\text{S}^2 \!, \, p_i=0} = \sum_a
	\left( \frac{1}{2} \gamma^{ki} \gamma^{lj} Q_{a ij} \delta_a \right)_{;kl} \,,
\end{equation}
but no further contributions to $\mathcal{H}^{\text{matter}}_i$ arise.
Though there is no difference to (\ref{HSLO}), this source term is
now valid to next-to-leading order for the case $p_i=0$.

However, we need the source terms for the case $\hat{p}_i=0$ and not
for $p_i=0$. Also position and spin variables are not yet the canonical
ones and we must discuss whether the variable redefinitions will have an
impact on the source terms in the static case. As there are no contributions
to $\mathcal{H}^{\text{matter}}_i$ at quadratic level in spin for $p_i=0$,
no further \emph{static} contributions to the redefinition of spin and momentum
variables can arise from the conditions (\ref{defP}) and (\ref{Scons}).
Though static contributions to $z^i_{\Delta a(4)}$ could be necessary,
they can be removed by a canonical transformation with generator $- \hat{p}_i z^i_{\Delta a(4)}$.
(Notice that in the case $\hat{p}_i=0$ this transformation only
changes the position variable.) Finally only the redefinitions found
at the linear order in spin are relevant and only (\ref{NWmom}) gives
contributions in the static case. The result for the static source
finally reads
\begin{equation}
\begin{split}\label{HS2cov}
\mathcal{H}^{\rm matter}_{\text{S}^2 \!, \, \hat{p}_i=0} &= \sum_a \bigg[
	\left( \frac{1}{2} \gamma^{ki} \gamma^{lj} \hat{Q}_{a ij} \hat{\delta}_a \right)_{; kl}
	+ \frac{1}{8 m_a} \gamma_{mn} \gamma^{pj} \gamma^{ql} {\gamma^{mi}}_{,p} {\gamma^{nk}}_{,q} \hat{S}_{1 ij} \hat{S}_{1 kl} \hat{\delta}_a \nlq
	+ \frac{1}{4m_a} \left( \gamma^{ij} \gamma^{mn} {\gamma^{kl}}_{,m} \hat{S}_{a ln} \hat{S}_{a jk} \hat{\delta}_a \right)_{,i}
\bigg] \,,
\end{split}
\end{equation}
where
\begin{equation}\label{Qcan}
\hat{Q}_{a ij} = \frac{C_{Qa}}{m} \left( \gamma^{kl} \hat{S}_{aik} \hat{S}_{ajl}
	- \frac{1}{3} \gamma_{ij} \gamma^{kl} \gamma^{mn} \hat{S}_{akm} \hat{S}_{aln} \right) \,.
\end{equation}
Equation (\ref{HS2cov}) was found for the black hole case $C_{Qa} = 1$ in \cite{Steinhoff:Hergt:Schafer:2008:1} from
a 3-dimensional covariant ansatz for $\mathcal{H}^{\rm matter}_{\text{S}^2 \!, \, p_i=0}$
containing four coefficients. Two of these coefficients were fixed by
matching to the Kerr metric, but the other two gave no contribution
to the Hamiltonian or to the center of mass vector. One of the latter two
coefficients would also arise here if we would have kept the trace part of
the mass quadrupole, ${Q^{\rho}}_{\rho}$. The ansatz in \cite{Steinhoff:Hergt:Schafer:2008:1}
was generalized to arbitrary $C_{Qa}$ in \cite{Hergt:Steinhoff:Schafer:2010:1}.

The derivation given in this {\secname} is quite involved and it would thus
be desirable to give a more coherent one with the help of the action
approach in the future. This would also facilitate further investigations
of quadrupole or higher multipole effects with the help of canonical
methods.

\mychapter{Results for Hamiltonians\label{result}}
In this {\chapname} the obtained canonical formalism is applied to
calculations within the post-Newtonian approximation.
In particular, the next-to-leading order spin corrections
to the conservative Hamiltonian are derived. The Hamiltonians
are checked with the help of the global Poincar\'e algebra.

In this {\chapname} we make use of xTensor \cite{MartinGarcia:2002}, a free package
for Mathematica \cite{Wolfram:2003}, especially of its fast index
canonicalizer based on the package xPerm \cite{MartinGarcia:2008}.

\mysection{Post-Newtonian Expansion}
The post-Newtonian expansion of the ADM Hamiltonian has been well studied
for nonspinning objects, for the second post-Newtonian level see
\cite{Kimura:Toiya:1972,*Ohta:Okamura:Kimura:Hiida:1974,*Ohta:Kimura:Hiida:1975,*Schafer:1985,*Damour:Schafer:1985,*Damour:Schafer:1988,*Ohta:Kimura:1989},
and up to and including the 3.5 post-Newtonian order see
\cite{Jaranowski:Schafer:1997,*Jaranowski:Schafer:1999,Jaranowski:Schafer:1998,Damour:Jaranowski:Schafer:2000,Damour:Jaranowski:Schafer:2001}.
From this expanded Hamiltonian the approximate equations of motion
can be derived in a straightforward way.
In this {\secname} we derive general formulas for the ADM Hamiltonian
up to and including the formal second post-Newtonian order,
which will then be applied to calculate spin corrections to the
Hamiltonian in \Sec{spinham}. Another interesting application
would be to obtain spin corrections to the post-Minkowskian Hamiltonian,
see, e.g., \cite{Ledvinka:Schafer:Bicak:2008} for the nonspinning case.

Besides the ADM formalism, there are various other methods available
for post-Newtonian calculations.
The equations of motion at the first post-Newtonian order are due to
Einstein, Infeld, and Hoffmann \cite{Einstein:Infeld:Hoffmann:1938},
obtained with the help of a surface integral approach.
This method got further developed and applied up to and including the
third post-Newtonian level, see, e.g., \cite{Futamase:Itoh:2007}.
A further important method uses point-masses in harmonic gauge,
which also succeeded to derive the third post-Newtonian order equations of motion;
for a review see \cite{Blanchet:2006}.
This method has advantages for flux and waveform calculations, which succeeded up to
the third post-Newtonian order
\cite{Arun:Blanchet:Iyer:Qusailah:2008,*Blanchet:Faye:Iyer:Sinha:2008,*Arun:Blanchet:Iyer:Sinha:2009}
(corresponding to the knowledge of the equations of motion at the 5.5 post-Newtonian level, which
seem to be impossible to obtain directly). Another approach in the harmonic gauge
is the direct integration of the relaxed Einstein equations, see, e.g., \cite{Pati:Will:2000,*Pati:Will:2002}.
More recently also methods inspired by quantum field theory were
developed, see, e.g, \cite{Goldberger:Rothstein:2006,Gilmore:Ross:2008,*Kol:Smolkin:2009}.
An advantage of these methods is that some of the very sophisticated and systematic
techniques for perturbative calculations used in high energy physics
can be applied in a straightforward way.

\mysubsection{Review of the Formalism}
We now give a short summary of the calculation of the ADM Hamiltonian.
First the field constraints
\begin{equation}\label{fconstraints}
	\frac{1}{16\pi\sqrt{\gamma}} \left[ \gamma \Riem
		+ \frac{1}{2} \left( \gamma_{ij} \pi^{ij} \right)^2
		- \gamma_{ij} \gamma_{k l} \pi^{ik} \pi^{jl}\right]
		= \mathcal{H}^{\text{matter}} \,, \qquad
	- \frac{1}{8\pi} \gamma_{ij} \pi^{jk}_{~~ ; k} = \mathcal{H}^{\text{matter}}_i \,,
\end{equation}
have to be solved within the ADM transverse traceless gauge, which
for the metric leads to the decomposition
\begin{equation}
	\gamma_{ij} = \left( 1 + \frac{\phi}{8} \right)^4 \delta_{ij} + h^{\text{TT}}_{ij} \,, \label{gdecomp2}
\end{equation}
at least to linear order in spin.
Such a solution can in general only be found in some approximation scheme and
we consider the post-Newtonian one here. Having the decomposition (\ref{gdecomp2})
one can solve the Hamilton constraint for $\phi$ (this will become obvious in
the next {\secname}). Then we can calculate the ADM Hamiltonian
\begin{equation}\label{HADM}
H_{\text{ADM}} = - \frac{1}{16\pi} \int \dd^3 x \, \Delta \phi \,,
\end{equation}
which must be expressed in terms of the canonical variables. It is
suitable to already express the source terms $\mathcal{H}^{\text{matter}}$ and
$\mathcal{H}^{\text{matter}}_i$ in terms of the canonical matter
variables, which is done in \Sec{Hmattercan}. Then no further redefinition of the matter variables is
necessary.

However, it seems to be simpler to perform the redefinition of the
field momentum after solving the constraints. As the gauge condition
at linear order in spin now reads $\hat{\pi}^{ii} = 0$, or, with
(\ref{pican}), (\ref{defpia}), and $B^{kl}_{ij} \delta_{kl} = 0$,
\begin{equation}\label{newgauge}
\pi^{ii} = - 16 \pi \sum_a \pi^{ii}_a \hat{\delta}_a
	= - 16 \pi \sum_a \delta_{ij} \gamma^{ik} \gamma^{jl} \frac{m_a \hat{p}_{a k} nS_{al }}{2 n\hat{p}_a (m_a - n\hat{p}_a)} \hat{\delta}_a \,,
\end{equation}
the decomposition (\ref{pidecomp}) is not valid any more.
But we can still use the general decomposition
\begin{equation}\label{decomp}
\pi^{ij} = \pi^{ij\text{TT}} + \tilde{\pi}^{ij} + \breve{\pi}^{ij} \,,
\end{equation}
with
\begin{gather}
\pi^{ij\text{TT}} = \delta^{\text{TT}ij}_{kl} \pi^{kl} \,, \qquad
\breve{\pi}^{ij} = \frac{1}{2} \left( \delta_{ij} - \partial_i \partial_j \Delta^{-1} \right) \pi^{kk} \,, \label{pidecomp2} \\
\tilde{\pi}^{ij} = \tilde{\pi}^i{}_{, j} + \tilde{\pi}^j{}_{, i}
	- \frac{1}{2} \delta_{ij} \tilde{\pi}^k{}_{, k}
	- \frac{1}{2} \Delta^{-1} \tilde{\pi}^k{}_{, ijk} \,, \label{pisolve}
\end{gather}
and the vector potential is still $\tilde{\pi}^i = \Delta^{-1} {\pi^{ij}}_{,j}$.
This can be shown by inserting (\ref{pidecomp2}, \ref{pisolve}) and (\ref{TTproj}) into
(\ref{decomp}), which then turns into an identity. The new part
$\breve{\pi}^{ij}$ can immediately be obtained using (\ref{newgauge}).
After the constraints have been solved using this decomposition,
we go over to the canonical field momentum $\hat{\pi}^{ij \text{TT}}$ by
\begin{equation}
\pi^{ij\text{TT}} = \hat{\pi}^{ij\text{TT}} - 16 \pi \sum_a \delta^{\text{TT}ij}_{kl} \pi^{kl}_a \hat{\delta}_a \,.
\end{equation}
No redefinition of $h^{\text{TT}}_{ij}$ is needed at the linear order in spin.

\mysubsection{Expansion of the Constraints\label{expansion}}
Now we expand the constraints according to the formal post-Newtonian counting
rules introduced in \Sec{count}. Notice that only the field parts $\phi$,
$\tilde{\pi}^{ij}$, and $\breve{\pi}^{ij}$ are expanded, but not
$h^{\text{TT}}_{ij}$ and $\pi^{ij\text{TT}}$. The latter are still
dynamical variables in the ADM Hamiltonian and can be expanded only after
their equations of motion were obtained and solved.
For the Hamilton constraint we get
\begin{align}
- \frac{1}{16\pi} \Delta \phi_{(2)} &= \mathcal{H}^{\text{matter}}_{(2)} \,, \qquad
- \frac{1}{16\pi} \Delta \phi_{(4)} = \mathcal{H}^{\text{matter}}_{(4)}
	- \frac{1}{8} \mathcal{H}^{\text{matter}}_{(2)} \phi_{(2)} \,, \label{Hconst2} \\
\begin{split}
- \frac{1}{16\pi} \Delta \phi_{(6)} &= \mathcal{H}^{\text{matter}}_{(6)}
	- \frac{1}{8} \left( \mathcal{H}^{\text{matter}}_{(4)} \phi_{(2)}
		+ \mathcal{H}^{\text{matter}}_{(2)} \phi_{(4)} \right)
	+ \frac{1}{64} \mathcal{H}^{\text{matter}}_{(2)} \phi_{(2)}^2 \nl
	+ \frac{1}{16\pi} \left[ \left( \tilde{\pi}^{i j}_{(3)} \right)^2
	- \frac{1}{2} \left( \phi_{(2)} h^{\text{TT}}_{i j} \right)_{, i j} \right] \,, \label{Hconst6}
\end{split}\\
\begin{split}
- \frac{1}{16\pi} \Delta \phi_{(8)} &=
\frac{1}{16\pi}\left[ \frac{1}{8} \phi_{(2)} \left(\tilde{\pi}^{i j}_{(3)}\right)^2
	+ 2 \tilde{\pi}^{i j}_{(3)} \tilde{\pi}^{i j}_{(5)}
	- \frac{1}{16} \phi_{(2) , i} \phi_{(2) , j} h^{\text{TT}}_{i j}
	+ \frac{1}{4} \left(h^{\text{TT}}_{i j , k}\right)^2 \right] \nl
+ \mathcal{H}^{\text{matter}}_{(8)}
- \frac{1}{8} \left( \mathcal{H}^{\text{matter}}_{(6)} \phi_{(2)}
		+ \mathcal{H}^{\text{matter}}_{(4)} \phi_{(4)}
		+ \mathcal{H}^{\text{matter}}_{(2)} \phi_{(6)} \right) \nl
+ \frac{1}{64} \left( \mathcal{H}^{\text{matter}}_{(4)} \phi_{(2)}^2
		+ 2 \mathcal{H}^{\text{matter}}_{(2)} \phi_{(2)} \phi_{(4)} \right)
- \frac{1}{512} \mathcal{H}^{\text{matter}}_{(2)} \phi_{(2)}^3
+ (\text{td}) \,, \label{Hconst8}
\end{split}
\end{align}
up to and including the formal second post-Newtonian order.
These equations can be solved iteratively for $\phi$
by applying an inverse Laplacian to them. The ADM Hamiltonian (\ref{HADM})
results from an integration over the right-hand sides of these equations.
It was used that $\breve{\pi}^{ij} = \Order{(c^{-9})}$
at linear order in spin.

However, we also have to solve the momentum constraint as $\pi^{ij}$ appears
on the right-hand side of the Hamilton constraint.
The expansion of the momentum constraint immediately follows from the exact formula
\begin{equation}
\tilde{\pi}^{ij}_{~~,j} =
	- 8\pi \mathcal{H}^{\text{matter}}_{i}
	+ {B^{ij}}_{,j}
	+ C^i
	- \Delta \left( V^k h_{ki}^{\text{TT}} \right)
	+ \frac{1}{2} \pi^{jk\text{TT}} h_{jk,i}^{\text{TT}}
	- ( \pi^{jk\text{TT}} h_{ki}^{\text{TT}} )_{,j} \,,
\end{equation}
with
\begin{gather}
B^{ij} = \left[ 1 - \left( 1 + \tfrac{1}{8} \phi \right)^4 \right]
		( \tilde{\pi}^{ij} + \pi^{ij\text{TT}} )
	+ V^k ( h_{ki,j}^{\text{TT}} + h_{kj,i}^{\text{TT}} - h_{ij,k}^{\text{TT}} )
	- \frac{1}{3} {V^k}_{,k} h_{ij}^{\text{TT}} \,, \label{defBij} \\
C^i = \frac{1}{2} \breve{\pi}^{jk} \gamma_{jk,i} - \breve{\pi}^{jk} \gamma_{ij,k} \,, \label{defCi}
\end{gather}
which is analogous to (\ref{cmom}).
Here we introduced the alternative vector potential
\begin{equation}
V^i = \left( \delta_{ij} - \frac{1}{4} \partial_i \partial_j \Delta^{-1} \right) \tilde{\pi}^j \,, \label{Vtrans}
\end{equation}
for which it holds
\begin{equation}\label{pidecompV}
\tilde{\pi}^{ij} = {V^i}_{, j} + {V^j}_{, i} - \frac{2}{3} \delta_{ij} {V^k}_{, k} \,.
\end{equation}
To the considered order we thus have
\begin{equation}\label{momCexp}
	\tilde{\pi}^{i j}_{(3) , j} =
		- 8\pi \mathcal{H}^{\text{matter}}_{(3) i} \,, \qquad
	\tilde{\pi}^{i j}_{(5) , j} =
		- 8\pi \mathcal{H}^{\text{matter}}_{(5) i}
		- \frac{1}{2} \left( \phi_{(2)} \tilde{\pi}^{i j}_{(3)} \right)_{, j} \,.
\end{equation}
With the help of $\tilde{\pi}^i = \Delta^{-1} \tilde{\pi}^{ij}_{~~,j} \,,$
the expanded momentum constraint can be solved iteratively for
$\tilde{\pi}^i$ by applying an inverse Laplacian to it.
$\tilde{\pi}^{ij}$ and $V^i$ then follow from (\ref{pisolve}) and (\ref{Vtrans}).

\mysubsection{Formulas for Hamiltonians\label{matHam}}
The first contribution to the ADM Hamiltonian (\ref{HADM}) results from an
integration over the first relation in (\ref{Hconst2}) as
\begin{equation}
H_{0} = \int \dd^3 x \, \mathcal{H}^{\text{matter}}_{(2)} \,.
\end{equation}
Notice that $\mathcal{H}^{\text{matter}}_{(2)}$ is just the Newtonian mass
density, so $H_{0}$ is the constant energy belonging
to the total Newtonian mass. Similarly, from the second relation in (\ref{Hconst2}) we obtain the
Newtonian Hamiltonian
\begin{equation}
H_{\text{N}} = \int \dd^3 x \, \left[ \mathcal{H}^{\text{matter}}_{(4)}
	- \frac{1}{8} \phi_{(2)} \mathcal{H}^{\text{matter}}_{(2)} \right] \,.
\end{equation}
$\phi_{(2)}$ results from (\ref{Hconst2}) as
\begin{equation}\label{phi2}
\phi_{(2)} = - 16\pi \Delta^{-1} \mathcal{H}^{\text{matter}}_{(2)} \,,
\end{equation}
and agrees up to a factor with the Newtonian gravitational potential
of the mass distribution $\mathcal{H}^{\text{matter}}_{(2)}$.
$\mathcal{H}^{\text{matter}}_{(4)}$ is the Newtonian kinetic energy density.

Next we proceed to the Hamiltonian at the first post-Newtonian order.
However, we first apply the partial integration formulas
\begin{equation}
\begin{split}
\frac{1}{8} \phi_{(4)} \mathcal{H}^{\text{matter}}_{(2)}
	&= \frac{1}{8} \left( \mathcal{H}^{\text{matter}}_{(4)}
		- \frac{1}{8} \phi_{(2)} \mathcal{H}^{\text{matter}}_{(2)} \right) \phi_{(2)} + (\text{td}) \,, \\
\frac{1}{16\pi} \left( \tilde{\pi}^{ij}_{(3)} \right)^2
	&= V^i_{(3)} \mathcal{H}^{\text{matter}}_{(3) i} + (\text{td}) \,,
\end{split}
\end{equation}
to the right-hand side of (\ref{Hconst6}), following
from the formal solution
\begin{equation}\label{phi4}
\phi_{(4)} = - 16\pi \Delta^{-1} \left[ \mathcal{H}^{\text{matter}}_{(4)}
	- \frac{1}{8} \phi_{(2)} \mathcal{H}^{\text{matter}}_{(2)} \right] \,,
\end{equation}
of the Hamilton constraint and from (\ref{pidecompV}, \ref{momCexp}).
$V^i_{(3)}$ is determined by (\ref{Vtrans}) and
\begin{equation}\label{pi3}
\tilde{\pi}_{(3)}^i = - 8\pi \Delta^{-1} \mathcal{H}^{\text{matter}}_{(3) i} \,.
\end{equation}
Finally we get for the first post-Newtonian (PN) order Hamiltonian
\begin{equation}\label{H1PN}
H_{\text{1PN}} = \int \dd^3 x \, \left[ \mathcal{H}^{\text{matter}}_{(6)}
	- \frac{1}{4} \phi_{(2)} \mathcal{H}^{\text{matter}}_{(4)}
	+ \frac{1}{32} \phi_{(2)}^2 \mathcal{H}^{\text{matter}}_{(2)}
	+ V^i_{(3)} \mathcal{H}^{\text{matter}}_{(3) i} \right] \,.
\end{equation}
Notice that all terms in the Hamiltonian involve matter source terms $\mathcal{H}^{\text{matter}}$
or $\mathcal{H}^{\text{matter}}_i$ and are thus integrations over
delta distributions only.
Further only the Newtonian potential $\phi_{(2)}$ and the leading order vector
potential $V^i_{(3)}$ need to be determined (lapse $N$ and shift $N^i$ are
not even needed at any higher order).
This shows the efficiency of the ADM formalism
in calculating the conservative post-Newtonian dynamics.

In the same way one can obtain a formula for the second post-Newtonian Hamiltonian
\begin{equation}
\begin{split}\label{H2PN}
H_{\text{2PN}}^{\text{ADM}} &= \int \dd^3 x \, \bigg[
	\mathcal{H}^{\text{matter}}_{(8)}
	- \frac{1}{8} \left( 2 \phi_{(2)} \mathcal{H}^{\text{matter}}_{(6)}
		+ \phi_{(4)} \mathcal{H}^{\text{matter}}_{(4)} \right)
	- \frac{1}{256} \phi_{(2)}^3 \mathcal{H}^{\text{matter}}_{(2)} \nl
	+ \frac{1}{64} \left( 2 \phi_{(2)}^2 \mathcal{H}^{\text{matter}}_{(4)}
		+ 3 \phi_{(2)} \phi_{(4)} \mathcal{H}^{\text{matter}}_{(2)} \right)
	+ 2 V^i_{(3)} \mathcal{H}^{\text{matter}}_{(5) i} \nl
	+ \frac{1}{16\pi} \bigg( - \phi_{(2)} \left( \tilde{\pi}_{(3)}^{ij} \right)^2
		- \frac{1}{8} \phi_{(2),i} \phi_{(2),j} h_{ij}^{\text{TT}}
		+ \frac{1}{4} \left( h_{ij,k}^{\text{TT}} \right)^2 \bigg)
\bigg] \,,
\end{split}
\end{equation}
where the partial integrations
\begin{align}
\begin{split}
\phi_{(6)} \mathcal{H}^{\text{matter}}_{(2)} &=
	\phi_{(2)} \mathcal{H}^{\text{matter}}_{(6)}
	- \frac{1}{8} \left( \phi_{(2)}^2 \mathcal{H}^{\text{matter}}_{(4)}
		+ \phi_{(2)} \phi_{(4)} \mathcal{H}^{\text{matter}}_{(2)} \right)
	+ \frac{1}{64} \phi_{(2)}^3 \mathcal{H}^{\text{matter}}_{(2)} \nl
	+ \frac{1}{16\pi} \left[ \phi_{(2)} \left( \tilde{\pi}^{i j}_{(3)} \right)^2
	+ \frac{1}{2} \phi_{(2),i} \phi_{(2),j} h^{\text{TT}}_{i j} \right]
	+ (\text{td}) \,,
\end{split}\\
\tilde{\pi}^{ij}_{(3)} \tilde{\pi}^{ij}_{(5)} &=
	16 \pi V^i_{(3)} \mathcal{H}^{\text{matter}}_{(5) i}
	- \frac{1}{2} \phi_{(2)} \left( \tilde{\pi}_{(3)}^{ij} \right)^2
	+ (\text{td}) \,,
\end{align}
were used. Notice that $\phi_{(6)}$ and $\tilde{\pi}^{ij}_{(5)}$ were eliminated
from the Hamiltonian by these partial integrations. Therefore no
solutions to the constraints besides (\ref{phi2}), (\ref{phi4}),
and (\ref{pi3}) have to be determined explicitly.
The Hamiltonian $H_{\text{2PN}}^{\text{ADM}}$ has the additional label ADM as it still
depends on the dynamical field variable $h_{ij}^{\text{TT}}$. The elimination of $h_{ij}^{\text{TT}}$
from $H_{\text{2PN}}^{\text{ADM}}$ leads to the matter-only Hamiltonian
$H_{\text{2PN}}$ and is discussed in the next {\secname}.
Further $\pi^{ij\text{TT}}$ first appears at the formal third
post-Newtonian level.

Notice that the obtained formulas are valid for quite general source
expressions $\mathcal{H}^{\text{matter}}$ and $\mathcal{H}^{\text{matter}}_i$,
not only to the ones linear in spin.

\mysubsection{Matter-Only Hamiltonian\label{matteronly}}
In the last {\secname} the ADM Hamiltonian $H_{\text{ADM}}$ was expanded as
\begin{equation}
H_{\text{ADM}} = H_0 + H_{\text{N}} + H_{\text{1PN}} + H_{\text{2PN}}^{\text{ADM}} + \cdots \,.
\end{equation}
The conservative matter-only Hamiltonian results from plugging the solution
for $h^{\text{TT}}_{i j}$ and $\hat{\pi}^{ij\text{TT}}$ into the \emph{action},
\Eq{ADMspinaction} (and a subsequent elimination of emerging higher order time
derivatives of the matter variables), see \cite{Jaranowski:Schafer:1998}.
As $\hat{\pi}^{ij\text{TT}}$ is neglected at the considered order, the first term in (\ref{ADMspinaction})
does not contribute here. Therefore $H_{\text{2PN}}^{\text{ADM}}$ turns into the
matter-only Hamiltonian $H_{\text{2PN}}$ by simply inserting the solution for $h^{\text{TT}}_{i j}$
into $H_{\text{2PN}}^{\text{ADM}}$.

The field evolution can be obtained from the ADM Hamiltonian by
\begin{equation}
\begin{split}
\frac{\partial h^{\text{TT}}_{i j}}{\partial t} &= \{ h^{\text{TT}}_{i j} , H_{\text{ADM}} \}
	= 16\pi \delta^{\text{TT} i j}_{k l} \frac{\delta H_{\text{ADM}}}{\delta \hat{\pi}^{k l \text{TT}}} \,, \\
\frac{\partial \hat{\pi}^{i j \text{TT}}}{\partial t} &= \{ \hat{\pi}^{i j \text{TT}} , H_{\text{ADM}} \}
	= - 16\pi \delta^{\text{TT} i j}_{k l} \frac{\delta H_{\text{ADM}}}{\delta h^{\text{TT}}_{k l}} \,. \label{wave}
\end{split}
\end{equation}
However, as the $\hat{\pi}^{i j \text{TT}}$-contributions are of higher
order here, we formally just have
\begin{equation}
0 = \delta^{\text{TT} i j}_{k l} \frac{\delta H_{\text{2PN}}^{\text{ADM}}}{\delta h^{\text{TT}}_{k l}} \,,
\end{equation}
or explicitly, given that $\mathcal{H}^{\rm matter}_{(8)}$ has contributions linear in $h^{\text{TT}}_{ij}$,
\begin{equation} \label{htt4}
\Delta h^{\text{TT}}_{i j} = 2 \delta^{\text{TT} k l}_{i j} f_{(4)kl} \,, \qquad \text{with} \;
f_{(4)ij} = 16\pi \frac{\delta \left( \int{ \dd^3 x \, \mathcal{H}^{\rm matter}_{(8)} } \right)}{\delta h^{\text{TT}}_{ij}}
		- \frac{1}{8} \phi_{(2),i} \phi_{(2),j} \,.
\end{equation}
Using the formal solution $h^{\text{TT}}_{i j} = 2 \delta^{\text{TT} k l}_{i j} \Delta^{-1} f_{(4)kl}$,
\emph{all} contributions of $h^{\text{TT}}_{i j}$ to $H_{\text{2PN}}$ can be collected as
\begin{equation}\label{httpart}
+ \frac{1}{16\pi} \int \dd^3 x \, \frac{1}{4} h^{\text{TT}}_{i j} \Delta h^{\text{TT}}_{i j}
	= + \frac{1}{16\pi} \int \dd^3 x \, \frac{1}{2} h^{\text{TT}}_{i j} f_{(4) i j} \,.
\end{equation}
If one is interested in the spin contribution of this integral only, one can
obviously perform a partial integration in (\ref{httpart}) in such a way
that only the spin part of $h^{\text{TT}}_{i j}$ is needed, see also \cite{Steinhoff:Schafer:Hergt:2008}.
This is desirable as the spin-dependent part of $h^{\text{TT}}_{i j}$ is much simpler than
the spin-independent part.

Though the discussion of $h^{\text{TT}}_{i j}$ was straightforward here, it is
quite subtle to obtain the post-Newtonian expansion of (\ref{wave}) at higher orders.
Indeed, it is not easy to correctly implement the boundary conditions into
the solution of the first order equations (\ref{wave}). At higher
orders (\ref{wave}) can be converted into a (second order) wave equation for $h^{\text{TT}}_{i j}$,
with source terms expanded according to the post-Newtonian counting rules.
In \cite{Jaranowski:Schafer:1997,Jaranowski:Schafer:1998} this wave equation
is then solved order by order using a near zone expansion of the retarded solution
up to the 3.5 post-Newtonian order,
corresponding to the boundary condition of no incoming gravitational waves;
see also, e.g., \cite{Blanchet:2006} for other aspects like tails.
Equation (\ref{htt4}) is indeed the leading order near zone approximation
of the wave equation for $h^{\text{TT}}_{i j}$. The solution for
$h^{\text{TT}}_{i j}$ at higher orders is responsible for the
half post-Newtonian orders in the matter-only Hamiltonian, starting
at the 2.5 post-Newtonian order.

\mysection{Spin Corrections to the Hamiltonian\label{spinham}}
By now there are a lot of results regarding spin effects at the conservative
orders in the post-Newtonian approximation. The main goal of this {\secname}
is to derive the next-to-leading order spin effects within the developed
formalism, which were tackled only recently.
Even higher post-Newtonian orders linear in spin were derived recently in
\cite{Barausse:Racine:Buonanno:2009} for test spinning objects in
the Kerr metric.
Also Hamiltonians of cubic and higher order in spin were
obtained for binary black holes \cite{Hergt:Schafer:2008:2,Hergt:Schafer:2008,Barausse:Racine:Buonanno:2009}.
The calculation of the leading order dissipative spin-orbit and spin(1)-spin(2)
Hamiltonians was prepared in \cite{Steinhoff:Wang:2009}. The corresponding
equations of motion were already obtained \cite{Will:2005,*Wang:Will:2007,*Zeng:Will:2007};
see also the considerations in terms of orbital elements in \cite{Gergely:Perjes:Vasuth:1998,*Gergely:1999,*Gergely:2000}.

More work needs to be done for an application of the Hamiltonians derived in this {\secname}
to gravitational wave astronomy. In particular the spin contributions to the
next-to-leading order radiation field are only known for the spin-orbit case
\cite{Blanchet:Buonanno:Faye:2006,*Blanchet:Buonanno:Faye:2006:err},
but not yet for the spin(1)-spin(2) and spin(1)-spin(1) cases
(for the latter case the stress-energy tensor derived in \Chap{higher} is needed).
Further, it would be useful to find a parametrization of the orbits by solving
the equations of motion, i.e., extending the solutions from
\cite{Konigsdorffer:Gopakumar:2005,Tessmer:2009} at least to some
of the new Hamiltonians.
Finally, one should consider to incorporate the new Hamiltonians
into the very successful effective one-body approach \cite{Damour:Nagar:2009:2,*Buonanno:etal:2009,*Damour:Nagar:2009:3},
which already succeeded for the leading order spin Hamiltonians \cite{Damour:2001} as well as
for the next-to-leading order spin-orbit Hamiltonian \cite{Damour:Jaranowski:Schafer:2008:3,*Pan:etal:2009,*Barausse:Buonanno:2009}.

Formulas and regularization procedures for the integrals that
need to be solved in this {\secname} are given in, e.g., \cite{Jaranowski:1997,Jaranowski:Schafer:1997,Jaranowski:Schafer:1998}.
Some parts of the calculations were also checked using Riesz
kernels in arbitrary dimension, see, e.g., \cite{Damour:Jaranowski:Schafer:2008:2}.

\mysubsection{Field Constraints in Canonical Variables\label{Hmattercan}}
Before starting the calculation of the Hamiltonians, it is suitable to
express the source terms of the constraints $\mathcal{H}^{\text{matter}}$ and
$\mathcal{H}^{\text{matter}}_i$ in terms of the canonical matter variables.
Then the formulas provided in \Sec{matHam} automatically give the
Hamiltonian (the redefinition of $\pi^{ij\text{TT}}$ is not necessary here).
Applying the variable redefinitions from \Sec{vartrans} to (\ref{Hcov}, \ref{Hicov})
leads to
\begin{gather}
\begin{split}
\mathcal{H}^{\text{matter}} &= \sum_a \bigg[ - n\hat{p}_a \hat{\delta}_a
	- \frac{1}{2} \bigg( \frac{\hat{S}_{ali} \hat{p}_{aj}}{n\hat{p}_a}
		+ \gamma^{mn} \frac{\hat{S}_{ami} \hat{p}_{aj} \hat{p}_{an} \hat{p}_{al}}{(n\hat{p}_a)^2(m_a-n\hat{p}_a)}
	\bigg) \gamma^{kl} \gamma^{ij}_{~~,k} \hat{\delta}_a \nlq
	+ \frac{\hat{p}_{aj} \gamma^{ji}}{n\hat{p}_a} \hat{A}^{kl}_a e_{(m)k} {e^{(m)}}_{l,i} \hat{\delta}_a
	- \bigg( \frac{\hat{p}_{al}}{m_a-n\hat{p}_a}\gamma^{ij}\gamma^{kl} \hat{S}_{ajk} \hat{\delta}_a \bigg)_{,i} \, \bigg] \,, \label{cansource1}
\end{split}\\
\mathcal{H}^{\text{matter}}_i = \sum_a \bigg[ \hat{p}_{ai} \hat{\delta}_a
	- \hat{A}^{kl}_a e_{(m)k} {e^{(m)}}_{l,i} \hat{\delta}_a
	+ \frac{1}{2} \left( s_a^{ij} \hat{\delta}_a \right)_{,j} \bigg] \,, \label{Hicans}
\end{gather}
where
\begin{equation}
s_a^{ij} = \gamma^{jk} \hat{S}_{aik}
	+ \gamma^{jk} \gamma^{lp} \frac{2 \hat{p}_{al} \hat{p}_{a(i} \hat{S}_{ak)p}}{n\hat{p}_a (m_a - n\hat{p}_a)} \,, \label{cansource3}
\end{equation}
and $\hat{A}^{kl}$ given by (\ref{defAhat}). These source expressions are valid in general,
also within the test-spin Hamiltonian (\ref{Htest}). In the spatial symmetric gauge
it holds
% \begin{equation}
$\hat{A}^{kl}_a e_{(m)k} {e^{(m)}}_{l,\mu} = \pi^{kl}_a \gamma_{kl,\mu}$,
% \end{equation}
where (\ref{eder}) and (\ref{defpiaA}) were used and $\pi^{kl}_a$ is given by (\ref{defpia}).
Notice that the variable redefinitions from the action approach leading to these expressions have
been checked up to and including the formal 3.5 post-Newtonian order by the
symmetry generator approach \cite{Steinhoff:Wang:2009}. This includes the
formal third post-Newtonian or next-to-next-to-leading order linear in spin,
which for maximal spin is at the 3.5 post-Newtonian order in the spin-orbit case
and at the fourth post-Newtonian order in the spin(1)-spin(2) case.
It was shown in \cite{Steinhoff:Wang:2009} as a further check to the same level of
approximation that the wave equation for $h^{\text{TT}}_{i j}$ following
from the ADM Hamiltonian agrees with the Einstein equations, which
again verifies that the used variables are canonical.

The expansion of $\mathcal{H}^{\text{matter}}$ sufficient for the formal second post-Newtonian Hamiltonian reads
\begin{align}
\mathcal{H}^{\text{matter}}_{(2)} &= \sum_a m_a \hat{\delta}_a \,, \qquad
\mathcal{H}^{\text{matter}}_{(4)} = \sum_a \left[ \frac{\hat{\vct{p}}^2_a}{2 m_a} \hat{\delta}_a
	+ \frac{1}{2 m_a} \hat{p}_{a i} \hat{S}_{a (i) (j)} \hat{\delta}_{a , j} \right] \,, \\
\begin{split}
\mathcal{H}^{\text{matter}}_{(6)} &= \sum_a \bigg[
	- \frac{(\hat{\vct{p}}^2_a)^2}{8 m_a^3} \hat{\delta}_a
	- \frac{\hat{\vct{p}}^2_a}{4 m_a} \phi_{(2)} \hat{\delta}_a
	+ \frac{1}{4 m_a} \hat{p}_{a i} \hat{S}_{a (i) (j)} \phi_{(2) , j} \hat{\delta}_a \nlq
	- \frac{\hat{\vct{p}}^2_a}{8 m^3_a} \hat{p}_{a i} \hat{S}_{a (i) (j)} \hat{\delta}_{a , j}
	- \frac{1}{4 m_a} \hat{p}_{a i} \hat{S}_{a (i) (j)} ( \phi_{(2)} \hat{\delta}_a )_{, j} \bigg] \,,
\end{split}\\
\begin{split}
\mathcal{H}^{\text{matter}}_{(8)} &= \sum_a \bigg[
	\frac{(\hat{\vct{p}}^2_a)^3}{16 m^5_a} \hat{\delta}_a
	+ \frac{(\hat{\vct{p}}^2_a)^2}{8 m^3_a} \phi_{(2)} \hat{\delta}_a
	+ \frac{5 \hat{\vct{p}}^2_a}{64 m_a} \phi_{(2)}^2 \hat{\delta}_a
	- \frac{\hat{\vct{p}}^2_a}{4 m_a} \phi_{(4)} \hat{\delta}_a
	- \frac{1}{2 m_a} \hat{p}_{a i} \hat{p}_{a j} h^{\text{TT}}_{i j} \hat{\delta}_a \nlq
	- \frac{\hat{\vct{p}}_a^2}{8 m_a^3} \hat{p}_{a i} \hat{S}_{a (i) (j)} \phi_{(2) , j} \hat{\delta}_a
	- \frac{5}{32 m_a} \hat{p}_{a i} \hat{S}_{a (i) (j)} \phi_{(2)} \phi_{(2) , j} \hat{\delta}_a \nlq
	+ \frac{1}{4 m_a} \hat{p}_{a i} \hat{S}_{a (i) (j)} \phi_{(4) , j} \hat{\delta}_a
	+ \frac{1}{2 m_a} \hat{p}_{a i} \hat{S}_{a (j) (k)} h^{\text{TT}}_{i j , k} \hat{\delta}_a \bigg]
	+ (\text{td}) \,,
\end{split}
\end{align}
where $\hat{\vct{p}}_a = (\hat{p}_{ai})$.
The expansion of the source $\mathcal{H}^{\text{matter}}_{i}$ is given by (\ref{Hicans}) and
\begin{equation}
	s_{a(3)}^{ij} = \hat{S}_{a(i)(j)} \,, \qquad
	s^{ij}_{a (5)} = - \frac{1}{2 m_a^2} \hat{p}_{a k} ( \hat{p}_{a i} \hat{S}_{a(j)(k)} + \hat{p}_{a j} \hat{S}_{a(i)(k)} ) \,.
\end{equation}
Notice that the triad terms in (\ref{Hicans}) do not contribute at the
considered order.

The expansion of the static source terms needed at the spin(1)-spin(1)
order follow from (\ref{HS2cov}) as
\begin{align}
\mathcal{H}^{\rm matter}_{(4) \, \text{S}^2 \!, \, \hat{p}_i=0} &= \sum_a
	\frac{1}{2} \hat{Q}_{a(i)(j)} \hat{\delta}_{a , ij} \,, \label{H2exp1} \\
\begin{split}\label{H2exp2}
\mathcal{H}^{\rm matter}_{(6) \, \text{S}^2 \!, \, \hat{p}_i=0} &= \sum_a \bigg[
	\frac{1}{4} \hat{Q}_{a(i)(j)} ( \phi_{(2) , i} \hat{\delta}_a )_{, j}
	- \frac{1}{4} \hat{Q}_{a(i)(j)} ( \phi_{(2)} \hat{\delta}_a )_{, ij} \nlq
	+ \frac{1}{8 m_a} \hat{S}_{a(i)(k)} \hat{S}_{a(j)(k)} ( \phi_{(2) , i} \hat{\delta}_a )_{, j} \bigg] \,,
\end{split}\\
\begin{split}\label{H2exp3}
\mathcal{H}^{\rm matter}_{(8) \, \text{S}^2 \!, \, \hat{p}_i=0} &= - \sum_a
	\frac{1}{32 m_a} \hat{S}_{a(i)(k)} \hat{S}_{a(k)(j)} \phi_{(2),i} \phi_{(2),j} \hat{\delta}_a
	+ (\text{td}) \,,
\end{split}
\end{align}
and it holds
\begin{equation}
% I_{a(i)(j)} = - \frac{m_a}{C_{Qa}} \hat{Q}_{a(i)(j)}
% 	= \hat{S}_{a(i)(k)} \hat{S}_{a(k)(j)} + \frac{2}{3} \delta_{ij} \hat{\vct{S}}_a^2 \,.
\hat{Q}_{a(i)(j)} = \frac{C_{Qa}}{m_a}
	\left( \hat{S}_{a(i)(k)} \hat{S}_{a(j)(k)} - \frac{2}{3} \delta_{ij} \hat{\vct{S}}_a^2 \right) \,.
\end{equation}
Here $\hat{\vct{S}}_a = (\hat{S}_{a(i)})$ and $\hat{S}_{a(i)} = \frac{1}{2} \epsilon_{ijk} \hat{S}_{a(j)(k)}$.
No further contributions to $\mathcal{H}^{\text{matter}}_i$ arise
in the spin(1)-spin(1) case.

\mysubsection{Leading Order\label{HLO}}
The leading order spin effects are at the formal first post-Newtonian
order and their Hamiltonian can be obtained from (\ref{H1PN}),
which of course gives the first post-Newtonian Hamiltonian in the nonspinning case.
The needed solutions of the constraints read
\begin{gather}
\phi_{(2)} = 4 \sum_a \frac{m_a}{\hat{r}_a} \,, \qquad
\tilde{\pi}^i_{(3)} = \sum_a \left[ 2 \frac{\hat{p}_{a i}}{\hat{r}_a}
	+ \hat{S}_{a (i) (j)} \left( \frac{1}{\hat{r}_a} \right)_{,j} \right] \,, \\
% \Rightarrow \quad
V^i_{(3)} = \sum_a \left[ 2 \frac{\hat{p}_{a i}}{\hat{r}_a} - \frac{1}{4} \hat{p}_{a j} \hat{r}_{a,ij}
	+ \hat{S}_{a (i) (j)} \left( \frac{1}{\hat{r}_a} \right)_{,j} \right] \,,
\end{gather}
where $\hat{r}_a = | \vct{x} - \hat{\vct{z}}_a |$ and $\hat{\vct{z}}_a = (\hat{z}_a^i)$.
The leading order (LO) spin-orbit (SO) Hamiltonian follows as
\begin{equation}
	H_{\text{SO}}^{\text{LO}} =
 		\sum_a \sum_{b \neq a}
		\frac{1}{\hat{r}_{a b}^2} (\hat{\mathbf{S}}_a \times \hat{\mathbf{n}}_{ab})
			\cdot \left[ \frac{3 m_b}{2 m_a} \hat{\mathbf{p}}_a - 2 \hat{\mathbf{p}}_b \right] \,,
\end{equation}
where $\hat{r}_{ab} = | \hat{\vct{z}}_a - \hat{\vct{z}}_b |$ and
$\vct{\hat{n}}_{ab} = ( \hat{\vct{z}}_a - \hat{\vct{z}}_b ) / \hat{r}_{ab}$.
This Hamiltonian is at the 1.5 post-Newtonian order for maximal spins.
Further, the leading order spin($a$)-spin($b$), or S$_a$S$_b$, Hamiltonian results as
\begin{equation}
	H_{\text{S}_a\text{S}_b}^{\text{LO}} =
		\sum_a \sum_{b \neq a}
			\frac{1}{2 \hat{r}_{a b}^3}
			\left[ 3 (\hat{\mathbf{S}}_a \cdot \hat{\mathbf{n}}_{ab})
				(\hat{\mathbf{S}}_b \cdot \hat{\mathbf{n}}_{ab})
			- (\hat{\mathbf{S}}_a \cdot \hat{\mathbf{S}}_b) \right] \,.
\end{equation}
For maximal spins this Hamiltonian is at the second post-Newtonian level.
Finally, the leading order spin($a$)-spin($a$), or S$_a^2$, Hamiltonian is given by
\begin{equation}
	H_{\text{S}_a^2}^{\text{LO}} =
		\sum_a \sum_{b \neq a}
			\frac{C_{Qa} m_b}{2 m_a \hat{r}_{a b}^3}
			\left[ 3 (\hat{\mathbf{S}}_a \cdot \hat{\mathbf{n}}_{ab})^2
			- \hat{\mathbf{S}}_a^2 \right] \,,
\end{equation}
which is also at the second post-Newtonian order for maximal spins.
All Hamiltonians in this {\secname} are valid for arbitrary many spinning objects.
The Poisson brackets are the standard canonical ones, i.e.,
\begin{equation}
\{ \hat{z}^i_a, \hat{p}_{a j} \} = \delta_{ij} \,, \qquad
\{ \hat{S}_{a (i)}, \hat{S}_{a (j)} \} = \epsilon_{ijk} \hat{S}_{a (k)} \,,
\end{equation}
zero otherwise.

The leading order spin effects derived here are well-known for black holes ($C_Q=1$), see, e.g.,
\cite{Barker:OConnell:1975,*DEath:1975,*Thorne:Hartle:1985,Barker:OConnell:1979}.
For the leading order $C_Q$-dependence see \cite{Barker:OConnell:1979,Poisson:1998}.

\mysubsection{Next-to-Leading Order\label{HNLO}}
Now we proceed to the formal second post-Newtonian Hamiltonian (\ref{H2PN}),
which includes the next-to-leading order spin effects.
There we also need the functions
\begin{align}
\phi_{(4)} &= \sum_a \bigg[ \frac{2 \hat{\vct{p}}_a^2}{m_a \hat{r}_a}
	- \sum_{b \neq a} \frac{2 m_a m_b}{\hat{r}_{ab} \hat{r}_a}
	+ \frac{2 \hat{p}_{a i} \hat{S}_{a (i) (j)}}{m_a} \left( \frac{1}{\hat{r}_a} \right)_{,j}
	+ 2 \hat{Q}_{a(i)(j)} \left( \frac{1}{\hat{r}_a} \right)_{,ij}
\bigg] \,, \label{phi4sol} \\
\begin{split}\label{pi3sol}
\tilde{\pi}^{ij}_{(3)} &= \sum_a \bigg[
	2 \hat{p}_{a i} \left( \frac{1}{\hat{r}_a} \right)_{,j}
	+ 2 \hat{p}_{a j} \left( \frac{1}{\hat{r}_a} \right)_{,i}
	- \delta_{ij} \hat{p}_{a k} \left( \frac{1}{\hat{r}_a} \right)_{,k}
	- \frac{1}{2} \hat{p}_{a k} \hat{r}_{a,ijk} \nlq
	- \hat{S}_{a (k) (i)} \left( \frac{1}{\hat{r}_a} \right)_{, k j}
	- \hat{S}_{a (k) (j)} \left( \frac{1}{\hat{r}_a} \right)_{, k i} \bigg] \,.
\end{split}
\end{align}
Notice that there are $C_{Qa}$-contributions in (\ref{phi4sol}).
We restrict to two spinning objects in this {\secname}.
The results provided here complete the knowledge of spin corrections to the Hamiltonian
up to and including the third post-Newtonian order for
maximal spins.

\mysubsubsection{Next-to-Leading Order Spin-Orbit}
Following the method developed here, the next-to-leading order (NLO)
spin-orbit Hamiltonian results as \cite{Steinhoff:Schafer:Hergt:2008}
\begin{align}
% \begin{split}
	H_{\text{SO}}^{\text{NLO}} &=
		- \frac{\picSin}{\hat{r}_{12}^3} \left[
			  \frac{11 m_2}{2} + \frac{5 m_2^2}{m_1}
		\right] %\nonumber \nl
		+ \frac{\piicSin}{\hat{r}_{1 2}^3} \left[
			  6 m_1 + \frac{15 m_2}{2}
		\right] \nonumber \nl
		- \frac{\picSin}{\hat{r}_{1 2}^2}
			\Bigg[ \frac{5 m_2 \pipi}{8 m_1^3}
			+ \frac{3 \pipii}{4 m_1^2}
			- \frac{3 \piipii}{4 m_1 m_2}
			+ \frac{3 \pin \piin}{4 m_1^2} \nonumber \nl
			+ \frac{3 \piin^2 }{2 m_1 m_2} \Bigg]
		+ \frac{\piicSin}{\hat{r}_{12}^2}
			\left[ \frac{\pipii}{m_1 m_2} + \frac{3 \pin \piin }{m_1 m_2} \right] \nonumber\nl
		+ \frac{\picSipii}{\hat{r}_{12}^2}
			\left[ \frac{2 \piin}{m_1 m_2} - \frac{3 \pin}{4 m_1^2} \right]
	+ (1 \leftrightarrow 2) \,, \label{HSO}
% \end{split}
\end{align}
where $(1 \leftrightarrow 2)$ indicates an exchange of
particle labels, and is identical to the one derived earlier in \cite{Damour:Jaranowski:Schafer:2008:1}.
The next-to-leading order spin-orbit case was first tackled on the level of the
equations of motion in \cite{Tagoshi:Ohashi:Owen:2001} and
was later rederived and improved in \cite{Faye:Blanchet:Buonanno:2006}
(both in the harmonic gauge). Within the ADM canonical formalism
the Hamiltonian $H_{\text{SO}}^{\text{NLO}}$ corresponding to these equations
of motion was obtained in \cite{Damour:Jaranowski:Schafer:2008:1}
from the spin equation of motion (\ref{eom}).
The linear-in-$G$ part of $H_{\text{SO}}^{\text{NLO}}$ was also
derived in \cite{Hergt:Schafer:2008:2} from corresponding source
terms of the constraints, similar to the approach used here
(however, in \cite{Hergt:Schafer:2008:2} the source terms were
obtained from the approximate Kerr metric in the ADM transverse traceless gauge).
Very recently derivations within the effective field theory approach
also succeeded \cite{Levi:2010,*Porto:2010,*Perrodin:2010}.

\mysubsubsection{Next-to-Leading Order Spin(1)-Spin(2)}
The spin(1)-spin(2), or S$_1$S$_2$, Hamiltonian reads \cite{Steinhoff:Hergt:Schafer:2008:2}
\begin{align}
% \begin{split}
	H_{\text{S}_1\text{S}_2}^{\text{NLO}} &= \frac{1}{2 m_1 m_2 \hat{r}_{1 2}^3} [
			\tfrac{3}{2} \picSin \piicSiin
			+ \tfrac{1}{2} \SiSii \pipii \nonumber \nl
			+ 6 \piicSin \picSiin
			- \tfrac{1}{2} \Sipii \Siipi \nonumber \nl
			- 15 \Sin \Siin \pin \piin
			+ \Sipi \Siipii \nonumber \nl
			- 3 \Sin \Siin \pipii
			+ 3 \Sipii \Siin \pin \nonumber \nl
			+ 3 \Siipi \Sin \piin
			+ 3 \Sipi \Siin \piin \nonumber \nl
			+ 3 \Siipii \Sin \pin
			- 3 \SiSii \pin \piin
		] \nonumber \nl
		+ \frac{3}{2 m_1^2 \hat{r}_{1 2}^3} [
			- \picSin \picSiin
			+ \SiSii \pin^2 \nonumber \nl
			- \Sin \Siipi \pin
		]
		+ \frac{3}{2 m_2^2 \hat{r}_{1 2}^3} [
			\SiSii \piin^2 \nonumber\nl
			- \piicSiin \piicSin
			- \Siin \Sipii \piin
		] \nonumber \nl
		+ \frac{6 ( m_1 + m_2 )}{\hat{r}_{1 2}^4} [ \SiSii - 2 \Sin \Siin ] \,, \label{HS1S2}
% \end{split}
\end{align}
and was confirmed by \cite{Porto:Rothstein:2008:1,Levi:2008}.
Notice that no agreement with the result in \cite{Porto:Rothstein:2006}
could be found, see \cite{Steinhoff:Hergt:Schafer:2008:2}. Indeed,
the result in \cite{Porto:Rothstein:2006} turned out to be incomplete
\cite{Steinhoff:Hergt:Schafer:2008:2,Porto:Rothstein:2008:1}.

\mysubsubsection{Next-to-Leading Order Spin(1)-Spin(1)\label{NLOS2}}
A nonreduced potential (i.e., with the spin supplementary condition not
eliminated on the level of the potential) for the next-to-leading order
spin(1)-spin(1), or S$_1^2$, dynamics is given in \cite{Porto:Rothstein:2008:2,Porto:Rothstein:2008:2:err}.
Within the method described in \Sec{S2calc} an equivalent Hamiltonian
$H_{\text{S$_1^2$}}^{\text{NLO}}$ will be derived here.
This Hamiltonian was first given only for the black hole case ($C_{Q1} = 1$)
in \cite{Steinhoff:Hergt:Schafer:2008:1} and then generalized to
arbitrary $C_{Q1}$ later \cite{Hergt:Steinhoff:Schafer:2010:1}.
However, the comparison with
\cite{Porto:Rothstein:2008:2,Porto:Rothstein:2008:2:err} was quite cumbersome.
First agreement with \cite{Porto:Rothstein:2008:2} could not even
be found in the spin precession equation \cite{Steinhoff:Hergt:Schafer:2008:1},
however, this finally succeeded after identifying a sign typo in
\cite{Porto:Rothstein:2008:2}, see \cite{Steinhoff:Schafer:2009:1}
(all for the case $C_{Q1} = 1$).
After a further correction \cite{Porto:Rothstein:2008:2:err} full
agreement was finally found in \cite{Hergt:Steinhoff:Schafer:2010:1}, now
also for arbitrary $C_{Q1}$. For this comparison
the potential from \cite{Porto:Rothstein:2008:2,Porto:Rothstein:2008:2:err} was first
transformed into a fully reduced Hamiltonian in \cite{Hergt:Steinhoff:Schafer:2010:1} by a Legendre transformation
and an elimination of the spin supplementary condition using Dirac brackets (\ref{DB}).
Then a canonical transformation leading to our result in
\cite{Hergt:Steinhoff:Schafer:2010:1} was searched for and found.

The result for general compact objects (including neutron stars) is \cite{Hergt:Steinhoff:Schafer:2010:1}
\begin{align}
% \begin{split}
H_{\text{S$_1^2$}}^{\text{NLO}} &=
\frac{m_{2}}{m_{1}^3 \hat{r}_{12}^3} \bigg[
	\left(-\frac{5}{4}+\frac{3}{2}C_{Q1}\right) \! \Sipi^2
	+ \left(-\frac{21}{8}+\frac{9}{4}C_{Q1}\right) \! \pipi\Sin^2 \nonumber \nl
	+ \left(\frac{15}{4}-\frac{9}{2}C_{Q1}\right) \! \pin\Sin\Sipi
	+ \left(\frac{5}{4}-\frac{5}{4}C_{Q1}\right) \! \pipi\SiSi \nonumber \nl
	+ \left(-\frac{9}{8}+\frac{3}{2}C_{Q1}\right) \! \pin^2\SiSi
\bigg]
+ \frac{C_{Q1}}{m_{1}m_{2}\hat{r}_{12}^3} \bigg[
	\frac{9}{4} \piipii\Sin^2 - \frac{3}{4} \piipii\SiSi
\bigg] \nonumber \nl
+\frac{1}{m_{1}^2 \hat{r}_{12}^3} \bigg[
	\left(-\frac{3}{2}+\frac{9}{2}C_{Q1}\right) \! \piin\Sin\Sipi \nonumber \nl
	- \frac{15}{4}C_{Q1} \pin\piin\Sin^2
	+ \left(\frac{3}{2}-\frac{3}{2}C_{Q1}\right) \! \Sipi\Sipii \nonumber \nl
	+ \left(-3+\frac{3}{2}C_{Q1}\right) \! \pin\Sin\Sipii
	+ \left(-\frac{3}{2}+\frac{9}{4}C_{Q1}\right) \! \pipii\SiSi \nonumber \nl
	+ \left(\frac{3}{2}-\frac{3}{4}C_{Q1}\right) \! \pin\piin\SiSi
	+ \left(3-\frac{21}{4}C_{Q1}\right) \! \pipii\Sin^2
\bigg] \nonumber\nl
+ \frac{m_{2}}{\hat{r}_{12}^4} \bigg[
	\left(-3-\frac{3}{2}C_{Q1}\right) \! \Sin^2
	+ \left(2+\frac{1}{2}C_{Q1}\right) \! \SiSi
\bigg] \nonumber \nl
+ \frac{m_{2}^2}{m_{1}\hat{r}_{12}^4} \bigg[
	(1+2C_{Q1}) \SiSi
	+ (-1-6C_{Q1}) \Sin^2
\bigg] \,. \label{HS1S1}
% \end{split}
\end{align}
The corresponding spin(2)-spin(2) Hamiltonian $H_{\text{S}_2^2}^{\text{NLO}}$
simply results from an exchange of particle labels.
According to \Sec{S2calc} the linear-in-$G$ part was derived
with the help of the Poincar\'e algebra method from \cite{Hergt:Schafer:2008}, while
the $G^2$ part (the last two lines) results from the source
expressions (\ref{H2exp1}--\ref{H2exp3}) derived in the present thesis.

Notice that for black holes ($C_{Q1} = 1$) this Hamiltonian was already found in
\cite{Steinhoff:Hergt:Schafer:2008:1}, for the first time including the correct center of mass motion.
Further, the earlier result for the general case in
\cite{Porto:Rothstein:2008:2,Porto:Rothstein:2008:2:err}
is not a fully reduced Hamiltonian. The Hamiltonian
presented here is on a higher level of sophistication
with advantages for applications, e.g., the spin vectors
appearing in our Hamiltonian have a constant length and it
is easier to obtain all equations of motion in terms of
these ``good'' spin variables.

\mysubsection{Center of Mass and Poincar\'e Algebra\label{poincarecheck}}
The post-Newtonian expansion of the center of mass vector
\begin{equation}
G^i = - \frac{1}{16\pi} \int \dd^3x \, x^i \Delta \phi
	= G_{\text{N}}^i + G_{\text{1PN}}^i + G_{\text{2PN}}^i + \cdots \,,
\end{equation}
can be obtained from the expanded Hamilton constraint (\ref{Hconst2}, \ref{Hconst6}).
To the formal second post-Newtonian order this leads to
\begin{gather}
G_{\text{N}}^i = \int \dd^3x \, x^i \, \mathcal{H}^{\text{matter}}_{(2)} \,, \qquad
G_{\text{1PN}}^i = \int \dd^3x \, x^i \left[ \mathcal{H}^{\text{matter}}_{(4)}
	- \frac{1}{8} \mathcal{H}^{\text{matter}}_{(2)} \phi_{(2)} \right] \,, \\
\begin{split}
G_{\text{2PN}}^i &= \int \dd^3x \, \bigg[ x^i \bigg( \mathcal{H}^{\text{matter}}_{(6)}
	- \frac{1}{8} \left( \mathcal{H}^{\text{matter}}_{(4)} \phi_{(2)}
		+ \mathcal{H}^{\text{matter}}_{(2)} \phi_{(4)} \right) \nl
	+ \frac{1}{64} \mathcal{H}^{\text{matter}}_{(2)} \phi_{(2)}^2
	+ V^i_{(3)} \mathcal{H}^{\text{matter}}_{(3) i} \bigg)
	+ \frac{1}{16\pi} \frac{5}{2} V^i_{(3)} \tilde{\pi}^k_{(3),k} \bigg] \,.
\end{split}
\end{gather}
For the formula for $G_{\text{2PN}}^i$ partial integrations were applied,
similar as in \Sec{matHam}.
Notice that $\tilde{\pi}^k_{(3),k}$ is spin-independent.
For results in the nonspinning case see \cite{Damour:Jaranowski:Schafer:2000}.

The contributions to the center of mass vector corresponding to the leading order spin Hamiltonians
follow from $\vct{G}_{\text{1PN}} = (G_{\text{1PN}}^i)$ as
\begin{equation}
	\mathbf{G}_{\text{SO}}^{\text{LO}} =
		\sum_a \frac{1}{2 m_a} (\hat{\mathbf{p}}_a \times \hat{\mathbf{S}}_a) \,,
	\qquad \mathbf{G}_{\text{S}_1\text{S}_2}^{\text{LO}} = 0 \,,
	\qquad \mathbf{G}_{\text{S}_1^2}^{\text{LO}} = 0 \,.
\end{equation}
From $\vct{G}_{\text{2PN}} = (G_{\text{2PN}}^i)$ the next-to-leading order parts result as
\begin{align}
\begin{split}\label{GSO}
	\mathbf{G}_{\text{SO}}^{\text{NLO}} &=
	\sum_a \sum_{b \neq a} \frac{m_b}{4 m_a \hat{r}_{ab}} \bigg[
			((\hat{\mathbf{p}}_a \times \hat{\mathbf{S}}_a) \cdot \hat{\mathbf{n}}_{ab})
				\frac{5\hat{\mathbf{z}}_a+\hat{\mathbf{z}}_b}{\hat{r}_{ab}}
			- 5 (\hat{\mathbf{p}}_a \times \hat{\mathbf{S}}_a)
		\bigg] \nl
	+ \sum_a \sum_{b \neq a} \frac{1}{\hat{r}_{ab}} \bigg[
			\frac{3}{2} (\hat{\mathbf{p}}_b \times \hat{\mathbf{S}}_a)
			- \frac{1}{2} (\hat{\mathbf{n}}_{ab} \times \hat{\mathbf{S}}_a)
				(\hat{\mathbf{p}}_b \cdot \hat{\mathbf{n}}_{ab}) \nlq
			- ((\hat{\mathbf{p}}_b \times \hat{\mathbf{S}}_a) \cdot \hat{\mathbf{n}}_{ab})
				\frac{\hat{\mathbf{z}}_a+\hat{\mathbf{z}}_b}{\hat{r}_{ab}}
		\bigg]
	- \sum_a \frac{\hat{\mathbf{p}}_a^2}{8 m_a^3} (\hat{\mathbf{p}}_a \times \hat{\mathbf{S}}_a) \,,
\end{split}\\
	\mathbf{G}_{\text{S}_1\text{S}_2}^{\text{NLO}} &=
	\frac{1}{2} \sum_a \sum_{b \neq a} \bigg[
	\left(3(\hat{\mathbf{S}}_{a}\cdot\hat{\mathbf{n}}_{ab})(\hat{\mathbf{S}}_{b}\cdot\hat{\mathbf{n}}_{ab})
		-(\hat{\mathbf{S}}_{a}\cdot\hat{\mathbf{S}}_{b})\right)
		\frac{\hat{\mathbf{z}}_{a}}{\hat{r}_{ab}^3}
	+ (\hat{\mathbf{S}}_{b}\cdot\hat{\mathbf{n}}_{ab})
		\frac{\hat{\mathbf{S}}_{a}}{\hat{r}_{ab}^2} \bigg] \,, \label{GS1S2} \\
\begin{split}\label{GS1S1}
\vct{G}_{\text{S}_{1}^2}^{\text{NLO}} &= \frac{m_{2}}{m_{1}} \bigg[
	C_{Q1} \left( 3\Sin^2 - \SiSi \right) \frac{\hat{\vct{z}}_{1}+\hat{\vct{z}}_{2}}{4 \hat{r}_{12}^3}
	+ (1+C_{Q1}) \SiSi \frac{\hat{\vct{n}}_{12} }{2 \hat{r}_{12}^3} \nlq
	- (1+3 C_{Q1}) \Sin \frac{\hat{\vct{S}}_1}{2\hat{r}_{12}^2}
\bigg] \,.
\end{split}
\end{align}
Notice that (\ref{GSO}) and (\ref{GS1S2}) are valid for arbitrary many spinning objects,
while (\ref{GS1S1}) holds for two objects only.
For two objects $\mathbf{G}_{\text{SO}}^{\text{NLO}}$ was already found in \cite{Damour:Jaranowski:Schafer:2008:1}.
Further, $\vct{G}_{\text{S}_2^2}^{\text{NLO}}$ simply results from an exchange
of particle labels in (\ref{GS1S1})

Now one can check whether the Poincar\'e algebra (\ref{poinc1}, \ref{poinc2}) is fulfilled,
which is indeed the case (the Hamiltonian plays of course the role of
the energy $E$). At the spin-orbit level this was already shown in \cite{Damour:Jaranowski:Schafer:2008:1}.
At the spin(1)-spin(1) level this holds by construction,
as most terms of the Hamiltonian $H_{\text{S$_1^2$}}^{\text{NLO}}$ were obtained
from the Poincar\'e algebra via an ansatz in \cite{Hergt:Schafer:2008}.
However, the fulfillment of the Poincar\'e algebra provides a thorough check of
$H_{\text{SO}}^{\text{NLO}}$ and $H_{\text{S}_1\text{S}_2}^{\text{NLO}}$.

\mychapter{Conclusions and Outlook\label{conc}}
The first main goal of this thesis, the extension of the ADM
canonical formalism from nonspinning to spinning objects, succeeded
to linear order in spin via an action approach. The result was verified
by an independent order-by-order derivation. Even the extension to
higher orders in spin is well understood now, but somewhat more complicated
and requires further approximations, like the post-Newtonian one.
The second main goal of this thesis, the calculation of conservative
Hamiltonians for inspiralling binaries relevant for gravitational wave astronomy, was then
straightforward. The effort of first deriving the canonical formalism
was payed off by its efficiency in the calculation of these Hamiltonians.
New results are the next-to-leading order spin(1)-spin(2) and
spin(1)-spin(1) Hamiltonians, and the spin-orbit Hamiltonian derived
earlier by Damour, Jaranowski, and Sch\"afer was confirmed.
All Hamiltonians through the third post-Newtonian order for maximal spin are known.

The next most interesting Hamiltonian which could be calculated is
the conservative next-to-next-to-leading order spin-orbit one, which is
at the 3.5 post-Newtonian level for maximally rotating objects. Notice
that the verification of the canonical formalism given in this thesis via
the order-by-order construction already covered this case.
Leading order \emph{dissipative} Hamiltonians are also envisaged and its calculation
was already prepared in \cite{Steinhoff:Wang:2009}.
For maximal spins these Hamiltonians are even at the fourth post-Newtonian order
in the spin-orbit case and at the 4.5 post-Newtonian order in the spin(1)-spin(2) case.
The extension of a recent result within the \emph{post-Minkowskian} approximation \cite{Ledvinka:Schafer:Bicak:2008}
to spinning objects would also be desirable, as it could be applied to the gravitational scattering
of spinning bodies moving at relativistic speed.

Further, more work needs to be done for an application of the new Hamiltonians presented in this thesis
to gravitational wave astronomy. In particular the spin contributions to the
next-to-leading order radiation field are only known for the spin-orbit case
\cite{Blanchet:Buonanno:Faye:2006,*Blanchet:Buonanno:Faye:2006:err}.
This result should be extended to the spin(1)-spin(2) case, as well as to the
spin(1)-spin(1) case. For the latter the spin(1)-spin(1) contributions to
the stress-energy tensor given in this thesis are crucial. Also an
implementation of the new results given here into the very successful
effective one-body approach would be appealing.

Another though rather mathematical development for the future would be
to consider the full constraint algebra, gravitational field and
supplementary conditions, at different stages of gauge fixing,
as well as a treatment using Dirac brackets;
see also \cite{Nelson:Teitelboim:1978} for the case of Dirac fields.

% \ifpreprint
% \paragraph{Note Added}
% \currentpdfbookmark{Note Added}{note}
% \input{note}
% \fi

\paragraph*{Acknowledgments}
\currentpdfbookmark{Acknowledgments}{ack}
First I like to thank Prof.~Dr.~Gerhard Sch\"afer for the excellent supervision of this thesis.
He was aware of the elegance of the ADM formalism and its value for post-Newtonian calculations.
Especially my first two publications \cite{Steinhoff:Hergt:Schafer:2008:2,Steinhoff:Schafer:Hergt:2008}
gained pertinence and sophistication due to his influence and he substantially contributed
to almost all of my other publications. Much of the material in the present thesis is influenced by the many useful
discussions with him.

Another significant contribution to this thesis was given by Steven Hergt, as he often was a
coauthor of mine and I am grateful for the prolific collaboration with him.
Regarding the content of this thesis, he calculated the spin(1)-spin(1) center of mass
vector (\ref{GS1S1}) as well as the linear-in-$G$ part of the spin(1)-spin(1)
Hamiltonian (\ref{HS1S1}) and provided the calculations regarding the Poincar\'e algebra in \Sec{poincarecheck}.
He also did most of the calculations for the full comparison (i.e., including the center of mass motion)
of the spin(1)-spin(1) Hamiltonian (\ref{HS1S1}) with the result in \cite{Porto:Rothstein:2008:2,Porto:Rothstein:2008:2:err}, see \cite{Hergt:Steinhoff:Schafer:2010:1}.
He further often checked my calculations, in particular the ones for the
spin-orbit and spin(1)-spin(2) Hamiltonians (\ref{HSO}, \ref{HS1S2})
as well as the $G^2$ part of the spin(1)-spin(1) Hamiltonian (\ref{HS1S1}).

Also many other people influenced this thesis and broadened my field of interest through stimulating discussions.
I am very thankful for all these useful discussions, particularly
with Dr.~Dirk Puetzfeld on the multipole formalism in general relativity,
with Dr.~Han Wang on radiation reaction effects,
with Manuel Tessmer on data analysis and parametrization of orbits,
with Johannes Hartung on data analysis and Mathematica related issues,
and with Dr.~David Brizuela on the handling of xPert and xTensor.
Further I gratefully acknowledge useful comments on this thesis by
Prof.~Dr.~Ji\v{r}{\'i} Bi\v{c}{\'a}k, Jan Sperrhake, and David Hilditch.
Finally, the German Research Foundation
(Deutsche Forschungsgemeinschaft, DFG) made
this work possible by providing both financial support
through the SFB/TR7 ``Gravitational Wave Astronomy''
and advanced education through the GRK 1523 ``Quantum and Gravitational Fields.''

Further, this work was encouraged by my family and friends.
They placed me in a sociable environment that helped to keep my mind clear.
I also gratefully acknowledge the pleasant atmosphere within my workgroup.
Last but not least I thank my girlfriend Julia Damm for her understanding
that the spare time I could spent with her diminished while writing this thesis.

% \newpage
\appendix
\def\columnseprule{0.5pt}
\mychapter{Symbols\label{symbols}}
\begin{multicols}{2}
\begin{tabbing}
% \hspace{1.2cm} \= \kill
% Symbol \> Definition \\
% \line(1,0){5cm} \\
\hspace{1.1cm} \= \kill
% $a$ \> dimensionless Kerr parameter, $a = 0 \dots 1$ \\
% $a^i$ \> constant coordinate translation \\
% $A^{ij}$ \> defined by (\ref{defA}) \\
$\hat{A}^{ij}$ \> defined by (\ref{defAhat}) \\
% $B^{ij}$ \> defined by (\ref{defBij}) \\
% $\hat{B}^{ij}$ \> defined by (\ref{Bijcan}) \\
$B^{ij}_{kl}$ \> defined by (\ref{Bedef}) \\
% $B^{(4)}_{\mu\nu}$ \> magnetic part of $C^{(4)}_{\mu\nu\alpha\beta}$, see (\ref{EMWeyl}) \\
$c$ \> speed of light, usually $c=1$ here \\
% $C^i$ \> defined by (\ref{defCi}) \\
$C_Q$ \> mass-quadrupole parameter, see (\ref{Qansatz}) \\
% $C^{(4)}_{\mu\nu\alpha\beta}$ \> 4-dimensional Weyl tensor, see (\ref{Weyl}) \\
$\delta$ \> defined as $\delta = \delta(x^i - z^i)$ \\
$\hat{\delta}$ \> defined as $\hat{\delta} = \delta(x^i - \hat{z}^i)$ \\
$\delta_{(4)}$ \> defined as $\delta_{(4)} = \delta(x^{\mu} - z^{\mu})$ \\
$\delta_{ij}$ \> Kronecker delta, $(\delta_{ij}) = \text{diag}(1, 1, 1)$ \\
$\delta^{\text{TT}kl}_{ij}$ \> transverse traceless projector, see (\ref{TTproj}) \\
$\Delta$ \> Laplace operator, $\Delta = \partial_i \partial_i$ \\
$\Delta^{-1}$ \> inverse of $\Delta$ for usual boundary conditions \\
$e_{I\mu}$ \> tetrad field, $g_{\mu\nu} = e_{I\mu} {e^I}_{\nu}$ \\
$e_{ij}$ \> triad in the symmetric gauge, see (\ref{egauge}) \\
$\hat{e}^{ij}$ \> defined as $\hat{e}^{ij} \equiv \frac{1}{2} (e^{i(j)} - e^{j(i)})$ \\
$\hat{e}^{ij}_a$ \> defined as $\hat{e}^{ij}_a \equiv \hat{e}^{ij}(\hat{z}_a^k)$ \\
$\epsilon_{ijk}$ \> 3-dimensional Levi-Civita symbol \\
% $\epsilon^{(4)}_{\mu\nu\alpha\beta}$ \> 4-dimensional Levi-Civita symbol \\
$E$ \> energy of the system, see (\ref{Gint}) \\
$E^{(4)}_{\mu\nu}$ \> electric part of $C^{(4)}_{\mu\nu\alpha\beta}$, see (\ref{EMWeyl}, \ref{Weyl}) \\
$f^{\mu}$ \> timelike vector in conditions (\ref{covSSC2}, \ref{Lssc}) \\
% $f_{ij}$ \> source of the wave equation (\ref{htt4}) \\
$\phi$ \> trace part of the induced metric, see (\ref{gdecomp2}) \\
% $\varphi_i$ \> angle variables parametrizing a rotation \\
$g$ \> defined as $ g = \det( g_{\mu\nu} )$ \\
$g_{\mu\nu}$ \> 4-dimensional metric \\
$\gamma$ \> defined as $\gamma = \det( \gamma_{ij} )$ \\
$\gamma_{\mu\nu}$ \> projector (\ref{nprojector}), contains induced metric \\
$G$ \> gravitational constant, usually $G=1$ here \\
$G^i$ \> center of mass vector, see (\ref{Gint}) \\
% $\vct{G}$ \> vector notation for $G^i$ \\
$\Gamma_{kij}$ \> 3-dim.\ Christoffel symbol of first kind \\
$\Gamma^{(4)}_{\alpha\mu\nu}$ \> 4-dim.\ Christoffel symbol of first kind \\
$h_{ij}^{\text{TT}}$ \> transverse traceless part of $\gamma_{ij}$, see (\ref{gdecomp2}) \\
$H$ \> general symbol for a Hamiltonian \\
$H_{\text{ADM}}$ \> ADM Hamiltonian, see (\ref{HADMspin}) \\
% $H_G$ \> field Hamiltonian, see (\ref{Dham}) \\
% $H_M$ \> test-body Hamiltonian, see (\ref{Htest}) \\
% $H_{M\tau}$ \> see (\ref{HM}) or (\ref{HMquad}) \\
$\mathcal{H}$ \> Hamilton constraint, see (\ref{fconstraints}) \\
$\mathcal{H}_i$ \> momentum constraint, see (\ref{fconstraints}) \\
$\mathcal{H}^{\text{field}}$ \> field part of $\mathcal{H}$, see (\ref{HHifield}) \\
$\mathcal{H}^{\text{field}}_i$ \> field part of $\mathcal{H}_i$, see (\ref{HHifield}) \\
$\mathcal{H}^{\text{matter}}$ \> matter part of $\mathcal{H}$, see \Sec{Hmattercan} \\
$\mathcal{H}^{\text{matter}}_i$ \> matter part of $\mathcal{H}_i$, see \Sec{Hmattercan} \\
$\mathcal{H}^{\pi \text{matter}}_i$ \> defined by (\ref{HSpi}) \\
% $\mathcal{H}^{\text{matter}}_{\text{S}^2}$ \> quadratic-in-spin part of $\mathcal{H}^{\text{matter}}$ \\
$I$ \> moment of inertia of a spherical top \\
$J^{\mu\nu}$ \> total angular momentum, see (\ref{boost}, \ref{PJsum}) \\
% $J_i$ \> defined as $J_i = \tfrac{1}{2} \epsilon_{ijk} J^{jk}$ \\
$J^{\mu\nu\alpha\beta}$ \> Dixon's quadrupole moment \\
% $J_{a[i][j]}^{\text{body}}$ \> see (\ref{Jbody}) \\
% $J_{ij}^{\text{field}}$ \> field part of $J_{ij}$, see (\ref{Jfield}) \\
% $J_{ij}^{\text{matter}}$ \> matter part of $J_{ij}$, see (\ref{Jmatter}) \\
% $k_i$ \> Fourier space coordinate belonging to $x^i$ \\
% $\vct{k}$ \> vector notation for $k_i$ \\
$K_{ij}$ \> extrinsic curvature, see (\ref{Kdef}) \\
$\lambda$ \> Lagrange multiplier \\
% $\lambda_{ij}$ \> Lagrange multiplier, see (\ref{Htriad}) \\
$L_M$ \> matter Lagrangian \\
$\mathcal{L}_G$ \> field Lagrangian density (\ref{EHaction}, \ref{EH31}, \ref{WADM}) \\
$\mathcal{L}_M$ \> matter Lagrangian density \\
$\Lambda^{AI}$ \> Lorentz matrix, see \Sec{mincoup} \\
$\hat{\Lambda}^{[i](j)}$ \> canonical rotation matrix defined by (\ref{Lcan}) \\
$m$ \> constant mass-like parameter, see (\ref{mtom0}) \\
$m_0$ \> constant mass-like parameter, see (\ref{Lspin}) \\
$m_p$ \> dynamical mass, $m_p^2 = - p_{\mu}p^{\mu}$ \\
$M$ \> mass of the system, $M^2 = - P_{\mu} P^{\mu}$ \\
$n^{\mu}$ \> normal vector for (3+1)-split, see (\ref{ndef}) \\
$\hat{\vct{n}}_{ab}$ \> defined by $\vct{\hat{n}}_{ab} = ( \hat{\vct{z}}_a - \hat{\vct{z}}_b ) / \hat{r}_{ab}$ \\
$np$ \> defined by $np = n^{\mu} p_{\mu}$, see (\ref{npDef}) or (\ref{npQuad}) \\
$n\hat{p}$ \> defined by $n\hat{p} = - \sqrt{m^2 + \gamma^{ij} \hat{p}_i \hat{p}_j}$ \\
$nS_i$ \> defined by $nS_i = n^{\mu} S_{\mu i}$, see (\ref{nSDef}, \ref{zredef}) \\
$N$ \> lapse function, see (\ref{ndef}) \\
$N^i$ \> shift vector, see (\ref{ndef}) \\
% $\omega^{ij}$ \> antisymmetric matrix parametrizing a constant rotation \\
% $\omega^{[i][j]}_a$ \> as $\omega^{ij}$, but rotates the body-fixed frame, see (\ref{bodysym}) \\
${\omega_{\mu}}^{IJ}$ \> Ricci rotation coefficients \\
$\Omega^{\mu\nu}$ \> angular velocity tensor, see (\ref{Odef}) \\
$\hat{\Omega}^{(i)(j)}$ \> angular velocity
	$\hat{\Omega}^{(i)(j)} = \hat{\Lambda}_{[k]}{}^{(i)} \dot{\hat{\Lambda}}^{[k](j)}$  \\
$p_{\mu}$ \> linear momentum, see (\ref{PUrel}, \ref{PUrelQuad}) \\
$\hat{p}_i$ \> canonical momentum conjugate to $\hat{z}^i$ \\
% $\hat{\vct{p}}$ \> vector notation for $\hat{p}_i$ \\
% $p_i^{\varphi}$ \> canonical conjugate to the angle variables $\varphi_i$ \\
$P_{\mu}$ \> total linear momentum, $P_0 = - E$, (\ref{PJsum}) \\
% $P_i^{\text{field}}$ \> field part of the total linear momentum, see (\ref{Pfield}) \\
% $P_i^{\text{matter}}$ \> matter part of the total linear momentum, see (\ref{Pmatter}) \\
$P_{\mu\nu}$ \> the projector $P_{\mu\nu} = g_{\mu\nu} - \frac{1}{f_{\rho}f^{\rho}} f^{\mu} f^{\nu}$ \\
% $\mathcal{P}_{ij}$ \> defined by (\ref{Pcal}) \\
$\pi^{ij}$ \> defined by (\ref{pfield}) \\
$\pi^{ij\text{TT}}$ \> transverse traceless part of $\pi^{ij}$, (\ref{decomp}, \ref{pidecomp2}) \\
$\pi^{ij}_a$ \> spin correction to $\hat{\pi}^{ij\text{TT}}$, see (\ref{pican}) \\
$\tilde{\pi}^{i}$ \> vector potential for $\tilde{\pi}^{ij}$, see (\ref{pisolve}) \\
$\tilde{\pi}^{ij}$ \> vector potential part of $\pi^{ij}$, see (\ref{decomp}, \ref{pisolve}) \\
$\breve{\pi}^{ij}$ \> trace part of $\pi^{ij}$, see (\ref{decomp}, \ref{pidecomp2}) \\
$\hat{\pi}^{ij}$ \> canonical field momentum, see (\ref{NWpi}) \\
$\hat{\pi}^{ij\text{TT}}$ \> transverse traceless part of $\hat{\pi}^{ij}$, see (\ref{picandecomp}) \\
$\hat{\tilde{\pi}}^{i}$ \> vector potential for $\hat{\tilde{\pi}}^{ij}$, see (\ref{picandecomp}) \\
$\hat{\tilde{\pi}}^{ij}$ \> longitudinal part of $\hat{\pi}^{ij}$, see (\ref{picandecomp}) \\
% $\bar{\pi}^{(i)j}$ \> canonical conjugate to the triad, see (\ref{pecan}) \\
$Q^{\mu\nu}$ \> mass quadrupole part of $J^{\mu\nu\alpha\beta}$, see (\ref{Jdecomp}) \\
$\hat{Q}_{ij}$ \> defined by (\ref{Qcan}) \\
% $Q_{\mu\nu}^{\text{STF}}$ \> traceless part of $Q^{\mu\nu}$, see (\ref{Qdecomp}) \\
% $Q^{\mu\nu\alpha}$ \> flow quadrupole part of $J^{\mu\nu\alpha\beta}$, see (\ref{Jdecomp}) \\
% $Q^{\mu\nu\alpha\beta}$ \> stress quadrupole part of $J^{\mu\nu\alpha\beta}$, see (\ref{Jdecomp}) \\
% $r$ \> distance of the components of a binary \\
$\hat{r}_a$ \> defined by $\hat{r}_a = | \vct{x} - \hat{\vct{z}}_a |$ \\
$\hat{r}_{ab}$ \> defined by $\hat{r}_{ab} = | \hat{\vct{z}}_a - \hat{\vct{z}}_b |$ \\
$\Riem$ \> 3-dimensional Ricci scalar, $\Riem = \gamma^{ij} \Riem_{ij}$ \\
$\Riem^{(4)}$ \> 4-dimensional Ricci scalar, $\Riem = g^{\mu\nu} \Riem^{(4)}_{\mu\nu}$ \\
$\Riem_{ij}$ \> 3-dim.\ Ricci tensor, $\Riem_{ij} = \gamma^{kl} \Riem_{ikjl}$ \\
$\Riem^{(4)}_{\mu\nu}$ \> 4-dim.\ Ricci tensor, $\Riem^{(4)}_{\mu\nu} = g^{\alpha\beta} \Riem^{(4)}_{\mu\alpha\nu\beta}$ \\
$\Riem_{ijkl}$ \> 3-dim.\ Riemann tensor, sign as in (\ref{Rdef}) \\
$\Riem^{(4)}_{\mu\nu\alpha\beta}$ \> 4-dim.\ Riemann tensor, see (\ref{Rdef}) \\
$s_a^{ij}$ \> defined by (\ref{defP}), see also (\ref{cansource3}) \\
$S$ \> spin length, $2 S^2 = S^{\mu\nu} S_{\mu\nu}$ \\
$S^{\mu\nu}$ \> spin tensor, usually restricted to (\ref{pSSC}) \\
$\hat{S}^{\mu\nu}$ \> canonical spin tensor, see (\ref{PBScan}) \\
$\hat{S}_{(i)}$ \> canonical spin vector, $\hat{S}_{(i)} = \frac{1}{2} \epsilon_{ijk} \hat{S}_{(j)(k)}$ \\
$\hat{\vct{S}}$ \> canonical spin vector, $\hat{\vct{S}} = (\hat{S}_{(i)})$ \\
% $\tilde{S}^{\mu\nu}$ \> spin tensor belonging to the condition (\ref{comSSC}) \\
$t$ \> time coordinate, $x^0 \equiv t$ or $z^0 \equiv t$ \\
$\tau$ \> worldline parameter \\
$t^{\mu\nu\dots}$ \> multipole moments, see (\ref{Texp}) \\
$\delta \theta^{IJ}$ \> variation for $\Lambda^{AI}$, $\delta \theta^{IJ} = {\Lambda_A}^I \delta \Lambda^{AJ}$ \\
$T^{\mu\nu}$ \> stress-energy tensor \\
$(\text{td})$ \> denotes a total divergence \\
$u^{\mu}$ \> 4-velocity, $u^{\mu} = \frac{\dd z^{\mu}}{\dd \tau}$ \\
% $v$ \> relative velocity of a binary \\
$V^i$ \> vector potential for $\tilde{\pi}^{ij}$, see (\ref{Vtrans}, \ref{pidecompV}) \\
$\hat{V}^i$ \> vector potential for $\hat{\tilde{\pi}}^{ij}$, see (\ref{Vcan}, \ref{picanV}) \\
$W$ \> full action, $W = W_G + W_M$ \\
$W_G$ \> Einstein-Hilbert action, see (\ref{EHaction}) \\
$W_M$ \> matter part of the action $W$ \\
$x^{\mu}$ \> spacetime coordinates, $x^0 \equiv t$ \\
% $\vct{x}$ \> vector notation for $x^i$ \\
$z^{\mu}$ \> worldline function, $z^0 \equiv t$ \\
$\hat{z}^i$ \> canonical position variable \\
% $\hat{\vct{z}}$ \> vector notation for $\hat{z}^i$ \\
$z^i_{\Delta}$ \> possible correction to $\hat{z}^i$, see (\ref{NWpos})
% $Z^{\mu}$ \> center belonging to condition (\ref{covSSC}) \\
% $\tilde{Z}^{\mu}$ \> center belonging to condition (\ref{comSSC}) \\
% $\hat{Z}^{\mu}$ \> center belonging to condition (\ref{NWSSC})
\end{tabbing}
\end{multicols}

\setlength{\bibsep}{0pt}
% \bibliographystyle{utphys}
% \bibliography{references}
\providecommand{\href}[2]{#2}\begingroup\raggedright\endgroup

\addcontentsline{toc}{section}{\refname}

\end{document}